\documentclass[twocolumn,aps,prx,showkeys,longbibliography]{revtex4-1}             
\usepackage{graphics}
\usepackage{hyperref}
\hypersetup{
     colorlinks=true,       
    linkcolor=blue,          
    citecolor=blue,        
    filecolor=magenta,      
    urlcolor=cyan           
}
\usepackage{epsfig}
\usepackage{amssymb}
\usepackage{amsmath}
\usepackage{bbm}

\begin{document}

\title{Nonlinear quantum transport of light in a cold atomic cloud}
\author{Tobias Binninger}
\altaffiliation[Present address:]{ IBM Research-Zurich, S\"aumerstr. 4, CH-8803 R\"uschlikon}
\affiliation{Physikalisches Institut, Albert-Ludwigs-Universit\"at, Hermann-Herder-Str. 3, D-79104 Freiburg, Germany}
\author{Vyacheslav N. Shatokhin}
\affiliation{Physikalisches Institut, Albert-Ludwigs-Universit\"at, Hermann-Herder-Str. 3, D-79104 Freiburg, Germany}
\author{Andreas Buchleitner}
\affiliation{Physikalisches Institut, Albert-Ludwigs-Universit\"at, Hermann-Herder-Str. 3, D-79104 Freiburg, Germany}
\author{Thomas Wellens}
\affiliation{Physikalisches Institut, Albert-Ludwigs-Universit\"at, Hermann-Herder-Str. 3, D-79104 Freiburg, Germany}

\begin{abstract}
We outline the non-perturbative theory of multiple scattering of resonant, intense laser light off a dilute cloud of cold atoms. A combination of master equation and diagrammatic techniques allows, for the first time, a quantitative description of nonlinear diffusive transport as well as of coherent backscattering of the injected electromagnetic field, notwithstanding the exponential growth of Hilbert space with the number of atomic scatterers. As an exemplary application, we monitor the laser light's intensity profile within the medium, the spectrum of the backscattered light and the coherent backscattering peak's height with increasing pump intensity. Our theory establishes a general, microscopic, scalable approach to nonlinear transport phenomena in complex quantum materials.
\end{abstract}


\date{\today}

\bibliographystyle{apsrev4-tw}

\maketitle

\section{Introduction}

Wave transport in disordered media is an important subject of research in many areas of physics, ranging, e.g., from the conductance of electrons in disordered metals to multiple scattering of photons in turbid samples \cite{akkermans}.
For three dimensional, linear media -- i.e., if the properties of the scattering medium are not modified by the scattered wave itself (a point to which we will come back below) -- one distinguishes two fundamentally different regimes of transport, which are usually referred to as the regime of \lq weak disorder\rq\ and \lq strong disorder\rq, respectively.

Weakly disordered media, defined by the condition $k\ell\gg 1$ (with wave number $k$ and mean free path $\ell$), essentially give rise to diffusive transport of the average wave intensity. Deviations from a purely diffusive behaviour, however, become visible when measuring the average intensity in the direction exactly opposite to the incident wave, where a coherent backscattering peak appears as a result of interference between wave amplitudes propagating along reversed scattering paths \cite{Albada:1985aa,Wolf:1985aa}. A detailed, microscopic understanding of these and other related effects of mesoscopic transport in weak disorder, such as weak localization \cite{Bergmann:1984aa} or universal conductance fluctuations \cite{Lee:1985aa}, is provided by diagrammatic multiple scattering theory \cite{Rossum:1999aa}, based on the  statistical properties of the disordered medium under study. 

The situation is different in the regime of strong disorder (i.e. $k\ell\simeq 1$ or smaller),
where complete suppression of diffusion due to Anderson localization \cite{Anderson:1958aa} is expected, and indeed has been observed in many different physical systems (e.g., sound waves \cite{Hu:2008aa} or matter waves \cite{Jendrzejewski:2012aa}). Whether it is possible to achieve Anderson localization of light is, according to present knowledge \cite{Skipetrov:2014aa,Sperling:2016aa} and due to the absence of a microscopic theory of multiple scattering in strong disorder,  an open question. 
Similarly, the understanding of recurrent or collective scattering effects \cite{Wiersma:1995aa,Schaefer:2016,Guerin:2017aa,Javanainen:2017aa}, which play an important role in the case of strong disorder, is far from complete.

A good candidate to study fundamental questions of multiple scattering theory is the scattering of light by cold atoms \cite{Labeyrie:1999aa,Kulatunga:2003aa,Jennewein:2016aa,Zhu:2016aa,Bromley:2016aa,Kupriyanov:2017oq}. The scattering properties of single atoms are well known and tunable, e.g., by changing the wave length or the intensity of the incident laser. Using atoms with a suitable level structure, it is furthermore possible to let the atom interact with several laser beams in a nonlinear way, such that  one beam can be used to control  a second one (e.g., to slow down its group velocity using the effect of electromagnetically induced transparency \cite{Fleischhauer:2000aa}).
Therefore, apart from its fundamental interest as a generic quantum transport scenario, 
 a precise understanding of multiple scattering effects in atomic gases is also 
desirable in view of applications such 
as quantum memories \cite{Chaneliere:2005aa}, random lasers \cite{Baudouin:2013aa} or photonic devices in disordered media \cite{Wiersma:2013aa}.

Coherent backscattering of light was experimentally observed for atoms with degenerate and non-degenerate ground states \cite{Labeyrie:1999aa,Bidel:2002aa}, at low temperatures, i.e. when the thermal motion of the atoms can be neglected. In the linear regime of small laser intensity, the results agree well with predictions of diagrammatic multiple scattering theory \cite{Labeyrie:2003aa,Muller:2002aa,kupriyanov03}.
The latter usually assumes that there exists a scattering matrix through which the outgoing field is linearly related to the incident field. This assumption, however, breaks down for larger laser intensity: first, the atomic response becomes nonlinear due to the saturation of the atomic transition. Second, the light scattered by near-resonant atoms exhibits fluctuations due to the quantum mechanical coupling of the atoms to the electromagnetic vacuum. 
These fluctuations are responsible for the incoherent or inelastic component of resonance fluorescence \cite{Mollow:1969aa}, where the frequencies of emitted photons differ from the frequency of the incident laser.

Whereas a decrease of the coherent backscattering interference peak with increasing saturation of the atomic transition was observed experimentally \cite{Chaneliere:2004aa,Balik:2005aa}, no satisfying theory so far exists for incorporating nonlinear and inelastic scattering into a multiple scattering approach. 
A theory for coherent backscattering by nonlinear, classical scatterers was presented in \cite{Wellens:2008aa,Wellens:2009aa}, but does not take into account any quantum fluctuations due to inelastic scattering.
A perturbative method based on the scattering matrix of two photons was proposed in \cite{Wellens:2004aa,Wellens:2006aa}, but is only valid if incident light intensity and optical thickness of the atomic medium are small. Similarly, approaches based on the truncation of a hierarchy of correlation functions \cite{Ruostekoski:1997aa,Lee:2016aa} fail for large laser intensities, because a large number of atoms will become correlated with each other  during a multiple scattering process involving many multi-photon scattering events. On the other hand, standard tools of quantum optics (master
equations, optical Bloch equations, etc.) are well adapted to describe the atom-field interaction for arbitrary intensities of the incident field, but are restricted to a small number of atoms coupled to each other by photon exchange \cite{shatokhin05,Shatokhin:2006,shatokhin07a}. This is due to the fact that the dimension of the atomic Hilbert space grows exponentially with the number of atoms. 
With all above methods having their limitations,
the problem of multiple inelastic scattering of intense laser light in cold atomic ensembles has hitherto  been considered as unsolvable.  

 However, as we here show, this problem can be overcome under the following two assumptions:
 (i) The atoms are placed independently from each other at {\em random positions}. Experimentally measurable quantities like the spectrum of the radiated light are averaged over the atomic positions. (ii) The atomic medium is {\em dilute}, i.e., the typical distance between neighbouring atoms is larger than the wave length of the incident laser. For near-resonant atomic scatterers, these assumptions correspond to the regime   $k\ell\gg 1$ of weak disorder mentioned above.
 Using a diagrammatic multiple scattering representation derived from the quantum optical $N$-atom master equation, we identify
 certain types of multiple scattering processes -- described by so-called ladder and crossed diagrams \cite{Rossum:1999aa} -- which survive the ensemble average over the atomic positions. The sum of all these diagrams leads to numerically solvable transport equations describing, both, nonlinear diffusive transport of photons in the atomic cloud as well as coherent backscattering.  

As already mentioned above, a similar diagrammatic approach has already been developed for nonlinear classical scatterers \cite{Wellens:2008aa,Wellens:2009aa}.
Within this model, the light is scattered purely elastically -- provided that a stationary scattering state is assumed in spite of the nonlinearity \cite{Paul:2005aa,Gremaud:2010aa}. The present paper fully takes into account inelastic scattering induced by the quantum-mechanical nature of the atom-field interaction. In contrast to the classical model, a unique stationary state is always reached in this case, as proven in Appendix~\ref{sec:appendixstat}.

The paper is organized as follows: In Sec.~\ref{sec:model}, we introduce our model consisting of $N$ two-level atoms at fixed, random positions, which are driven by a monochromatic laser and coupled to the electromagnetic vacuum. After tracing over the quantized radiation field, we arrive at a master equation describing the time evolution of atomic observables. 

In Sec.~\ref{sec:formal}, we rewrite the master equation as a generalized optical Bloch equation for $N$ atoms. Formal solutions for the stationary state of the generalized $N$-atom Bloch vector and the  corresponding power spectrum of the light emitted by the atoms in this stationary state are derived.  

On the basis of this formal solution, we introduce a diagrammatic multiple scattering representation in Sec.~\ref{sec:diagrams}. Using this representation, the radiation emitted by $N$ atoms is expressed in terms of single-atom building blocks.
We argue that,  in the case of a dilute atomic medium, 
only certain types of diagrams survive the ensemble average over the atomic positions:
ladder diagrams describing nonlinear diffusive transport, and crossed diagrams giving rise to coherent backscattering.

In Sec.~\ref{sec:ladder}, we perform the summation of all ladder diagrams after averaging over the atomic positions. We show that the light field incident on each single atom can be modelled as a stochastic polychromatic classical field. Thereby, the average power spectrum of the light emitted by a single atom -- as well as the refractive index determining the propagation of light in the effective atomic medium -- can be determined by solving the corresponding single-atom Bloch equations. Finally, the sum of all ladder diagrams corresponds to the solution of coupled transport equations for the laser amplitude $E_L^+({\bf r})$, on the one hand, and the average spectral irradiance  $\overline{I(\omega,{\bf r})}$ of the scattered fields, on the other hand.

In Sec.~\ref{sec:crossed}, the effect of coherent backscattering is quantified by the summation of crossed diagrams. For this purpose, we identify the building blocks out of which the crossed diagrams are composed, and give the rules according to which these building blocks are connected to each other. Thereby, we arrive at a \lq crossed transport equation\rq\ which describes the propagation of a pair of conjugate amplitudes along reversed scattering paths, which, in turn, gives rise to coherent backscattering, i.e., to an enhancement of the scattered intensity in the direction exactly opposite to the incident laser.

Results obtained by numerical solutions of the ladder and crossed transport equations are presented in Sec.~\ref{sec:results}. We consider a slab-like scattering geometry, where all atoms are confined (with uniform density) to a slab with finite length in the direction of the incident laser, and infinite extension in the perpendicular directions. We show and explain how increasing the incident laser intensity changes the  intensity profile of light propagating inside the slab, the spectrum of backscattered light and the height of the coherent backscattering peak. 

Finally, we provide conclusions and outlook in Sec.~\ref{sec:conclusions}. In the appendices, we prove 
that the generalized $N$-atom Bloch equation exhibits a unique stationary state (Appendix~\ref{sec:appendixstat}),
provide technical details concerning the calculation of partial derivatives with respect to probe fields of a given frequency (Appendix~\ref{sec:deriv}), verify that our ladder transport equations respect the property of flux conservation (Appendix~\ref{sec:flux}), and give the complete mathematical expressions of the crossed building blocks (Appendix~\ref{sec:crossedbb}). 

\section{Model}
\label{sec:model}

As described above, we consider an ensemble of $N$ two-level atoms at fixed positions
${\bf r}_1,\dots,{\bf r}_N$. These positions are assumed to be 
to be static on the time scale 
 of a typical multiple scattering process. This assumption is adequate if the atomic gas is cooled to sufficiently low temperature (in the range of $1~{\rm mK}$  \cite{Chaneliere:2004aa}) and if the recoil  induced by scattering of photons remains small throughout the experiment \cite{Chaneliere:2004aa}.
All atoms are driven by a monochromatic laser (which we treat classically) and coupled to the electromagnetic vacuum (which we treat quantum mechanically). In this article, we will, for the sake of clarity and simplicity, model the electromagnetic field as a scalar field. As further discussed in the conclusions, however, our theory can be generalized to vectorial fields and atoms with more complicated level structure.

\subsection{Hamiltonian}

The full Hamiltonian of our system decomposes as follows:  $H(t)=H_A(t)+H_F+H_{V}$,
where $H_A(t)$ refers to the atoms driven by the classical laser field, $H_F$ to the quantized electromagnetic field, and $H_V$ to the interaction between the atoms and the quantized field.
The atomic part reads:
\begin{equation}
H_A(t)  =  \sum_{j=1}^N \left[\hbar\tilde{\omega}_0 \sigma_j^+\sigma_j^-+d E_L({\bf r}_j,t) (\sigma_j^++\sigma_j^-)\right]
\end{equation}
with $\tilde{\omega}_0$ the (bare) atomic resonance frequency and $d$ the dipole moment of the atomic transition. Furthermore,
$\sigma_j^-=|1\rangle_j\langle 2|_j$ and $\sigma_j^+=|2\rangle_j\langle 1|_j$, with $|1\rangle_j$ and $|2\rangle_j$ the ground and excited states of atom $j$, denote the atomic lowering and raising operators.
The laser field 
\begin{equation}
E_L({\bf r},t)=E_L \cos(\omega_L t - {\bf k}_L\cdot{\bf r})
\end{equation}
describes  a plane, monochromatic wave with amplitude $E_L$, frequency $\omega_L$ and wavevector ${\bf k}_L$.

The field Hamiltonian $H_F$ can be expressed in terms of annihilation and creation operators $a_{\bf k}$ and $a^
\dagger_{\bf k}$ of electromagnetic field modes ${\bf k}$ (where, as mentioned above, the polarization degree of freedom is neglected):
\begin{equation}
H_F =   \sum_{\bf k} \hbar \omega_k a^\dagger_{\bf k}a_{\bf k}
\end{equation}
Finally, the interaction between the atoms and the quantized field in dipole approximation is given by:
\begin{equation}
H_V  =  \sum_{j=1}^N d \hat{E}({\bf r}_j)(\sigma_j^++\sigma_j^-)
\end{equation}
where the field operators
\begin{equation}
\hat{E}({\bf r})=\hat{E}^{+}({\bf r})+\hat{E}^{-}({\bf r})
\end{equation}
are split into the following positive- and negative-frequency components:
\begin{eqnarray}
\hat{E}^{+}({\bf r}) & = & i\sum_{\bf k}
\left(\frac{\hbar\omega_{\bf k}}{2\epsilon_0 {\mathcal V}}\right)^{\frac{1}{2}}
a_{\bf k} e^{i{\bf k}\cdot {\bf r}}\\
\hat{E}^{-}({\bf r}) & = & -i\sum_{\bf k}
\left(\frac{\hbar\omega_{\bf k}}{2\epsilon_0 {\mathcal V}}\right)^{\frac{1}{2}}
a^\dagger_{\bf k} e^{-i{\bf k}\cdot {\bf r}}
\end{eqnarray}
with quantization volume $\mathcal V$.

\subsection{Master equation for $N$ atoms}

By tracing over the quantized radiation field, and applying several standard approximations (i.e. rotating wave, Born-Markov and secular approximation) \cite{Lehmberg:1970aa}, it is possible to derive a  
master equation governing the evolution of the quantum-mechanical expectation value of an  arbitrary observable $Q$ of the $N$-atom system. The approximations mentioned above are fulfilled with very high accuracy, essentially due to the fact that the atomic resonance frequency is many orders of magnitude larger than all other relevant frequencies (such as the Rabi frequency $\Omega$, the atom-laser detuning $\delta$, or the spontaneous decay rate $\Gamma$, see below).
In the frame rotating at the laser frequency $\omega_L$, the expectation value
$\langle Q\rangle$ obeys the following equation of motion \cite{Lehmberg:1970aa}:
\begin{eqnarray}
\langle\dot{Q}\rangle & = & \sum_{j=1}^N\Bigl<-i \delta [\sigma_j^+\sigma_j^-,Q]-\frac{i}{2}[\Omega_j\sigma_j^++\Omega^*_j\sigma_j^-,Q]\Bigr.\nonumber\\
& & \Bigl.-\frac{\Gamma}{2}(\sigma_j^+\sigma_j^-Q+Q\sigma_j^+\sigma_j^--2\sigma_j^+Q\sigma_j^-\Bigr>\nonumber\\
& + & \sum_{j\neq k=1}^N \left<\frac{i}{2} T_{jk}[\sigma_j^+Q,\sigma_k^-]-\frac{i}{2}T_{kj}^*[\sigma_j^+,Q\sigma_k^-]\right>\label{eq:master}
\end{eqnarray}
Here, $\delta=\omega_L-\omega_0$ denotes the detuning of the laser frequency with respect to the atomic resonance frequency $\omega_0$. Due to the atom-field interaction, the latter is shifted with respect to the bare frequency $\tilde{\omega}_0$ \cite{Lehmberg:1970aa}.  Furthermore,
$\Omega_j=\Omega({\bf r}_j)$, with  $\Omega({\bf r})=\Omega e^{i{\bf k}_L\cdot {\bf r}}$ and $\Omega=dE_L/\hbar$,
 defines  the atomic Rabi frequency induced by the laser at position ${\bf r}_j$, and 
 \begin{equation}
 \Gamma=\frac{\omega_0^3 d^2}{2\pi\epsilon_0\hbar c^3}\label{eq:gamma}
 \end{equation}
the radiative decay rate of the excited state. The bottom line of Eq.~(\ref{eq:master}) describes the far-field 
dipole-dipole interaction between atoms due to exchange of real 
photons. This interaction is  determined by the complex couplings 
\begin{equation}
T_{jk}=T(|{\bf r}_k-{\bf r}_j|)
\end{equation}
which, in turn, depend on the distance  between the atoms $j$ and $k$ as follows:
\begin{equation}
T(r)=\Gamma \frac{e^{-ik_L r}}{k_Lr}\label{eq:T}
\end{equation}
The far-field approximation is adequate since, throughout this paper, we assume that the distances $r_{jk}=|{\bf r}_k-{\bf r}_j|$ between atoms fulfill $k_Lr_{jk}\gg 1$. On the other hand, we assume $r_{jk}\ll c/\Gamma$ such that the time delay due to the propagation of photons can be neglected as compared to the timescale $\Gamma^{-1}$ of the atomic evolution.

The electromagnetic field scattered by the atoms can be expressed as follows in terms of the atomic raising and lowering operators:
\begin{eqnarray}
E^{+}_{\rm sc}({\bf r},t) & = & 
\frac{\hbar}{2d}\sum_{j=1}^N T^*(|{\bf r}-{\bf r}_j|)\sigma_j^-(t)\label{eq:Eplus}
\\
E^{-}_{\rm sc}({\bf r},t) & = & 
\frac{\hbar}{2d}\sum_{j=1}^N T(|{\bf r}-{\bf r}_j|)\sigma_j^+(t)\label{eq:Eminus}
\end{eqnarray}
With the Wiener-Khinchine theorem \cite{loudon}, the spectrum (or spectral irradiance) of the scattered field finally results as:
\begin{eqnarray}
I(\omega,{\bf r},t) & = & \frac{c\epsilon_0}{\pi} \int_{-\infty}^\infty{\rm d}\tau~e^{-i\omega\tau} \nonumber\\
& & \!\!\!\!\!\!\!\!\!\!\times \left<E_{\rm sc}^-\left({\bf r},t+\frac{\tau}{2}\right)E_{\rm sc}^+\left({\bf r},t-\frac{\tau}{2}\right)\right>\label{eq:spectrum}
\end{eqnarray}
where, due to the rotating frame, $\omega$ denotes the detuning with respect to $\omega_L$, i.e. the detected frequency in the laboratory is given by $\omega_D=\omega_L+\omega$.

\section{Formal solution of the $N$-atom problem}
\label{sec:formal}

\subsection{Generalized optical Bloch equations for $N$ atoms}

To reformulate Eq.~(\ref{eq:master}) as a generalized  optical Bloch equation for $N$ dipole-dipole interacting atoms, we introduce the $4^N$-dimensional generalized Bloch vector 
\begin{equation}
\vec{S}=\langle \vec{\sigma}_1\otimes\dots\otimes \vec{\sigma}_N\rangle\label{eq:blochdef}
\end{equation}
which we write as the expectation value of the  tensor product of the single-atom vector operators
\begin{equation}
\vec{\sigma}_j=\left(\begin{array}{c} {\mathbbm 1}_j \\ \sigma_j^- \\ \sigma_j^+ \\ \sigma_j^z \end{array}\right)\label{eq:sigmaj}
\end{equation}
with $\sigma_j^z=|2\rangle_j\langle 2|_j-|1\rangle_j\langle 1|_j$.
The vector $\vec{S}$ completely characterizes the quantum state of the atomic system, and thus can be interpreted as an alternative representation of the $2^N\times 2^N$-dimensional atomic density matrix.

Evaluating the commutators in Eq.~(\ref{eq:master}), the time evolution of $\vec{S}$ can be written as:
\begin{equation}
\dot{\vec{S}} = L\vec{S}=(A+V)\vec{S}\label{eq:bloch}
\end{equation}
with $L=A+V$, where $A$ and $V$
describe the independent and interaction-induced evolution, respectively. Explicitly:    
\begin{eqnarray}
A & = & \sum_{j=1}^N A_j\label{eq:A}\\
 V & = & \sum_{j\neq k=1}^N \left(T_{jk} B_j^+ C_k^++T_{kj}^* B_k^-C_j^-\right)\label{eq:V}
\end{eqnarray}
where we introduced the $4\times 4$ matrices
\begin{equation}
A_j=\left(\begin{array}{cccc} 0 & 0 & 0 & 0 \\ 0 & i\delta-\frac{\Gamma}{2} & 0 & -i \Omega_j/2  \\ 0 & 0 & -i\delta-\frac{\Gamma}{2} & i \Omega^*_j/2 \\
- \Gamma & - i \Omega^*_j &  i \Omega_j & -\Gamma \end{array}\right)_j\label{eq:Aj}
\end{equation}
\begin{equation}
B_j^+=\left(\begin{array}{cccc} 0 & 0 & 1 & 0 \\ \frac{1}{2} & 0 & 0 &\frac{1}{2}  \\ 0 & 0 & 0 & 0 \\ 0 & 0 & -1 & 0 \end{array}\right)_j,\ 
B_j^-=\left(\begin{array}{cccc} 0 & 1 & 0 & 0 \\ 0 & 0 & 0 & 0  \\ \frac{1}{2} & 0 & 0 & \frac{1}{2} \\ 0 & -1 & 0 & 0 \end{array}\right)_j
\label{eq:B}
\end{equation}
\begin{equation}
C_j^+=\left(\begin{array}{cccc} 0 & 0 & 0 & 0 \\ 0 & 0 & 0 &0  \\ 0 & 0 & 0 & \frac{i}{2} \\ 0 & -i & 0 & 0 \end{array}\right)_j,\ 
C_j^-=\left(\begin{array}{cccc} 0 & 0 & 0 & 0 \\ 0 & 0 & 0 &-\frac{i}{2}  \\ 0 & 0 & 0 & 0 \\ 0 & 0 & i & 0 \end{array}\right)_j
\label{eq:C}
\end{equation}
acting only on the four-dimensional space associated with atom $j$, see Eq.~(\ref{eq:sigmaj}). 

Whereas $A_j$ describes the independent evolution of atom $j$ in presence of the laser field, $B_j^\pm$ and $C_j^\pm$ refer, respectively, to the emission and absorption of negative- ($B_j^+$ and $C_j^+$) or positive-frequency ($B_j^-$ and $C_j^-$) photons by atom $j$. 
The apparent asymmetry between the matrices  $B_j^\pm$ on the one hand and $C_j^\pm$ on the other hand originates from the fact that the complex coupling 
$T_{jk}$ describes, both, the reversible far-field 
dipole-dipole interaction and the irreversible collective decay, see also Eqs.~(\ref{eq:H}-\ref{eq:W}) in Appendix~\ref{sec:appendixstat}. Whereas the 
former corresponds to emission of a photon by atom $j$ and subsequent absorption of this photon by atom $k$, the latter can be interpreted as a photon exchange from atom $j$ to $k$ immediately followed by an irreversible decay of 
atom $k$ \cite{Smith:1992aa}. Both processes differ in their action on the second atom $k$, and the corresponding 
operator $C_k^\pm$ describes the sum of both processes. For simplicity, we will continue speaking of $C_k^\pm$ as describing \lq photon absorption\rq, keeping in mind that  this absorption may be accompanied by an irreversible decay.

For later convenience, let us note the following general properties of the above operators: 
$A_j$ has one eigenvalue $0$ and three eigenvalues  with negative real parts.  The real and imaginary parts of the latter correspond to the widths and positions, respectively, of the three peaks of the Mollow triplet describing the single-atom resonance fluorescence spectrum for strong enough driving field strengths 
\cite{Mollow:1969aa}.

The vector $(1,0,0,0)$ is left-eigenvector of $A_j$ associated with the eigenvalue $0$, i.e. $(1,0,0,0) A_j=(0,0,0,0)$. The corresponding right-eigenvector $\vec{s}^{\,(0)}_j$, defined by 
\begin{equation}
A_j\vec{s}^{\,(0)}_j=\vec{0}
\end{equation}
and the normalization condition $(1,0,0,0)\vec{s}^{\,(0)}_j=1$ denotes the stationary Bloch vector of a single atom driven only by the laser field.

 From the above, it follows that also $A=\sum_j A_j$, see Eq.~(\ref{eq:A}), has exactly one eigenvalue zero. The corresponding left-eigenvector is given by $(1,0,\dots,0)=(1,0,0,0)\otimes\dots\otimes (1,0,0,0)$, and the right-eigenvector by:
 \begin{equation}
 \vec{S}_0=\vec{s}^{\,(0)}_1\otimes\dots\otimes\vec{s}^{\,(0)}_N\label{eq:S0}
 \end{equation}
 which fulfills
 \begin{equation}
 A\vec{S}_0=\vec{0}\label{eq:AS0}
 \end{equation}
and $(1,0,\dots,0)\vec{S}_0=1$.
Finally, the vector $(1,0,\dots,0)$ is also left-eigenvector of $V$. This follows from the fact that the matrices $C_j^\pm$, see Eq.~(\ref{eq:C}), have only zero entries in the uppermost row. Therefore, $(1,0,\dots,0)$ is also left-eigenvector of
$L=A+V$, which governs the time evolution of the Bloch vector, see Eq.~(\ref{eq:bloch}), i.e.:
\begin{equation}
(1,0,\dots,0) L =(0,0,\dots,0)
\end{equation}
This property ensures conservation of the total norm. In other words: the expectation value of the identity operator must remain equal to one at all times. Using this property, it is possible to reduce the Bloch equation (\ref{eq:bloch}) to a $(4^N-1)$-dimensional equation for the remaining elements of the Bloch vector. In the following, however, we will continue working with the $4^N$-dimensional form of the Bloch equation, since this will allow us to exploit  the tensor product structure expressed in Eq.~(\ref{eq:blochdef}). 

\subsection{Stationary state}
\label{sec:stationary}

As shown in Appendix~\ref{sec:appendixstat}, under the condition that the distances between  all pairs of atoms are non-zero, 
the  generalized Bloch equation (\ref{eq:bloch}) has a unique stationary state defined by
\begin{equation}
L\vec{S}=\vec{0}\label{eq:Sdef}
\end{equation}
and the normalization condition $(1,0,\dots,0)\vec{S}=1$.
In the following, the symbol $\vec{S}$ will always refer to this stationary solution (unless indicated otherwise).
Moreover, we show in Appendix~\ref{sec:appendixstat} that
a formal solution for $\vec{S}$ is obtained as follows:
\begin{equation}
\vec{S}=\lim_{\epsilon\to 0}\left(\frac{1}{\epsilon-L}V+{\mathbbm 1}\right)\vec{S}_0\label{eq:Sstat}
\end{equation}
On the basis of this formal solution, a diagrammatic multiple scattering description is obtained by expanding the operator $(\epsilon-L)^{-1}$ in powers of the interaction $V$, see Sec.~\ref{sec:expansion} below.
 
\subsection{Spectrum emitted by $N$ atoms}
\label{sec:spectrum}

In the stationary state, the spectrum of the light emitted by $N$ atoms can be expressed in terms of the following spectral
correlation function \cite{Geiger_Phot_Nano_2010}:
\begin{equation}
P_{il}(\omega) = P^+_{il}(\omega)+P^-_{il}(\omega)\label{eq:Pjk}
\end{equation}
between the atomic raising and lowering operators for atom $i$ and $l$, respectively:
\begin{eqnarray}
 P^+_{il}(\omega) & = & \int_{0}^\infty \frac{{\rm d}\tau}{2\pi}e^{-i\omega\tau}\langle\sigma_i^+(\tau)\sigma_l^-(0)\rangle\\
P^-_{il}(\omega) & = & \int_{0}^\infty \frac{{\rm d}\tau}{2\pi}e^{i\omega\tau}\langle\sigma_i^+(0)\sigma_l^-(\tau)\rangle
\end{eqnarray}
where we assume that the stationary state is reached at time $t=0$.
The spectrum measured by a detector placed in the far field (distance $R$ from the atomic cloud) then results from Eqs.~(\ref{eq:Eplus}-\ref{eq:spectrum}) as follows:
\begin{equation}
I_D(\omega) = \frac{\hbar \omega_0\Gamma}{4\pi R^2}\sum_{i,l=1}^N  e^{i({\bf r}_i-{\bf r}_l)\cdot {\bf k}_D} P_{il}(\omega)\label{eq:ID}
\end{equation}
with $|{\bf k}_D|=k_L$, whereas the direction of ${\bf k}_D=k_L{\bf e}_D$ indicates the direction in which the detector is placed with respect to the atomic cloud.
To normalize the spectrum, we divide the outgoing flux (through a sphere with radius $R\to\infty$) by the incoming flux:
\begin{equation}
\gamma(\omega,{\bf e}_D)=\lim_{R\to\infty}\frac{4\pi R^2}{{\mathcal A}} \frac{I_D(\omega)}{I_L}\label{eq:bistatic}
\end{equation}
where $\mathcal A$ denotes the tranverse (with respect to the direction ${\bf k}_L$ of the incoming laser beam) area of the scattering medium, and $I_L=c\epsilon_0 E_L^2/2$ the incident laser intensity.  The total normalized intensity 
\begin{equation}
\gamma({\bf e}_D)=\int_{-\infty}^\infty{\rm d}\omega ~\gamma(\omega,{\bf e}_D)\label{eq:gammaDtot}
\end{equation}
 scattered into direction ${\bf e}_D$
is a dimensionless quantity also known as \lq bistatic coefficient\rq\ \cite{ishimaru1978wave}.

To calculate the spectra $P_{il}^\pm(\omega)$ of the atomic dipoles, we introduce the following vectors of correlation functions:
\begin{eqnarray}
\vec{D}_l^+(\tau)  & = &  \langle \left(\vec{\sigma}_1\otimes\dots\otimes\vec{\sigma}_{N}\right)(\tau) 
\sigma_l^-(0)\rangle\label{eq:Dp}\\
\vec{D}_i^-(\tau) & = & \langle\sigma_i^+(0)  \left(\vec{\sigma}_1\otimes\dots\otimes\vec{\sigma}_{N}\right)(\tau)\rangle\label{eq:Dm}
\end{eqnarray}
Due to the quantum regression theorem \cite{Lax:1963aa}, the correlation functions follow the same equation as the Bloch vector $\vec{S}$, recall Eq.~(\ref{eq:bloch}) above: 
\begin{equation}
\dot{\vec{D}}_l^\pm = L \vec{D}_l^\pm\label{eq:qregr}
\end{equation}
These equations must be solved with the initial conditions:
\begin{eqnarray}
\vec{D}_l^+(0) & = &  B_l^- \vec{S} \\
  \vec{D}_i^-(0) & = &  B_i^+ \vec{S}
\end{eqnarray}
resulting from Eqs.~(\ref{eq:Dp},\ref{eq:Dm}) evaluated at $\tau=0$, together with the operator identities
$\langle\vec{\sigma}_l\sigma^-_l\rangle=B_l^-\langle\vec{\sigma}_l\rangle$ and $\langle\sigma_i^+\vec{\sigma}_i\rangle=B_i^+\langle\vec{\sigma}_i\rangle$ [which were also  used in the derivation of Eqs.~(\ref{eq:bloch},\ref{eq:V}) from Eq.~(\ref{eq:master})]. Solving Eq.~(\ref{eq:qregr}) by means of Laplace transform, the spectrum is obtained as:
\begin{eqnarray}
P^+_{il}(\omega)  & = &  \lim_{\epsilon\to 0} \frac{1}{2\pi} 
\left[B_i^+\frac{1}{i\omega-L+\epsilon} B_l^- \vec{S}\right]_{1}
\label{eq:Pjkp}\\
 P^-_{il}(\omega) & = & 
 \lim_{\epsilon\to 0} \frac{1}{2\pi} 
 \left[B_i^-\frac{1}{-i\omega-L+\epsilon} B_l^+ \vec{S}\right]_{1}
 \label{eq:Pjkm}
 \end{eqnarray}
 where $\epsilon>0$ ensures the existence of the Laplace transform, and 
$[\dots]_1$ refers to the first vector component, i.e. the one which, for an 
arbitrary $4^N$-dimensional vector $\vec{\mathcal S}$ (not necessarily the stationary Bloch vector), is defined as the scalar product with the vector $(1,0,\dots,0)$, i.e.
$[\vec{\mathcal S}]_1=(1,0,\dots,0)\vec{\mathcal S}$.

The elastic component of the spectrum (i.e. the component emitted at the same frequency as the laser frequency)
originates from the eigenvalue $0$ of $L$ in Eqs.~(\ref{eq:Pjkp},\ref{eq:Pjkm}). Using the corresponding left- and right-eigenvectors $(1,0,\dots,0)$ and
$\vec{S}$, 
we obtain:
\begin{eqnarray}
P_{il}^{({\rm el})}(\omega) & = & \frac{1}{2\pi}\lim_{\epsilon\to 0}\left(\frac{1}{i\omega+\epsilon}+\frac{1}{-i\omega+\epsilon}\right) \left[B_i^+\vec{S}\right]_{1} \left[B_l^-\vec{S}\right]_{1}\nonumber\\
& = & \delta(\omega) 
\langle \sigma^+_i\rangle\langle \sigma^-_l\rangle
\label{eq:Pjkel}
\end{eqnarray}
where we used $[B_i^\pm\vec{S}]_1 = \langle \sigma^\pm_i\rangle$ in the second line.

\section{Representation of the formal $N$-atom solution in terms of diagrams}
\label{sec:diagrams}

\subsection{Expansion of the formal $N$-atom solution}
\label{sec:expansion}

The formal solution (\ref{eq:Pjkp},\ref{eq:Pjkm}) for the spectral function $P_{il}(\omega)$ obtained in Sec.~\ref{sec:spectrum} can be expanded in powers of the interaction $V$ using the relation
\begin{equation}
\frac{1}{i\omega-L+\epsilon}=G(\omega)+G(\omega) V G(\omega)+G(\omega) V G(\omega)V G(\omega)+\dots\label{eq:expansion}
\end{equation}
with $L=A+V$ (see above) and, hence,
\begin{equation}
G(\omega)=\frac{1}{i\omega-A+\epsilon}
\end{equation} 
In view of Eqs.~(\ref{eq:Sstat},\ref{eq:Pjkp}), a typical term of the resulting series has the following form:
\begin{eqnarray}
P_{il}^+(\omega)  & = & \dots + \frac{1}{2\pi} \left[B_i^+G(\omega)V G(\omega) VG(\omega)\right. \nonumber\\
& & \left. \ \ \times B_l^- G(0)V G(0) V G(0) V\vec{S}_0\right]_1 + \dots \label{eq:series}
\end{eqnarray}
and similarly for $P_{il}^-$. (From now on, we will omit the limit $\epsilon\to 0$ and treat $\epsilon$ as an infinitesimally small positive quantity.)

\begin{figure}
\includegraphics[width=0.25\textwidth]{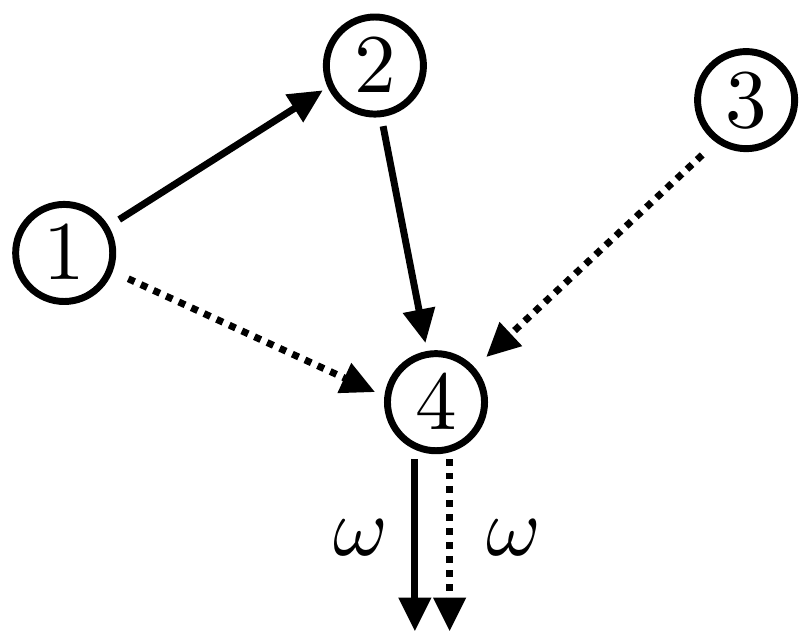}
\caption{Exemplary diagram contributing to the spectral function $P_{44}(\omega)$, see Eq.~(\ref{eq:Pjk}), of atom $4$. The latter 
is subject to radiation emitted by atoms 1, 2, and 3. The contribution of this diagram is evaluated by expanding the formal solution given by Eqs.~(\ref{eq:Sstat},\ref{eq:Pjk},\ref{eq:Pjkp},\ref{eq:Pjkm}) in first order of the couplings $T_{14}$, $T_{34}$ (dotted arrows from 1 to 4 and from 3 to 4) and $T_{12}^*$, $T_{24}^*$ (solid arrows from 1 to 2 and from 2 to 4) indicated in the diagram. 
\label{fig:example_ladder}}
\end{figure}

Let us now consider the operators $V$ appearing in the expansion (\ref{eq:series}). Each of them corresponds to a sum over all atom pairs $(j,k)$, see Eq.~(\ref{eq:V}). In the following, we adopt the following convention: for each factor $T_{jk}B_j^+ C_k^+$, we draw a dotted line from atom $j$ to atom $k$. Similarly, for each factor $T^*_{kj}B_k^- C_j^-$, we draw a solid line from atom $k$ to atom $j$. Thereby, the exchange of negative-frequency (dotted lines) and positive-frequency (solid lines) photons between individual atoms can be visualized in form of a diagram. To specify the indices $i$ and $l$ of the spectral function $P_{il}(\omega)$, we attach an outgoing dotted arrow to atom $i$ and a solid arrow to atom $l$, which are both labeled by the frequency  $\omega$.
An example of a diagram contributing to the spectral function $P_{44}(\omega)$  of atom 4 is shown in Fig.~\ref{fig:example_ladder}. This diagram contains 4 photon exchanges, represented by the solid and dashed arrows pointing from atom 1 to atom 2, and from atoms 1, 2 and 3 to atom 4. Note that the diagram does not specify the order in which the respective interaction terms occur in the series (\ref{eq:series}). Any diagram such as the one depicted in Fig.~\ref{fig:example_ladder} thus implicitly contains 
a sum over all possible orderings. We will come back to this point below.

\subsection{Decomposition into single-atom evolutions}
\label{sec:decomposition}

To set up a diagrammatic multiple scattering theory for $N$ atoms, our aim is to express the $N$-atom signal given by Eq.~(\ref{eq:ID})  in terms of quantities involving only single atoms. For this purpose, let us look at an arbitrary term of the series (\ref{eq:series}), where the photon emission and absorption events occur in a given order, see the example presented in Fig.~\ref{fig:contract} and Eq.~(\ref{eq:example_ladder}) below.
Both, the state $\vec{S}_0$ defined by Eqs.~(\ref{eq:S0},\ref{eq:AS0}), as well as the interaction $V$ are already given in terms of single-atom Bloch vectors $\vec{s}_j^{\,(0)}$ or single-atom operators
$B_j^\pm$ and $C_j^\pm$. 
Furthermore,  also 
$G(\omega)$ can be decomposed into single-atom contributions,
since  it describes the evolution of independent (non-interacting) atoms. For this purpose, it is most convenient to switch to the time domain:
\begin{equation}
G(\omega)=\int_0^\infty {\rm d}t~e^{-(i\omega+\epsilon) t} e^{A t}\label{eq:time}
\end{equation}
and then use:
\begin{equation}
e^{A t}=\prod_{j=1}^N e^{A_j t}\label{eq:factorize}
\end{equation}
see Eq.~(\ref{eq:A}), due to the fact that the operators $A_j$ commute with each other (since they act on different atoms). The operator $e^{A_j t}$ expresses the time evolution of the Bloch vector for a single atom $j$ driven only by the laser with Rabi frequency $\Omega_j$.

We can now explore (\ref{eq:time},\ref{eq:factorize}) for each $G$ occuring in (\ref{eq:series}). 
The resulting expression can be further simplified by using the following rules valid for each single atom $j$: 
\begin{eqnarray}
e^{A_j t_2}e^{A_j t_1} & = & e^{A_j (t_2+t_1)}\label{eq:product}\\
e^{A_j t} \vec{s}_j^{\,(0)} & = & \vec{s}_j^{\,(0)}\label{eq:rightcontract}\\
\left[e^{A_j t}  \vec{s}_j\right]_1 & = & \left[\vec{s}_j\right]_1\label{eq:leftcontract}
\end{eqnarray}
Eqs.~(\ref{eq:rightcontract},\ref{eq:leftcontract}) result from the fact that, as discussed above, $\vec{s}_j^{\,(0)}$ and $(1,0,0,0)$ are right- and left-eigenvectors of $A_j$ with eigenvalue $0$, respectively. Eq.~(\ref{eq:leftcontract}) is valid for an arbitrary four-dimensional vector $\vec{s}_j$. 
Using these rules, the evolution of each single atom in a given diagram can be expressed as a sequence of photon absorption and emission events (described by $B_j^+$, $B_j^-$, $C_j^+$ or $C_j^-$) with single-atom propagators ($e^{A_j t}$) sandwiched in between, see also the example (\ref{eq:F3long}) presented below.

Finally, we switch back to the frequency domain by applying 
\begin{equation}
\int_{-\infty}^\infty \frac{{\rm d}\omega}{2\pi} e^{i\omega t}G_j(\omega) = \left\{\begin{array}{cl} e^{A_j t} &
\text{for }t>0 \\ 0 & \text{for }t<0\end{array}\right.
\label{eq:frequencydomain}
\end{equation}
with 
\begin{equation}
G_j(\omega)=\frac{1}{i\omega-A_j+\epsilon},
\end{equation}
 to each single-atom time propagator $e^{A_j t}$. All of them are evaluated at $t>0$, see Eq.~(\ref{eq:time}).
 Since all poles of $G_j(\omega)$
  are located in the upper half of the complex plane (due to the fact that $\epsilon>0$ and that the eigenvalues of $A_j$ exhibit zero or negative real parts, as mentioned above), 
Eq.~(\ref{eq:frequencydomain}) vanishes for $t<0$. Therefore,
we may extend the limits of integration to the entire real axis (from $-\infty$ to $+\infty$), for each time variable. Doing so
amounts to considering the sum of all terms that arise from the original one by permuting the order of emission and absorption events in such a way that the  \lq local ordering\rq\ for each single atom is preserved, see the example shown in Fig.~\ref{fig:contract} below.
The time integrals  can then be performed using the rule $\int_{-\infty}^\infty {\rm d}t~\exp(i\omega t)=2\pi \delta(\omega)$. 

 Consider, e.g., a photon exchange event between atoms $j$ and $k$. If $\omega_{j}$ and $\omega_j'$ denote the frequencies of the single-atom evolutions $G_{j}(\omega_{j})$ and $G_{j}(\omega_{j}')$ before and after the photon exchange, respectively, and likewise for atom $k$, we obtain:
 \begin{equation}
 \int_{-\infty}^\infty {\rm d}t~e^{i(\omega_j+\omega_k-\omega_k'-\omega_k')t}=2\pi\delta(\omega_j+\omega_k-\omega_j'-\omega_k')
 \label{eq:deltafunctionrule}
 \end{equation}
We see that the frequencies of atom $j$ and $k$ change by the same amount, but with opposite sign:
 $\omega_j'-\omega_j=-(\omega_k'-\omega_k)$. This defines the frequency of the exchanged photon. We choose its sign such that each emission $B_j^\pm$ of a  photon $\omega$ changes  the frequency of atom $j$ by $\mp\omega$ and, correspondingly, each absorption $C_j^\pm$ 
by  $\pm\omega$. 

In summary, the contribution of each single diagram with given local orderings to (\ref{eq:series}) is determined as follows:
 
  (i) The contribution of each single atom $j$ is described by a sequence 
  $$\left[V_j^{(n_j)} G_j(\omega_j^{(n_j-1)}) V_j^{(n_j-1)}\dots
 V_j^{(2)}G_j(\omega_j^{(1)}) V_j^{(1)}\vec{s}_j^{\,(0)}\right]_1$$ of photon emission and absorption events $V_j^{(m)}\in\{B_j^+,B_j^-,C_j^+,C_j^-\}$, according to the given local ordering.  Since $[C_j^{\pm}\vec{s}_j]_1=0$ for an arbitrary four-dimensional vector $\vec{s}_j$, see Eq.~(\ref{eq:C}), the last event in this sequence must correspond to a photon emission event, i.e. $V_j^{(n_j)}\in \{B_j^+,B_j^-\}$.
 
 (ii) The arguments $\omega_j^{(m)}$  of the single-atom evolutions $G_j$ are related to the frequencies $\omega^{(m)}$ of the emitted or absorbed photons as follows:
 \begin{equation}
 \omega_j^{(m)}-\omega_j^{(m-1)}=\left\{\begin{array}{cl} +\omega^{(m)} & \text{if }V^{(m)}\in \{B_j^-,C_j^+\}\\
 -\omega^{(m)} & \text{if }V^{(m)}\in \{B_j^+,C_j^-\}\end{array}\right.\label{eq:frequencychange}
 \end{equation}
 where $\omega_j^{(0)}=\omega_j^{(n_j)}=0$ (since no single-atom evolution $G_i$ occurs before the first or after the last photon emission or absorption).  If $n_j=1$, i.e. if atom $j$ participates in only one single event (which then must be a photon emission), the frequency $\omega^{(1)}$ of the corresponding photon vanishes according to Eq.~(\ref{eq:frequencychange}).
 
 (iii) Multiply the contributions of each single atom, integrate over the frequencies (divided by $2\pi$) of the exchanged photons which are not determined by Eq.~(\ref{eq:frequencychange}), multiply the result by the prefactors $T_{jk}$ and $T_{jk}^*$ originating from the propagation of photons in the given diagram, and finally divide by $2\pi$, see Eq.~(\ref{eq:series}). 
 
\begin{figure}
\includegraphics[width=0.4\textwidth]{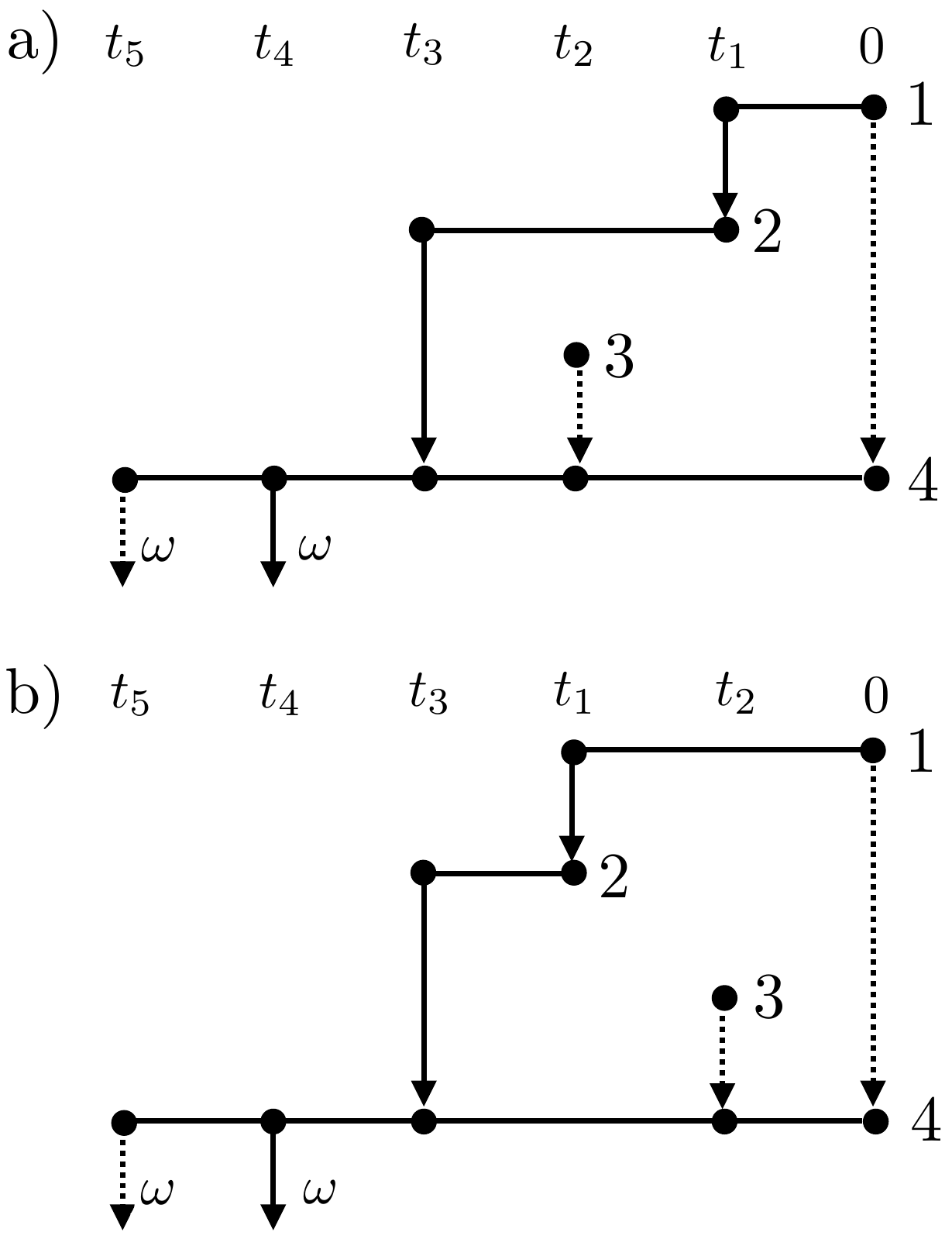}
\caption{a) Same process as shown in Fig.~\ref{fig:example_ladder} represented in a different way, where the order in which the photon exchanges occur is specified (i.e. $t_5>t_4>t_3>t_2>t_1>0$), see Eqs.~(\ref{eq:example_ladder},\ref{eq:F3long}). Vertical lines refer to exchange of photons between atoms and horizontal lines to single-atom evolutions. 
 b) Process with the same local orderings as in a) (i.e. $t_1>0$ for atom 1, $t_3>t_1$ for atom 2 and $t_5>t_4>t_3>t_2>0$ for atom 4), but a different global ordering (i.e. $t_1>t_2$ instead of $t_2>t_1$).
\label{fig:contract}}
\end{figure}
To illustrate the above general recipe, consider the example shown in Fig.~\ref{fig:contract}(a). It shows the same process 
as Fig.~\ref{fig:example_ladder}
in a different representation, where the ordering of the photon emission and absorption events is specified ($0<t_1<t_2<\dots <t_5$). 
As explained in Sec.~\ref{sec:expansion} above, this diagram corresponds to the following mathematical expression:
\begin{eqnarray}
P^{({\rm F\ref{fig:contract}a})}_{44}(\omega)& = & \frac{T_{14} T_{34} T_{12}^* T_{24}^*}{2\pi} \Biggl[B_4^+G(\omega)B_4^-G(0)B_2^-C_4^-\Biggr.\nonumber\\
& & \!\!\!\!\!\!\!\!\!\!\!\!\!\!    \left.\times G(0) B_3^+C_4^+ G(0) B_1^-C_2^-G(0) B_1^+C_4^+  \vec{S}_0\right]_{1}\label{eq:example_ladder}
\end{eqnarray}
Here (and in the following examples), $P^{({\rm F}X)}_{il}(\omega)$ denotes the contribution to  the spectral correlation function  $P_{il}(\omega)$ defined by the diagram shown in Fig.~$X$.

Next, we express each of the five terms $G$ representing the evolution of $N$ non-interacting atoms in presence of the laser driving in terms of single-atom evolutions by using Eqs.~(\ref{eq:time},\ref{eq:factorize}). This leaves us with the following five-fold integral over the time variables $t_1,\dots,t_5$:
\begin{eqnarray}
P^{({\rm F\ref{fig:contract}a})}_{44}(\omega)& = & \frac{T_{14}T_{34} T_{12}^* T_{24}^*}{2\pi} \int_0^\infty {\rm d}t_1 \int_{t_1}^\infty {\rm d}t_2 \dots \int_{t_4}^\infty {\rm d}t_5 \nonumber\\
& \times  &    \left[e^{(t_5-t_1)A_1}B_1^- e^{t_1 A_1} B_1^+\vec{s}_1^{\,(0)}
\right]_1 \nonumber\\
& \times &    \left[e^{(t_5-t_3)A_2} B_2^- e^{(t_3-t_1) A_2}  C_2^- e^{t_1 A_2} \vec{s}_2^{\,(0)}\right]_1 \nonumber\\
& \times &   \left[
e^{(t_5-t_2)A_3} B_3^+ e^{t_2 A_3}  \vec{s}_3^{\,(0)}\right]_1\nonumber\\
& \times &   \left[
B_4^+ e^{(t_5-t_4) A_4} B_4^-e^{(t_4-t_3) A_4} C_4^-
  e^{(t_3-t_2) A_4} C_4^+  \right.\nonumber\\
& &  \left.\times e^{t_2 A_4}  C_4^+\vec{s}_4^{\,(0)}\right]_1~e^{-i\omega (t_5-t_3)}\label{eq:F3long}
\end{eqnarray}
where we used the product rule (\ref{eq:product}).
Now, rule (\ref{eq:rightcontract}) allows us to 
eliminate the evolutions $e^{t_1A_2}$ and $e^{t_2 A_3}$ on the right-hand side of the expressions for atoms 2 and 3, respectively.
Similarly, rule (\ref{eq:leftcontract}) eliminates the propagators $e^{(t_5-t_1)A_1}$ etc. on the left-hand side for atoms 1, 2 and 3. We now express each of the remaining six single-atom time-propagators in frequency space, see Eq.~(\ref{eq:frequencydomain}), and 
extend the lower limits of all time integrations to $-\infty$. Since  
Eq.~(\ref{eq:frequencydomain}) determines only the local ordering of emission and absorption events  for each single atom
(i.e. $t_1>0$ for atom 1, $t_3>t_1$ for atom 2 and $t_5>t_4>t_3>t_2>0$ for atom 4),
we thereby obtain  an additional contribution from the process shown in Fig.~\ref{fig:contract}b),
which exhibits the same local ordering as Fig.~\ref{fig:contract}a), but a different global ordering (i.e. $t_1>t_2$ instead of $t_2>t_1$). Thus, the resulting expression directly yields the sum of both diagrams in Figs.~\ref{fig:contract}(a) and (b).

Applying the \lq $\delta$-function rule\rq\ [e.g., Eq.~(\ref{eq:deltafunctionrule})], for each time integral (which eliminates five among the six frequency integrations), we are left with the following integral over the frequency $\omega_{1}$:
\begin{eqnarray}
P^{({\rm F\ref{fig:contract}a})}_{44}(\omega)+P^{({\rm F\ref{fig:contract}b})}_{44}(\omega)& = & \frac{T_{14} T_{34} T_{12}^*T_{24}^*}{2\pi}\int_{-\infty}^\infty \frac{{\rm d}\omega_1}{2\pi} \nonumber\\
&  & \!\!\!\!\!\!\!\!\!\!\!\!\!\!\!\!\!\!\!\!\!\!\!\!\!\!\!\!\!\!\!\!\!\!\!\!\!\!\!\!\!\!\!\!\!\!\!\!\!\!\!\!\!\!\!\!\!\!\!\!\!\!\!\! \times \left[B_1^-G_1\left(-\omega_1\right)B_1^+ \vec{s}^{\,(0)}_1\right]_{1}
 \left[B_2^-G_2\left(-\omega_1\right)C_2^- \vec{s}^{\,(0)}_2\right]_{1}
\nonumber\\
&  &  \!\!\!\!\!\!\!\!\!\!\!\!\!\!\!\!\!\!\!\!\!\!\!\!\!\!\!\!\!\!\!\!\!\!\!\!\!\!\!\!\!\!\!\!\!\!\!\!\!\!\!\!\!\!\!\!\!\!\!\!\!\!\!\! \times
\left[B_3^+ \vec{s}^{\,(0)}_3\right]_{1}\left[B_4^+G_4\left(\omega\right)B_4^- 
G_4\left(0\right)\phantom{\vec{s}^{\,(0)}_4}\right.\nonumber\\
& & \!\!\!\!\!\!\!\!\!\!\!\!\!\!\!\!\!\!\!\!\!\!\!\!\!\!\!\!\! \times\left. C_4^-G_4\left(\omega_1\right)C_4^+G_4\left(\omega_1\right)C_4^+\vec{s}^{\,(0)}_4\right]_{1}
\label{eq:4bfinal}
\end{eqnarray}
The frequency $\omega_1$ is associated with the photon exchanges $T_{14}$, $T_{12}^*$ and $T_{24}^*$, whereas 
the photon exchanged between $3$ and $4$ carries frequency zero (i.e. the same frequency as the laser frequency). 
Analyzing the frequency arguments of the single-atom evolution operators $G_j$ ($j=1,2,4$), we verify the general  rule stated above, see Eq.~(\ref{eq:frequencychange}), according to which
each emission $B_j^-$ or $B_j^+$ of a positive or negative-frequency photon $\omega_1$  changes these arguments by $+\omega_1$ or $-\omega_1$, and each absorption $C_j^-$ or $C_j^+$ by $-\omega_1$ or $+\omega_1$, respectively.

\begin{figure}
\includegraphics[width=0.45\textwidth]{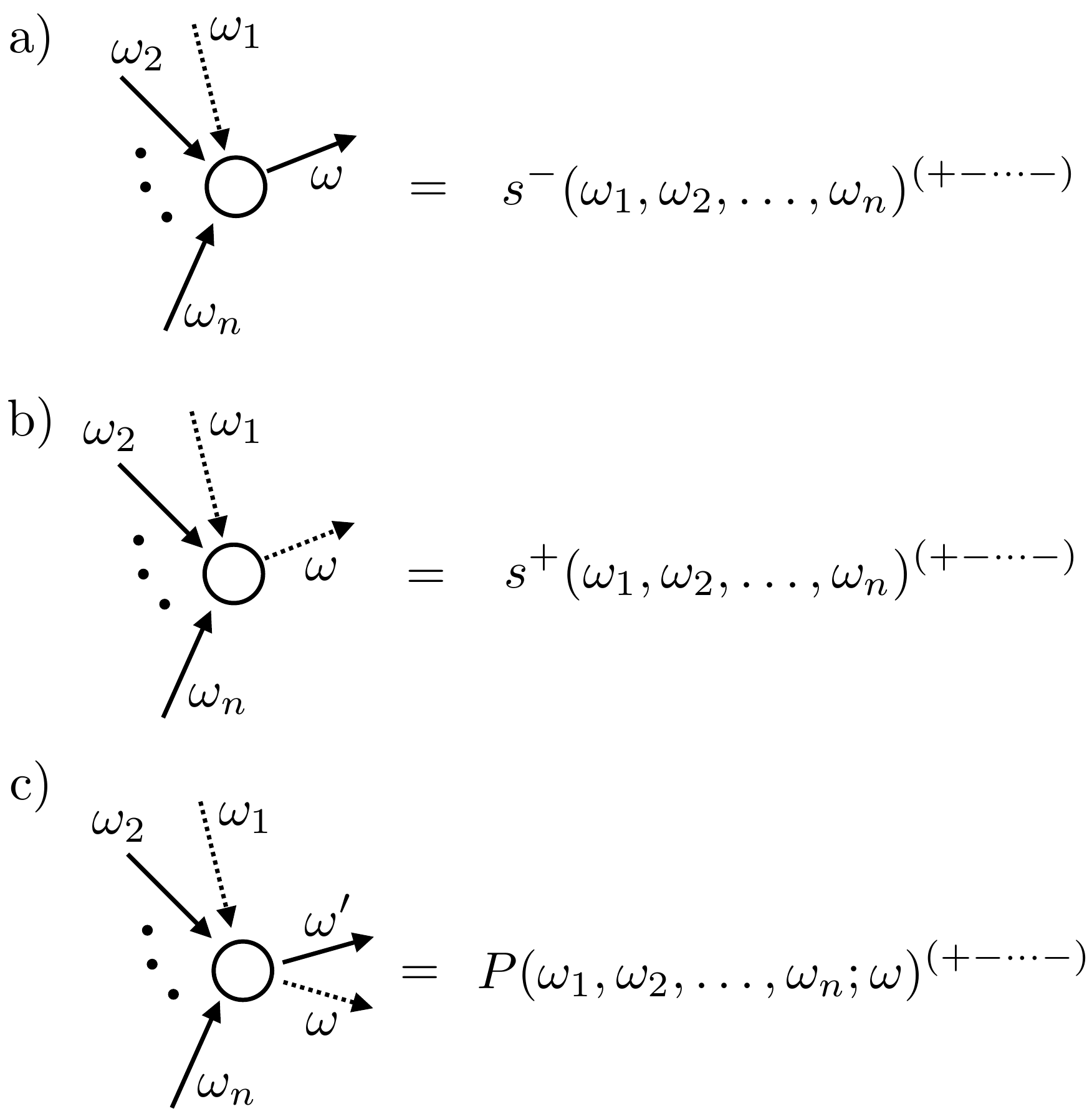}
\caption{Graphical representation of the single-atom building blocks which can be used to construct every relevant diagram in this article (i.e. ladder and crossed diagrams, see Sec.~\ref{sec:laddercrossed}). The incoming photons $\omega_1,\dots,\omega_n$ may carry negative or positive frequency (dotted or solid arrow), corresponding to superscript $\alpha_j=+$ or $-$ in the respective symbols $s^{(\alpha_1\dots \alpha_n)}$  and $P^{(\alpha_1\dots \alpha_n)}$. In a), the frequency of the outgoing photon is fixed to $\omega=- \sum_j \alpha_j \omega_j$, whereas
$\omega=\sum_j \alpha_j \omega_j$ in b) and $\omega-\omega'=\sum_j \alpha_j \omega_j$ in c), see Eq.~(\ref{eq:omegap}). All three building blocks can be calculated by solving the optical Bloch equations for a single atom driven by a polychromatic classical field, see Eqs.~(\ref{eq:sigmaclass}-\ref{eq:Pclassm}). 
 \label{fig:bblock}}
\end{figure}

\subsection{Single-atom building blocks}
\label{sec:bblocks}

The procedure outlined in Sec.~\ref{sec:decomposition} allows us to express the contribution of any diagram involving an arbitrary number of atoms in terms of single-atom evolutions which are coupled to each other through the frequencies of the exchanged photons.
As we will argue in the next section, to calculate the
{\em average} photodetection intensity $I_D(\omega)$, see Eq.~(\ref{eq:ID}), for the case of a dilute atomic medium, it is sufficient to consider diagrams where  each atom emits at most one photon, corresponding to the application of at most one photon emission operator $B_i^+$ for photons with negative frequencies, and at most one $B_i^-$ for photons with positive frequencies. Correspondingly, we obtain three different single-atom building blocks \cite{Shatokhin_Phys_Rev_A_2012}, which are depicted in Fig.~\ref{fig:bblock}.
Summing over all possible local orderings of the photon emission and absorption events, we obtain the corresponding expressions for the building blocks shown in Figs.~\ref{fig:bblock}(a) and (b):
\begin{eqnarray}
s_{{\bf r}_i}^\pm(\omega_1,\dots,\omega_n)^{(\alpha_1\dots \alpha_n)}  & = & \nonumber\\
& & \!\!\!\!\!\!\!\!\!\!\!\!\!\!\!\!\!\!\!\!\!\!\!\!\!\!\!\!\!\!\!\!\!\!\!\!\!\!\!\!\!\!\!\!\!\!\!\!\!\!\!\!\!\!\!\!\!\!    = \sum_{\pi(j_1,\dots,j_n)} 
\left[B_i^\pm G_i\left(\sum_{k=1}^n \alpha_{j_k}\omega_{j_k}\right) C_i^{\alpha_{j_n}}\dots G_i(\alpha_{j_1}\omega_{j_1}\right.\nonumber\\
& & \!\!\!\!\!\!\!\!\!\!\!\!\!\!\!\!\!\!\!\!\!\!\!\!\!\!\!\!\!\!\!\!\!\!\!\!\!\!\!\!\!\!\!\!\!\!\!\!\!\!\ 
\left. +\alpha_{j_2}\omega_{j_2})C_i^{\alpha_{j_2}} G_i\left(\alpha_{j_1}\omega_{j_1}\right) C_i^{\alpha_{j_1}} \vec{s}_i^{\,(0)}\right]_1\label{eq:bbsigma}
\end{eqnarray}
where $\pi(j_1,\dots,j_n)$ denotes $n!$ permutations of indices $j_1,\dots,j_n\in\{1,\dots,n\}$. 
This structure of the  building blocks $s_{{\bf r}_i}^\pm$ follows from the general rules (i) and (ii) established in Sec.~\ref{sec:decomposition}, see Eq.~(\ref{eq:frequencychange}). The photon emission event $B_i^\pm$ occurring at the end of the sequence determines the frequency of the emitted photon as $\omega=\pm \sum_j \alpha_j\omega_j$ according to Eq.~(\ref{eq:frequencychange}). The notation $s^\pm_{{\bf r}_i}$ with index ${\bf r}_i$ 
indicates that the dependence on $i$ enters only through the position ${\bf r}_i$, due to the position-dependent laser amplitudes $\Omega_i$. This will be important later when performing the average over the atomic positions.

Similarly, the expression of the building block shown in Fig.~\ref{fig:bblock}(c) involving two photon emission events $B_i^+$ and $B_i^-$ reads:
\begin{eqnarray}
P_{{\bf r}_i}(\omega_1,\dots,\omega_n;\omega)^{(\alpha_1\dots \alpha_n)} & = & \label{eq:Pblock}\\
& & \!\!\!\!\!\!\!\!\!\!\!\!\!\!\!\!\!\!\!\!\!\!\!\!\!\!\!\!\!\!\!\!\!\!\!\!\!\!\!\!\!\!\!\!\!\!\!\!\!\!\!\!\!\!\!\!\!\!\!\!\!\!\!\!\!\!\!\!\!\!\!\! = P^+_{{\bf r}_i}(\omega_1,\dots,\omega_n;\omega)^{(\alpha_1\dots \alpha_n)}+P^-_{{\bf r}_i}(\omega_1,\dots,\omega_n;\omega)^{(\alpha_1\dots \alpha_n)}\nonumber
\end{eqnarray}
where 
\begin{widetext}
\begin{eqnarray}
P^+_{{\bf r}_i}(\omega_1,\dots,\omega_n;\omega)^{(\alpha_1\dots \alpha_n)} & = & \frac{1}{2\pi}\sum_{\pi(j_1,\dots,j_n)} 
\Bigl[B_i^+G_i(\omega)C_i^{\alpha_{j_n}}\dots C_i^{\alpha_{j_2}} G_i(\omega'+\alpha_{j_1}\omega_{j_1}) C_i^{\alpha_{j_1}} G_i(\omega')B_i^-\vec{s}^{\,(0)}_i\Bigr.\nonumber\\
& & \ \ \ \ \ \ \ \ \ \ \ \ \ + B_i^+G_i(\omega)C_i^{\alpha_{j_n}}\dots C_i^{\alpha_{j_2}} G_i(\omega'+\alpha_{j_1}\omega_{j_1}) B_i^- G_i(\alpha_{j_1}\omega_{j_1}))C_i^{\alpha_{j_1}}\vec{s}^{\,(0)}_i \nonumber\\
& & \ \ \ \ \ \ \ \ \ \ \ \ \ +\dots +\Bigl. B_i^+G_i(\omega)B_i^-G_i(\omega-\omega')C_i^{\alpha_{j_n}}\dots C_i^{\alpha_{j_1}}\vec{s}^{\,(0)}_i\Bigr]_{1}\label{eq:Pblockp}\\
P^-_{{\bf r}_i}(\omega_1,\dots,\omega_n;\omega)^{(\alpha_1\dots \alpha_n)} & = & \frac{1}{2\pi}\sum_{\pi(j_1,\dots,j_n)} 
\Bigl[B_i^-G_i(-\omega')C_i^{\alpha_{j_n}}\dots C_i^{\alpha_{j_2}} G_i(-\omega+\alpha_{j_1}\omega_{j_1}) C_i^{\alpha_{j_1}} G_i(-\omega)B_i^+\vec{s}^{\,(0)}_i\Bigr.\nonumber\\
& & \ \ \ \ \ \ \ \ \ \ \ \ \ + B_i^-G_i(-\omega')C_i^{\alpha_{j_n}}\dots C_i^{\alpha_{j_2}} G_i(-\omega+\alpha_{j_1}\omega_{j_1}) B_i^+ G_i(\alpha_{j_1}\omega_{j_1}))C_i^{\alpha_{j_1}}\vec{s}^{\,(0)}_i \nonumber\\
& & \ \ \ \ \ \ \ \ \ \ \ \ \ +\dots +\Bigl. B_i^-G_i(-\omega')B_i^+G_i(\omega- \omega')C_i^{\alpha_{j_n}}\dots C_i^{\alpha_{j_1}}\vec{s}^{\,(0)}_i\Bigr]_{1}\label{eq:Pblockm}
\end{eqnarray}
\end{widetext}
with
\begin{equation}
\omega'=\omega-\sum_{j=1}^n \alpha_j\omega_j\label{eq:omegap}
\end{equation}
Again, we sum over all possible permutations of the $n$ absorption events $C_i^\pm$ and the two emission events $B_i^\pm$. As explained above, one of the two $B_i^\pm$'s must occur at the end of the sequence.

The building block $P$ as defined by the above equations exhibits, both, inelastic and elastic components. In this respect, it differs from the corresponding quantity introduced in \cite{Shatokhin_Phys_Rev_A_2012} which contains only the inelastic component.  

Arbitrary diagrams involving at most one photon emission $B_i^+$ and/or $B_i^-$ per atom can now be constructed by connecting these building blocks to each other. For this purpose, the outgoing arrow of one building block is identified with an incoming arrow of another building block. Each occurrence of a building block $P$ leads to an integral over the frequency of the exchanged photon, since only the difference between the outgoing frequencies $\omega$ and $\omega'$ is determined by the incoming frequencies in Fig.~\ref{fig:bblock}(c).

Remember, however, that the above building blocks involve a sum over all possible local orderings. When connecting different building blocks to each other, we must verify that these orderings are consistent with each other. For example,  if a positive- and a negative-frequency photon   are emitted by the same atom, see building block Fig.~\ref{fig:bblock}(c), and subsequently absorbed by another atom, the ordering of the absorption events must coincide with the ordering of the corresponding emission events. (Remember that time delays resulting from propagation of photons between atoms are neglected in our $N$-atom master equation.)
This condition may apparently be violated if one sums over all possible orderings independently for each building block. 
However, all \lq forbidden\rq\ combinations of terms (i.e. those which exhibit an inconsistent ordering) vanish identically since, in the time-domain representation, these combinations contain
 retarded single-atomic propagators evaluated at negative times, which vanish  due to Eq.~(\ref{eq:frequencydomain}).

For performing the ensemble average over the atomic positions (see below), the dependence of the building blocks
on ${\bf r}$ plays an important role. Using the structure of the matrices $A,B^\pm$ and $C^\pm$ defined in Eqs.~(\ref{eq:Aj}-\ref{eq:C}), it is possible to show that:
\begin{eqnarray}
s_{\bf r}^{\pm(\alpha_1\dots \alpha_n)}&= &e^{i \left(\alpha_1+\dots+\alpha_n\mp 1\right){\bf k}_L\cdot{\bf r}}
s_{\bf 0}^{\pm(\alpha_1\dots \alpha_n)}\label{eq:spositionphase}\\
P_{\bf r}^{(\alpha_1\dots \alpha_n)}& = & e^{i \left(\alpha_1+\dots +\alpha_n\right){\bf k}_L\cdot{\bf r}} P_{\bf 0}^{(\alpha_1\dots \alpha_n)}\label{eq:Ppositionphase}
\end{eqnarray}
As an example, let us illustrate the use of single-atom building blocks in the example shown in Fig.~\ref{fig:example_ladder}. The final result for the contribution of this diagram, including all possible orderings of  photon emission and absorption events, reads as follows:
\begin{eqnarray}
P^{({\rm F\ref{fig:example_ladder}})}_{44}(\omega) & = & T_{14} T_{34} T_{12}^*T_{24}^* \int_{-\infty}^\infty {\rm d}\omega_1~P_{{\bf r}_1}(\omega_1)\nonumber\\
&  \times  & s^-_{{\bf r}_2}(\omega_1)^{(-)}s^+_{{\bf r}_3}P_{{\bf r}_4}(\omega_1,\omega_1,0;\omega)^{(+-+)}\label{eq:4afinal}
\end{eqnarray}

\subsection{Single-atom Bloch equations}
\label{sec:bloch}

So far, we have shown that the spectrum emitted by $N$ laser-driven atoms can  be represented in terms of diagrams composed of single-atom building blocks.
This constitutes the first main result of the present paper, and generalizes the results previously established for the cases of $N=2$ \cite{Shatokhin_Chem_Phys_2010} and $N=3$ atoms \cite{Shatokhin_Phys_Rev_A_2012}  to an {\em arbitrary} number of atoms. 
In a second step, we now establish a method to perform  the sum over all relevant diagrams. 

To do so, we rely on the fact that
the above  building blocks can be calculated by solving single-atom optical Bloch equations for polychromatic driving fields representing the incoming photons.
Let us consider a field of the form
\begin{equation}
E(t)=e^{-i\omega_L t}E^+(t)+e^{i\omega_L t}E^-(t)
\end{equation}
with positive- and negative frequency components (in the frame rotating with frequency $\pm\omega_L$):
\begin{eqnarray}
E^+(t) & = & \sum_{j=1}^n  E_j^{+} e^{-i\omega_j t}\\
 E^-(t) & = & \sum_{j=1}^n E_j^{-} e^{i\omega_j t}
\end{eqnarray}
The time evolution of the atomic Bloch vector $\vec{s}=\langle\vec{\sigma}\rangle$ for an atom placed at position ${\bf r}$ driven by this field in addition to the laser field with associated Rabi frequency $\Omega({\bf r})$ is given by:
\begin{equation}
\dot{\vec{s}}_{\bf r}(t)=\left[A({\bf r})+C^+\frac{2 d}{\hbar}E^-(t)+C^- \frac{2 d}{\hbar}E^+(t)\right] \vec{s}_{\bf r}(t) \label{eq:blochclassical}
\end{equation}
where $A({\bf r})$, $C^+$ and $C^-$ are the same $4\times 4$-matrices as those in Eqs.~(\ref{eq:Aj}) and (\ref{eq:C}), but without subscripts, and $\Omega_j$ replaced by $\Omega({\bf r})$.

We consider the solution of Eq.~(\ref{eq:blochclassical}) starting from an arbitrary initial condition at time $t_0\ll -1/\Gamma$, such that a quasi-stationary state is reached at time $t=0$. Due to the time-dependence of the driving field, this state is not truly stationary, but quasi-stationary in the sense that it does not depend on the initial condition, thus being uniquely determined by the driving field.
Let us expand this quasi-stationary solution in a Taylor series with respect to the time-dependent driving fields:
\begin{eqnarray}
\vec{s}_{\bf r}(t) & = & \vec{s}_{\bf r}^{(0)} + \sum_{j=1}^n\sum_{\alpha_j=\pm} E^{\alpha_j}_j(t)  \frac{\partial \vec{s}_{\bf r}(t)}{\partial E_j^{\alpha_j}(t)}+\label{eq:taylor}\\
& & \!\!\!\!\!\!\!\!\!\!\!\!\!\!\!\!\!\!\!\!\!\!+ \frac{1}{2!}\sum_{j,k=1}^n\sum_{\alpha_j,\alpha_k=\pm} E^{\alpha_j}_j(t) E^{\alpha_k}_k(t) \frac{\partial^2\vec{s}_{\bf r}(t)}{\partial E^{\alpha_j}_j(t)\partial E^{\alpha_k}_k(t)}+\dots\nonumber
\end{eqnarray}
where $E^{\alpha_j}_j(t)=E^{\alpha_j}_j e^{-i\alpha_j\omega_jt}$, and the derivatives are evaluated at $E^{\pm}_1=\dots=E^{\pm}_n=0$. 
In the quasi-stationary regime (i.e. for $t\geq 0$), the partial derivatives thereby defined are independent of $t$.
Since the expansion of the quasi-stationary solution of Eq.~(\ref{eq:blochclassical}) in powers of the Rabi frequencies
$\Omega_j^\pm = 2 d E_j^\pm/\hbar$ induced by the
driving fields leads exactly to the same expression as given in Eq.~(\ref{eq:bbsigma}),
the building blocks $s_{\bf r}^\pm$  are obtained as the $n$-th fold partial derivative \cite{Ketterer:2014aa}:
\begin{equation}
s_{{\bf r}}^\pm(\omega_1,\dots,\omega_n)^{(\alpha_1\dots \alpha_n)} = \left(\frac{\hbar}{2d}\right)^n \frac{\partial^n s_{\bf r}^{\pm}(t)}{\partial E^{-\alpha_1}_1(t)\dots\partial E^{-\alpha_n}_n(t)}\label{eq:sigmaclass}
\end{equation}
evaluated at $E^{\pm}_1=\dots=E^{\pm}_n=0$. 
Note that the superscript $\alpha_i=\pm$ corresponds to a probe field with opposite sign $E_i^\mp$.
A similar rule applies to the third building block $P_{{\bf r}}$, if we 
apply the quantum regression theorem  \cite{Lax:1963aa} to calculate the atomic correlation functions $\langle \sigma^+(\tau)\sigma^-(0)\rangle_{\bf r}$ and 
$\langle \sigma^+(0)\sigma^-(\tau)\rangle_{\bf r}$, see Eqs.~(\ref{eq:regression1},\ref{eq:regression2}) in Appendix~\ref{sec:deriv}, and expand these in a Taylor series as above:
\begin{eqnarray}
P^+_{{\bf r}}(\omega_1,\dots,\omega_n;\omega)^{(\alpha_1\dots \alpha_n)}  & = & \nonumber\\
& & \!\!\!\!\!\!\!\!\!\!\!\!\!\!\!\!\!\!\!\!\!\!\!\!\!\!\!\!\!\!\!\!\!\!\!\!\!\!\!\!\!\!\!\!\!\!\!\!\!\!\!\!\!\!\!\!\!\!\!\!\!\!\!\!\!\!\!\!\!\!\!\!\!\!\!\!\!\! = \left(\frac{\hbar}{2d}\right)^n\int_0^\infty\frac{{\rm d}\tau}{2\pi} e^{-i\omega' \tau}\frac{\partial^n\langle \sigma^+(\tau)\sigma^-(0)\rangle_{\bf r}}{\partial E^{-\alpha_1}_1(\tau)\dots\partial E^{-\alpha_n}_n(\tau)}
\label{eq:Pclassp}\\
P^-_{{\bf r}}(\omega_1,\dots,\omega_n;\omega)^{(\alpha_1\dots \alpha_n)}  & = & \nonumber\\
& & \!\!\!\!\!\!\!\!\!\!\!\!\!\!\!\!\!\!\!\!\!\!\!\!\!\!\!\!\!\!\!\!\!\!\!\!\!\!\!\!\!\!\!\!\!\!\!\!\!\!\!\!\!\!\!\!\!\!\!\!\!\!\!\!\!\!\!\!\!\!\!\!\!\!\!\!\!\! = \left(\frac{\hbar}{2d}\right)^n\int_0^\infty\frac{{\rm d}\tau}{2\pi} e^{i\omega \tau}\frac{\partial^n\langle \sigma^+(0)\sigma^-(\tau)\rangle_{\bf r}}{\partial E^{-\alpha_1}_1(\tau)\dots\partial E^{-\alpha_n}_n(\tau)}
\label{eq:Pclassm}
\end{eqnarray}
Again, the partial derivatives are evaluated at $E_1^\pm=\dots=E_n^\pm=0$.
In principle, the Fourier transform of $\langle \sigma^+(\tau)\sigma^-(0)\rangle_{\bf r}$ with respect to $\tau$ yields the spectrum as a function of the frequency $\omega$ of the emitted negative-frequency photon in Eq.~(\ref{eq:Pclassp}). Due to the time dependence $\exp(i\sum_j \alpha_j \omega_j t)$ of the incident fields together with Eq.~(\ref{eq:omegap}), however, the frequency 
$\omega$ is shifted to  $\omega'$ in Eq.~(\ref{eq:Pclassp}), and vice versa (from $\omega'$ to $\omega$) in Eq.~(\ref{eq:Pclassm}).

\subsection{Ladder and crossed diagrams}
\label{sec:laddercrossed}

In Sec.~\ref{sec:bblocks}, we restricted ourselves to diagrams where each atom emits at most one photon. In this subsection, we justify this restriction, and further specify the types of diagrams considered in this paper.
For this purpose, we employ the assumption of a dilute medium, where the distances between atoms are much larger than the wave length of the scattered light. Furthermore, we restrict ourselves to calculating the ensemble average of the detected spectrum, Eq.~(\ref{eq:ID}), where the average is taken over the atomic positions ${\bf r}_j$.
The latter are assumed to be distributed independently from each other inside a certain volume $V$. This assumption requires the temperature of the atomic cloud to lie well above the threshold for Bose-Einstein condensation since, otherwise, quantum-statistical correlations between atomic positions become relevant \cite{Morice:1995aa}.

We now argue that, for the case of a dilute medium, only diagrams exhibiting a certain simple structure survive the ensemble average: For this purpose, we note that the couplings $T_{jk}$ between the atoms, see Eq.~(\ref{eq:T}), exhibit phase factors $e^{ikr}$ which sensitively depend on the distance $r$ between the respective atoms. Under the condition $kr\gg 1$ (dilute medium), the corresponding phase  is approximately uniformly distributed in the interval $[0,2\pi]$, such that it vanishes on average. Therefore, the only diagrams which survive the average are those where each coupling $T_{jk}$ is accompagnied by its complex conjugate $T_{jk}^*$ or $T_{kj}^*$, in order to compensate the random phase of the former. In some cases, the phase of $T_{jk}$ can also be compensated by the phases of the laser amplitudes $\Omega_j$ and $\Omega_k$, as further discussed below. 

\begin{figure}
\includegraphics[width=0.45\textwidth]{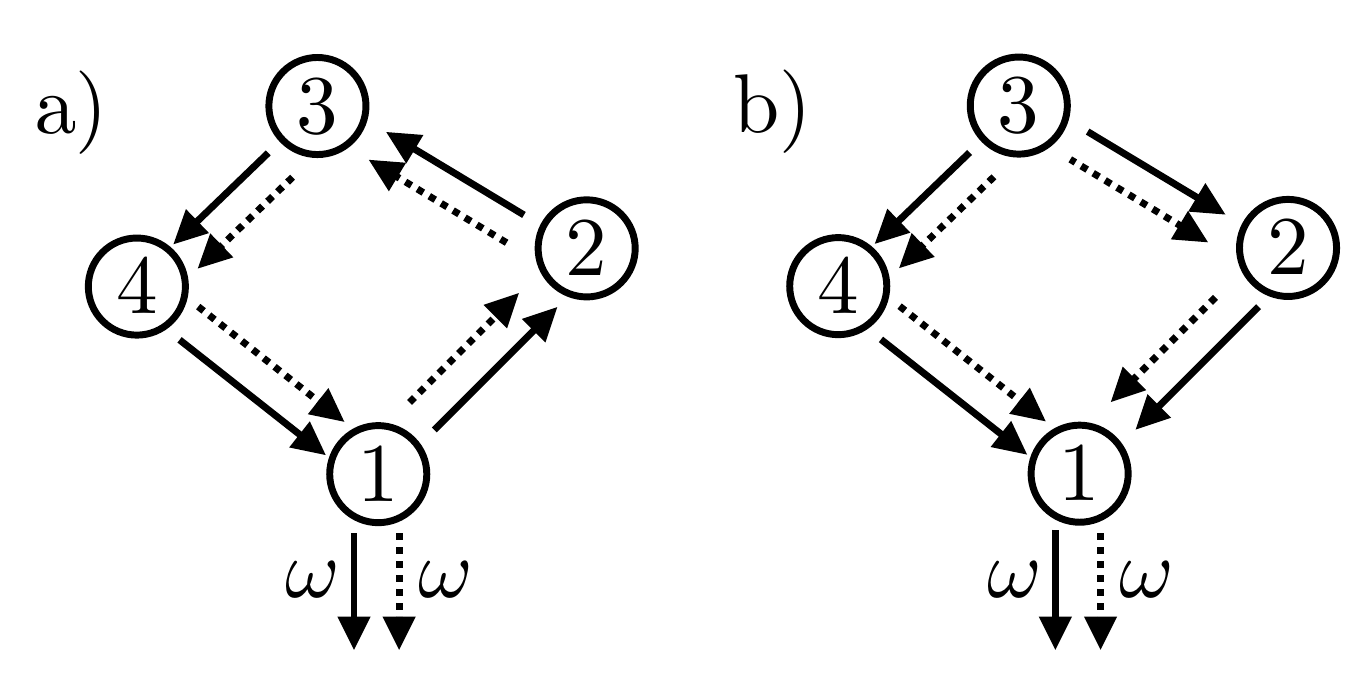}
\caption{Processes involving closed loops of photons which we neglect in our treatment. a) Atom 1 emits a photon, which is scattered by atoms 2, 3 and 4, and then reabsorbed by 1. This process leads, in principle, to a change of the atomic decay rate and resonance frequency as compared to an atom placed in vacuum \cite{Lehmberg:1970aa}. However, these changes are small for a dilute atomic cloud. b) Atom 3 emits a correlated pair of photons, which then meet again at atom 1. Similarly as in a), also this process involves a closed loop, and its weight hence tends to zero in the limit of decreasing atomic density. Both processes involve single-atom building blocks of higher order (i.e. with more outgoing arrows) than those depicted in Fig.~\ref{fig:bblock}, see atom 1 in a) and atom 3 in b).
 \label{fig:loops}}
\end{figure}

In addition to the condition of vanishing phases, we may furthermore neglect diagrams involving closed loops of photons, see Fig.~\ref{fig:loops}. These include, both, processes where a photon, described by a conjugate pair of solid and dotted arrows, is emitted by an atom, scattered by other atoms and then reabsorbed by the former atom, see Fig.~\ref{fig:loops}(a), and processes where an atom emits two photons  which then meet again at another atom, see Fig.~\ref{fig:loops}(b). Even if their phase vanishes, such processes can be neglected since, as known from the theory of multiple scattering for a single particle, the probability of \lq recurrent scattering\rq\ \cite{Wiersma:1995aa}, i.e. the probability of a photon returning to the same atom from which it has been emitted,  scales like $1/(k \ell)$ (with mean free path $\ell$) and thus can be neglected in the dilute regime $k\ell\gg 1$ (also called regime of \lq weak disorder\rq). In our case, this condition is fulfilled due to the assumption of the distances between atoms being much larger than the wave length of the scattered light. The neglect of closed loops allows us to restrict ourselves to those diagrams which can be constructed from the three single-atom building blocks defined above, see Fig.~\ref{fig:bblock}.

\begin{figure}
\includegraphics[width=0.45\textwidth]{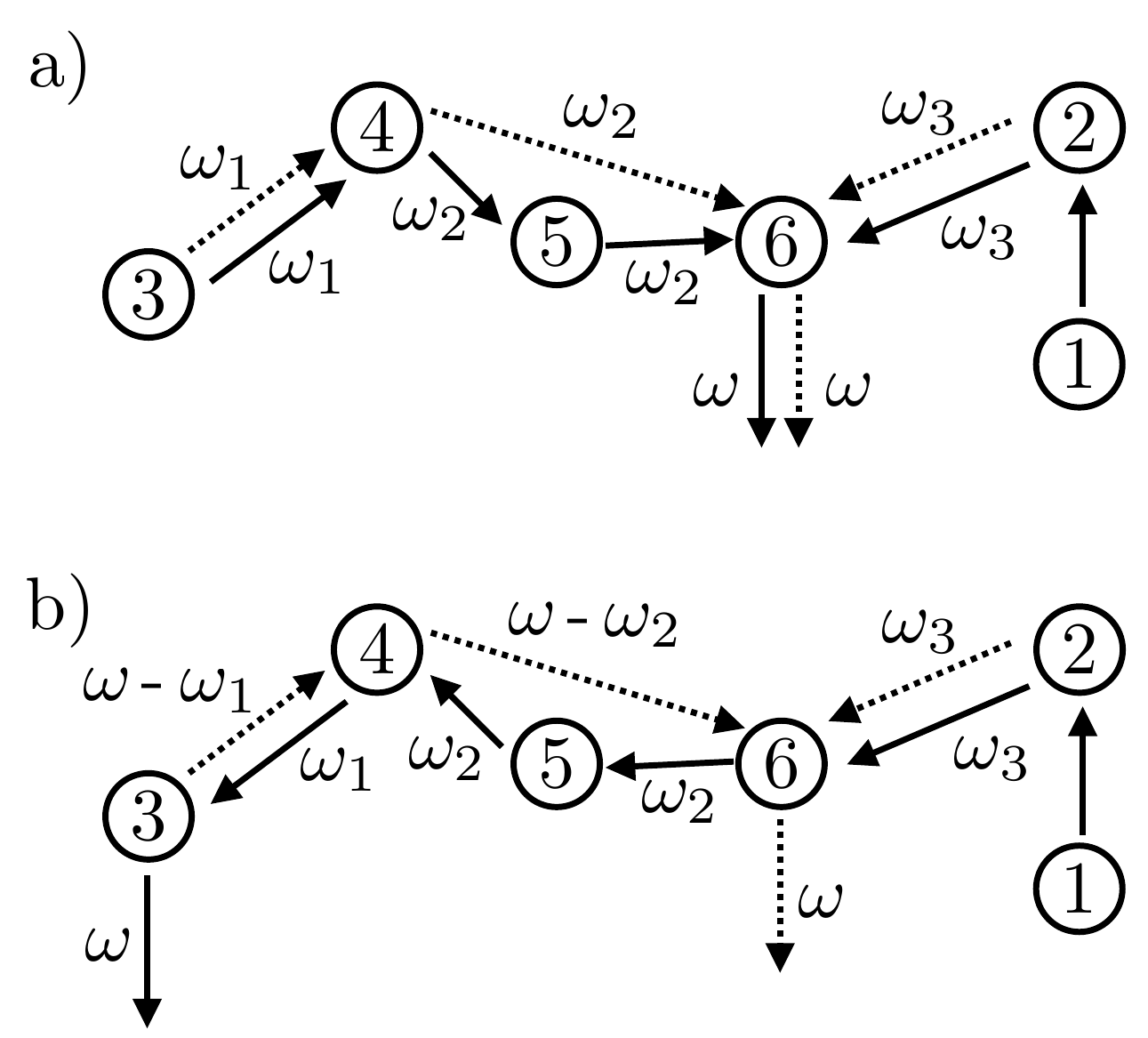}
\caption{a) Examplary ladder diagram describing light emitted by single atoms (here: atom 6). Atoms are irradiated either by single incident fields (here: atom 2 by atom 1) describing the attenuation of the laser amplitude inside the atomic cloud, or by pairs of positive- and negative frequency photon amplitudes (solid and dotted arrows) following the same path through the atomic medium. The scattering by atom 5 describes the refractive index of the atomic medium modifying propagation between atom 4 and 6 as compared to propagation in vacuum. b) Crossed diagram describing interference of light emitted from atoms 3 and 6 leading to coherent backscattering. This diagram is obtained from the ladder diagram by reversing the scattering sequence $3\to 4\to 5\to 6$ of the solid arrows in a). 
As argued in the main text, ladder and crossed diagrams 
describe the average intensity of
emitted light in the case of a dilute atomic medium, where the distances between atoms are much larger than the wave length of the incident laser.
\label{fig:ladder_meq}}
\end{figure}

Among the latter, the diagrams with vanishing phase can be divided into two classes called \lq ladder\rq\ and \lq crossed diagrams\rq\ in the following. Ladder diagrams are defined by the condition that two conjugate amplitudes (solid and dotted arrows) of a photon  emitted by one atom  \lq remain together\rq\ in the sense that they are absorbed by the same atom. In between, they may undergo an arbitrary sequence of scattering events described by the building blocks $s^\pm_{\bf r}$, which, as we will see later, describe the refractive index of the atomic medium. An  example is shown in  Fig.~\ref{fig:ladder_meq}(a). We see pairs of co-propagating conjugate photon amplitudes (solid and dotted arrows) from atom 2 to atom 6, from atom 3 to atom 4, and from atom 4 to atom 6, with intermediate scattering of the solid arrow at atom 5 in the latter case. 
Due to the condition of vanishing phases, this process contributes only if atom 5 is placed in the vicinity of  the straight line connecting atoms 4 and 6.
In addition, atom 2 is irradiated by a single photon amplitude from atom 1 which, as discussed below, describes the attenuation of the incident laser beam. If the line from atom 1 to 2 is parallel to ${\bf k}_L$, the phase  of the coupling $T_{21}^*$ is 
compensated by the phases of the laser fields acting on atom 1 and 2, since $s_{{\bf r}_1}^-\propto \exp(i{\bf k}_L\cdot{\bf r}_1)$ and $P_{{\bf r}_2}(0;\omega_3)^{(-)}\propto \exp(-i{\bf k}_L\cdot{\bf r}_2)$ according to Eqs.~(\ref{eq:spositionphase},\ref{eq:Ppositionphase}).

Crossed diagrams result from ladder diagrams by reversing the direction of single arrows, thus describing interference between counterpropagating amplitudes. The diagram shown in Fig.~\ref{fig:ladder_meq}(b), for example, contributes to the interference $P_{63}(\omega)$ between light emitted by atoms $3$ and $6$, respectively, with corresponding phase factor $\exp[i({\bf k}_D+{\bf k}_L)\cdot ({\bf r}_6-{\bf r}_3)]$, which follows from Eqs.~(\ref{eq:ID},\ref{eq:Ppositionphase}). Therefore, crossed diagrams contribute, on average, only in the vicinity of the backscattering direction ${\bf k}_D\simeq -{\bf k}_L$, giving rise to the coherent backscattering cone. In contrast, the ladder diagrams do not sensitively depend on the outgoing direction ${\bf k}_D$, and thus describe the diffusive background of the scattered light. 

The equations corresponding to the above exemplary diagrams are:
\begin{eqnarray}
P_{66}^{({\rm F\ref{fig:ladder_meq}a})}(\omega) & = & |T_{34}|^2T_{46} T_{45}^*T_{56}^* |T_{26}|^2 T^*_{12}\int_{-\infty}^\infty {\rm d}\omega_1{\rm d}\omega_2{\rm d}\omega_3\nonumber\\
& & \!\!\!\!\!\!\!\!\!\!\!\!\!\!\!\!\!\!\!\!\!\!\!\!\!\!\!\times P_{{\bf r}_3}(\omega_1) P_{{\bf r}_4}(\omega_1,\omega_1;\omega_2)^{(+-)}s^-_{{\bf r}_5}(\omega_2)^{(-)}\nonumber\\
& & \!\!\!\!\!\!\!\!\!\!\!\!\!\!\!\!\!\!\!\!\!\!\!\!\!\!\!\times s^-_{{\bf r}_1} P_{{\bf r}_2}(0;\omega_3)^{(-)}  P_{{\bf r}_6}(\omega_2,\omega_2,\omega_3,\omega_3;\omega)^{(+-+-)} \\
P_{63}^{({\rm F\ref{fig:ladder_meq}b})}(\omega) & = & T_{43}^*T_{34}T_{46} T^*_{54}T^*_{65} |T_{26}|^2 T^*_{12}\int_{-\infty}^\infty {\rm d}\omega_1{\rm d}\omega_2{\rm d}\omega_3\nonumber\\
& & \!\!\!\!\!\!\!\!\!\!\!\!\!\!\!\!\!\!\!\!\!\!\!\!\!\!\!\!\!\!\!\!\!\!\times  P_{{\bf r}_3}(\omega_1;\omega-\omega_1)^{(-)}
P_{{\bf r}_4}(\omega-\omega_1,\omega_2;\omega-\omega_2)^{(+-)} s^-_{{\bf r}_5}(\omega_2)^{(-)} \nonumber\\
& & \!\!\!\!\!\!\!\!\!\!\!\!\!\!\!\!\!\!\!\!\!\!\!\!\!\!\!\times s^-_{{\bf r}_1} P_{{\bf r}_2}(0;\omega_3)^{(-)} P_{{\bf r}_6}(\omega-\omega_2,\omega_3,\omega_3;\omega)^{(+-+)}
\end{eqnarray}
In accordance with the rules that determine the frequencies of the outgoing photons in  Fig.~\ref{fig:bblock}, we see that co-propagating photon pairs always carry the same frequency, i.e. $\omega_{1}$, $\omega_2$ or $\omega_3$ in Fig.~\ref{fig:ladder_meq}(a) and $\omega_3$ in Fig.~\ref{fig:ladder_meq}(b), whereas the frequencies $(\omega_i,\omega_i')$ of counterpropagating photons are related by $\omega_i'=\omega-\omega_i$, i.e. $(\omega_1,\omega-\omega_1)$ and 
$(\omega_2,\omega-\omega_2)$ in Fig.~\ref{fig:ladder_meq}(b). In the following, however, we will not be concerned with evaluating the contributions of individual diagrams such as the ones shown in Fig.~\ref{fig:ladder_meq}, but rather derive transport equations the solution of which yields the sum of all ladder and crossed diagrams.

\section{Summation of ladder diagrams}
\label{sec:ladder}


\subsection{Description of incident radiation as stochastic classical field}
\label{sec:classical}

\begin{figure}
\includegraphics[width=0.45\textwidth]{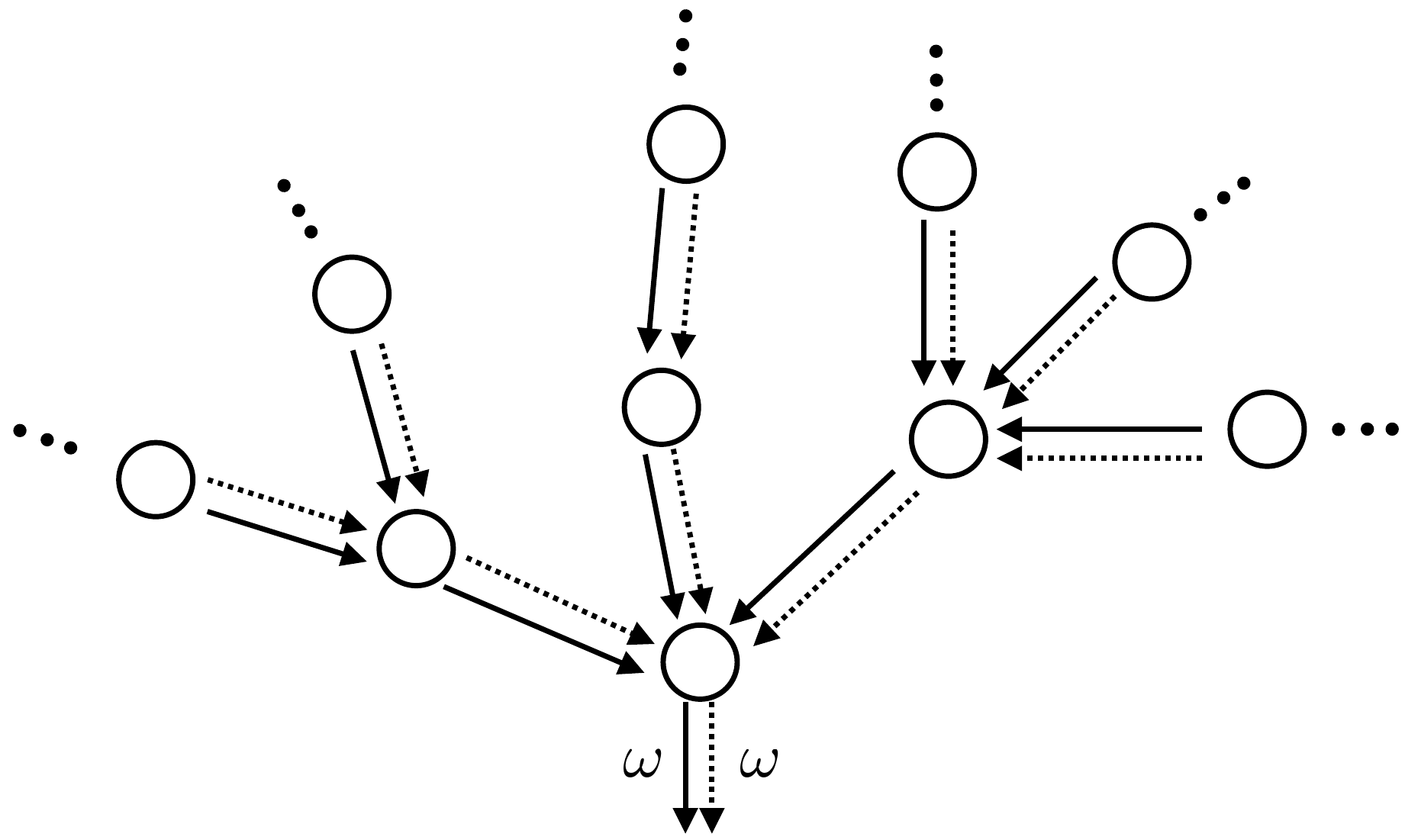}
\caption{Exemplary ladder diagram. Each atom is irradiated by the intensity emitted from other atoms (represented as pairs of solid/dotted arrows) which, in turn, are again irradiated by other atoms, and so on. Notice the tree-like structure leading to the fact that, for each single atom, the incident intensities are uncorrelated with each other.
\label{fig:example_ladder2}}
\end{figure}

Let us first, for simplicity, concentrate on diagrams describing nonlinear and inelastic scattering of light in the atomic cloud, while neglecting the effects of propagation in the atomic medium between two subsequent scattering events. An example of such a diagram is shown in Fig.~\ref{fig:example_ladder2}. To arrive at a complete summation of all these diagrams, we must take into account that each atom can be irradiated by an arbitrary number $n$ of other atoms, and sum over $n$. The spectrum radiated by one atom can then serve as incident spectrum for another atom, and so on. 

Taking the ensemble average over the atomic positions ${\bf r}_i$, the sum of all these ladder diagrams can be expressed as a nonlinear integral equation for the average spectral density
 \begin{equation}
{\mathcal P}({\omega},{\bf r})=\overline{\sum_{j=1}^N P_{jj}(\omega)\delta({\bf r}-{\bf r}_j)}\label{eq:Pav}
\end{equation}
 of the dipole correlation function of an atom placed at position ${\bf r}$:
\begin{eqnarray}
{\mathcal P}(\omega,{\bf r}) & = & {\mathcal N}({\bf r})\sum_{n=0}^\infty \frac{1}{n!}\int_V{\rm d}{\bf r}_1\dots {\rm d}{\bf r}_n\int_{-\infty}^\infty
{\rm d}\omega_1\dots{\rm d}\omega_n
\nonumber\\
& &  \times \left[\prod_{k=1}^n |T(|{\bf r}-{\bf r}_k|)|^2 {\mathcal P}(\omega_k,{\bf r}_k)\right]\nonumber\\
& & \times 
P_{\bf r}(\omega_1,\omega_1,\dots,\omega_n,\omega_n;\omega)^{(+-\dots+-)}\label{eq:laddersc}
\end{eqnarray}
Here, ${\mathcal N}({\bf r})$ denotes the density of atoms at ${\bf r}$, i.e. ${\mathcal N}({\bf r})=0$ for ${\bf r}\notin V$ 
and $\int_V{\rm d}{\bf r}~{\mathcal N}({\bf r})=N$.
The factor $1/n!$ arises from the selection of $n$ out of $N$ atoms (i.e. $\tbinom{N}{n}\simeq N^n/n!$ for $N\gg n$).
Eq.~(\ref{eq:laddersc}) states that the 
spectrum 
${\mathcal P}(\omega,{\bf r})$
of an atom at position ${\bf r}$ is influenced by the spectra emitted from arbitrarily many other atoms placed at ${\bf r}_1,\dots,{\bf r}_n$. Due to the tree-like structure of the ladder diagrams (with different branches referring to different atoms), see Fig.~\ref{fig:example_ladder2}, and due the fact that the atomic positions are distributed independently, these incident spectra are uncorrelated with each other. Therefore, the ensemble average over the product of these spectra can be factorized,
leaving us with a product of average spectra $\prod_k {\mathcal P}(\omega_k,{\bf r}_k)$  on the right-hand side of Eq.~(\ref{eq:laddersc}). 

We now show that the right-hand side of Eq.~(\ref{eq:laddersc}) can be expressed in terms of the spectrum radiated by a single atom under the influence of a classical, stochastic driving field (in addition to the laser field). For this purpose, we represent the continuous frequency variables  
on a discrete lattice of frequencies $\omega_j=j\Delta\omega$. The frequency spacing $\Delta\omega$  must be chosen small enough, such that it does not influence the final result presented below. (From our numerical calculations, we find that  $\Delta\omega\ll \Gamma$ is sufficiently small.) Let us now consider a classical field of the form:
\begin{equation}
E^{\pm}(t) = \sum_{j=-\infty}^\infty \sum_{k=1}^M E_{jk}^{\pm}  e^{\mp i\omega_j t}\label{eq:stochastic1}
\end{equation}
where
\begin{equation}
E_{jk}^{\pm}=\left(\frac{\Delta\omega~\overline{I(\omega_j,{\bf r})}}{2c\epsilon_0M}\right)^{1/2} e^{\pm i\phi_{jk}}\label{eq:components}
\end{equation}
The phases $\phi_{jk}$ are independent random variables uniformly distributed 
in the interval $[0,2\pi]$, and $M$ is a very large number approximately of the same order as the number $N$ of atoms. 
Furthermore, $\overline{I(\omega_j,{\bf r})}$ represents the average spectrum of the light emitted from all  atoms:
\begin{equation}
\overline{I(\omega_j,{\bf r})}=\frac{\hbar^2 c \epsilon_0}{2 d^2\Delta\omega}\int_{\omega_j-\frac{\Delta\omega}{2}}^{\omega_j+\frac{\Delta\omega}{2}} {\rm d}\omega \int_V{\rm d}{\bf r}'|T(|{\bf r}-{\bf r}'|)|^2 {\mathcal P}(\omega,{\bf r}')\label{eq:Ibar}
\end{equation}
 Our claim is that the average spectrum ${\mathcal P}(\omega,{\bf r})$ of the dipole correlation function of an atom at ${\bf r}$ can be calculated by modelling the fields emitted from all other atoms by the stochastic classical field given by Eq.~(\ref{eq:stochastic1}).
 To show this, we remind ourselves of the fact that the building block $P_{\bf r}$ gives the derivatives of the single-atom spectrum  with respect to an incident classical field, see Eqs.~(\ref{eq:Pclassp},\ref{eq:Pclassm}). Let us therefore 
 consider the single-atom spectrum
 \begin{eqnarray}
 P_{\bf r}(\omega) & = & \int_0^\infty\frac{{\rm d}\tau}{2\pi} e^{-i\omega\tau} \langle \sigma^+(\tau)\sigma^-(0)\rangle_{\bf r}\nonumber\\
 & & +\int_0^\infty\frac{{\rm d}\tau}{2\pi} e^{i\omega\tau} \langle \sigma^+(0)\sigma^-(\tau)\rangle_{\bf r}\label{eq:Pcl}
 \end{eqnarray}
 induced by the above stochastic classical  field (in addition to the laser field) in the quasi-stationary state, and expand the dipole correlation functions 
 $\langle \sigma^+(\tau)\sigma^-(0)\rangle_{\bf r}$ and $\langle \sigma^+(0)\sigma^-(\tau)\rangle_{\bf r}$
 in powers of the incident field amplitudes $E^{\pm}_{jk}$, in the same way as $\vec{s}_{\bf r}(t)$ in Eq.~(\ref{eq:taylor}). If we now take the average with respect to the random phases $\phi_{jk}$ (denoted by the overbar $\overline{\phantom{xx}}^{({\rm cl})}$ in the following), we see that only such terms survive the average where each derivative with respect to $E^{+}_{jk}$
  is counterbalanced by a  derivative with respect to the complex conjugate component $E^{-}_{jk}$. Due to the large number of different fields (remember that $M$ is very large), we can furthermore neglect double (and higher) derivatives with respect to the same field component.
Using Eqs.~(\ref{eq:Pclassp},\ref{eq:Pclassm}) (where $\omega=\omega'$),
the average of $P_{\bf r}(\omega)$ with respect to the random phases $\phi_{jk}$ is therefore given by:
 \begin{eqnarray}
\overline{P_{\bf r}(\omega)}^{({\rm cl})} & = & \sum_{n=0}^\infty \sum_{j_1,\dots,j_n=-\infty}^\infty \sum_{k_1,\dots,k_n=1}^{M}  \frac{1}{(2n)!} \left(\begin{array}{c} 2n \\ n\end{array}\right) n! \nonumber\\
& &  \times \left(\frac{2d}{\hbar}\right)^{2n}\frac{\Delta\omega~\overline{I(\omega_{j_1},{\bf r})}}{2c\epsilon_0M}\dots \frac{\Delta\omega~\overline{I(\omega_{j_n},{\bf r})}}{2c\epsilon_0M} \nonumber\\
& & \!\!\!\!\!\!\!\!\!\! \times P_{{\bf r}}(\omega_{j_1},\omega_{j_1},\dots,\omega_{j_n},\omega_{j_n};\omega)^{(+-\dots+-)}\label{eq:classicalspectrum}
\end{eqnarray}
The factor $1/(2n)!$ arises from the Taylor series (as prefactor for all terms involving a $2n$-fold derivative).
The binomial term $\tbinom{2n}{n}$ originates from selecting $n$ out of the $2n$ fields as fields with positive frequency.
Finally, the factor $n!$ describes all possible pairings of the $n$ positive frequency fields with  the $n$ negative frequency fields. Together, these factors yield $1/n!$ and thereby reproduce the corresponding term in Eq.~(\ref{eq:laddersc}).
Moreover,
the sums over $k_1,\dots,k_n$ drop out together with the denominators $1/M$. Together with Eq.~(\ref{eq:Ibar}), we see that the right-hand side of Eq.~(\ref{eq:classicalspectrum}) indeed reproduces, in the limit $\Delta\omega\to 0$, the right-hand side of Eq.~(\ref{eq:laddersc}) [apart from the factor ${\mathcal N}({\bf r})$, which arises in Eq.~(\ref{eq:Pav}) from the probability to find an atom at ${\bf r}$]. Therefore:
\begin{equation}
 {\mathcal P}(\omega,{\bf r})={\mathcal N}({\bf r})\overline{P_{\bf r}(\omega)}^{({\rm cl})}\label{eq:Pomfinal}
 \end{equation} 
which proves our above claim.

Before proceeding, we note that, for large $M$, the field defined by  Eq.~(\ref{eq:stochastic1}) can be simplified as follows: since the intensity resulting from a sum of many fields carrying the same frequency, but with random phases,  is known to fulfill Rayleigh statistics \cite{Goodman:1976aa}, we can rewrite Eq.~(\ref{eq:stochastic1}) as follows:
\begin{equation}
E^{\pm}(t) = \sum_{j=-\infty}^\infty \left(\frac{\Delta\omega I_j}{2c\epsilon_0}\right)^{1/2} e^{\mp i(\omega_j t-\phi_j)}\label{eq:stochastic}
\end{equation}
where, again, $\phi_j$ represents a random phase (uniformly distributed 
in $[0,2\pi]$), whereas $I_j\geq 0$ is a random variable with probability distribution $p(I_j)$ given by the Rayleigh law:
\begin{equation}
p(I_j)=\frac{1}{\overline{I(\omega_j,{\bf r})}} e^{-I_j/\overline{I(\omega_j,{\bf r})}}\label{eq:rayleigh}
\end{equation}
Thereby, the statistical properties of the stochastic classical field are completely characterized in terms of the average spectrum
$\overline{I(\omega_j,{\bf r})}$, see Eq.~(\ref{eq:Ibar}).

\subsection{Average refractive index of the atomic medium}
\label{sec:medium}
 
As already mentioned above, the sum of all ladder diagrams expressed in form of a nonlinear integral equation for the average spectral density ${\mathcal P}(\omega,{\bf r})$ of the atomic dipole correlation function, see Eq.~(\ref{eq:laddersc}), does not take into account effects due to the refractive index of the atomic medium. This becomes evident in the fact that the propagation of photons between atoms in Eq.~(\ref{eq:Ibar}) is described by the function $T(r)$, see Eq.~(\ref{eq:T}), which amounts to propagation in vacuum. 

\begin{figure}
\includegraphics[width=0.4\textwidth]{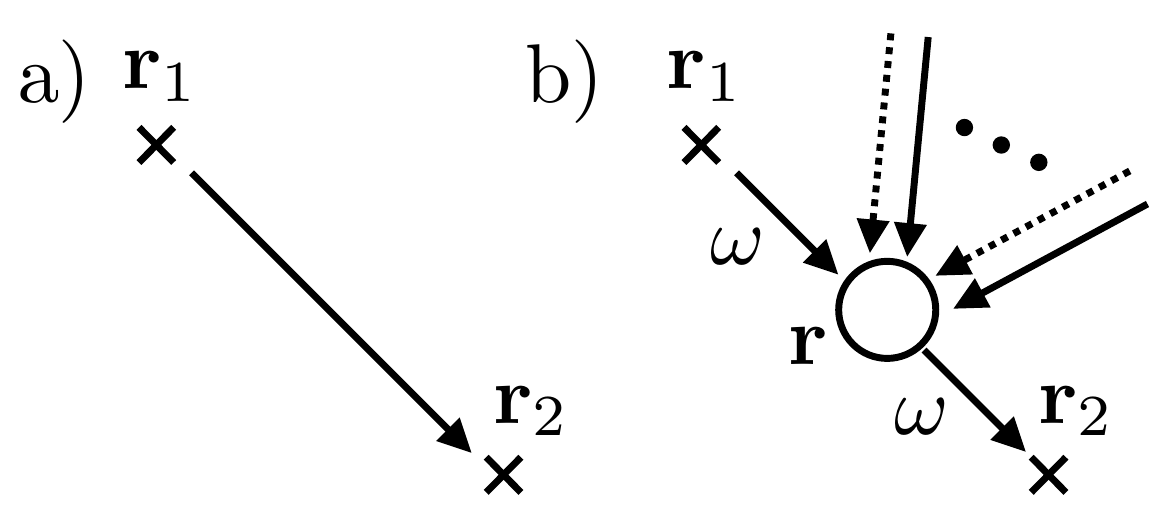}
\caption{a) The propagation of a positive-frequency photon (solid line) between ${\bf r}_1$ and ${\bf r}_2$ in vacuum is described by the coupling constant $T_{12}^*$, see Eq.~(\ref{eq:T}). b) Forward scattering by a single atom placed between ${\bf r}_1$ and ${\bf r}_2$ yields the first-order correction of propagation induced by the atomic medium, and thus determines the refractive index $n_\omega({\bf r})$. The latter, in turn, is influenced by the intensities radiated from other atoms.
 \label{fig:medium2}}
\end{figure}

For a dilute medium, it is possible to calculate the refractive index by considering scattering 
from just one single atom.
 Let us therefore examine the process depicted in Fig.~\ref{fig:medium2}(b). Here, propagation between ${\bf r}_1$ and ${\bf r}_2$ is modified by the presence of an atom at ${\bf r}$. 
 The corresponding single-atom building block can be calculated (summing over arbitrarily many incident intensities emitted from other atoms) in a similar way as above. It turns out that 
 the intensities emitted from the other atoms can again be represented by the 
stochastic classical field introduced in Sec.~\ref{sec:classical}, whereas the additional incident field emitted from ${\bf r}_1$ turns into a partial derivative $\partial/\partial E_\omega^+(t)$:
 \begin{eqnarray}
 \sum_{n=0}^\infty \frac{1}{n!}\int_V{\rm d}{\bf r}_1'\dots {\rm d}{\bf r}_n'\int_{-\infty}^\infty
{\rm d}\omega_1\dots{\rm d}\omega_n & & 
\nonumber\\
\times \left[\prod_{k=1}^n |T(|{\bf r}-{\bf r}_k'|)|^2 {\mathcal P}(\omega_k,{\bf r}_k')\right] & & \label{eq:sigmamedium}\\
\times 
s_{\bf r}^-(\omega_1,\omega_1,\dots,\omega_n,\omega_n,\omega)^{(+-\dots+--)} & = & \frac{\hbar}{2d}\overline{\frac{\partial s_{\bf r}^-(t)}{\partial E_\omega^+(t)}}^{\,({\rm cl})}\nonumber
\end{eqnarray}
To evaluate the right-hand side of Eq.~(\ref{eq:sigmamedium}), we consider the solution $s_{\bf r}^-(t)$ of the single-atom Bloch equation (\ref{eq:blochclassical}) in the presence of the stochastic field defined above, plus an additional weak probe field 
with frequency $\omega$. As explained in Appendix~\ref{sec:deriv}, the derivative $\partial s_{\bf r}^-(t)/\partial E_\omega^+(t)$ of $s_{\bf r}^-(t)$ with respect to this probe field fulfills an equation similar to the optical Bloch equation for $\vec{s}_{\bf r}(t)$. Finally, the solution of this equation must be averaged over many realizations of the  stochastic classical field. 
The fact that this average is reproduced by the diagrammatic expression given on the left-hand side of Eq.~(\ref{eq:sigmamedium}) 
can be proven by expanding $\partial s_{\bf r}^-(t)/\partial E_\omega^+(t)$ into a Taylor series with respect to the 
stochastic field components, using Eqs.~(\ref{eq:sigmaclass}) and (\ref{eq:Ibar}), and performing the average over the stochastic field in the same way as explained between Eqs.~(\ref{eq:Pcl}) and (\ref{eq:classicalspectrum}). Note that the quantity on the right-hand side of Eq.~(\ref{eq:sigmamedium}) is proportional to the average electric susceptibility of the atomic medium, since it describes 
the change of the atomic dipole $s_{\bf r}^-$ induced by a weak field $E_\omega^+$ with frequency $\omega$.

To determine how the atomic medium thereby affects the propagation of a photon  from ${\bf r}_1$ to ${\bf r}_2$,
 we perform the average over the position ${\bf r}$ in volume $V$ with density ${\mathcal N}({\bf r})$ in stationary phase approximation  \cite{roemer}, using the fact  that the average susceptibility defined on the right-hand side of Eq.~(\ref{eq:sigmamedium})
 and the atomic density ${\mathcal N}({\bf r})$ do not strongly vary when changing ${\bf r}$ by a distance of the order of the wave length:
 \begin{eqnarray}
 & &\int_V {\bf d}{\bf r}~  {\mathcal N}({\bf r})~T^*(|{\bf r}_2-{\bf r}|)  \frac{\hbar}{2d}\overline{\frac{\partial s_{\bf r}^-(t)}{\partial E_\omega^+(t)}}^{\,({\rm cl})} T^*(|{\bf r}-{\bf r}_1|)   =\nonumber\\
 & &  = \frac{i \Gamma d}{2 \epsilon_0}e^{ik_L r_{12}}\int_0^1{\rm d}s~{\mathcal N}({\bf r})\left.\overline{\frac{\partial s_{\bf r}^-(t)}{\partial E_\omega^+(t)}}^{\,({\rm cl})}\right|_{{\bf r}=s{\bf r}_2+(1-s){\bf r}_1}\label{eq:TT2}
 \end{eqnarray}
where the integral is taken over a straight line connecting ${\bf r}_1$ and ${\bf r}_2$. 
Let us  compare this expression with the one that we obtain when  introducing a position- and frequency-dependent refractive index $n_{\omega}({\bf r})$ into the definition (\ref{eq:T}) of the vacuum propagator $T(r)$:
\begin{eqnarray}
T^*_\omega({\bf r}_2,{\bf r}_1) & = & \frac{\Gamma}{k_L r_{12}} \exp\Biggl[ik_L r_{12} \Biggr.\nonumber\\
& & \ \ \ \ \times\Biggl.\int_0^1{\rm d}s~n_\omega\left(s{\bf r}_2+(1-s){\bf r}_1\right)\Biggr]\label{eq:Tom}
\end{eqnarray}
Expanding this function in $n_\omega$ around $n_\omega=1$, 
the resulting first-order term (since we consider scattering by only a single atom) 
coincides with
Eq.~(\ref{eq:TT2}) if
\begin{equation}
n_\omega({\bf r})= 1+\frac{d{\mathcal N}({\bf r})}{2\epsilon_0} \overline{\frac{\partial s_{\bf r}^-(t)}{\partial E_\omega^+(t)}}^{({\rm cl})}\label{eq:index}
\end{equation}
Terms of higher order in $n_\omega$, which are included in Eq.~(\ref{eq:Tom}), are described by diagrams where the photon $\omega$ is scattered by more than one atom while propagating from ${\bf r}_1$ to ${\bf r}_2$. 
The resulting refractive index then depends on  the frequency $\omega$ of the propagating photon, and -- through the stochastic field average -- also on the fields emitted from other atoms, which, in turn, depend on the position ${\bf r}$ inside the atomic cloud.

The imaginary part of the refractive index leads to an exponential damping in Eq.~(\ref{eq:Tom}), and thus yields the inverse of the scattering mean free path $\ell_\omega({\bf r})$:
\begin{equation}
 \frac{1}{\ell_\omega({\bf r})}=2 k_L {\rm Im}\left\{n_\omega({\bf r})\right\}\label{eq:ellom}
 \end{equation}
As already indicated in the title of this subsection, the refractive index described  in terms of ladder diagrams implies an average over the atomic positions. For a single realization, the elastic component of the intensity inside the atomic medium exhibits short-range \lq speckle\rq\ fluctuations, giving rise to fluctuations of the refractive index around its average value given by Eq.~(\ref{eq:index}) above. These fluctuations induce additional scattering processes \cite{Schwiete:2013aa}
which are ignored in our treatment, since they can be neglected in the case of a dilute medium.
 
\subsection{Attenuation of the incident laser beam}

\begin{figure}
\includegraphics[width=0.15\textwidth]{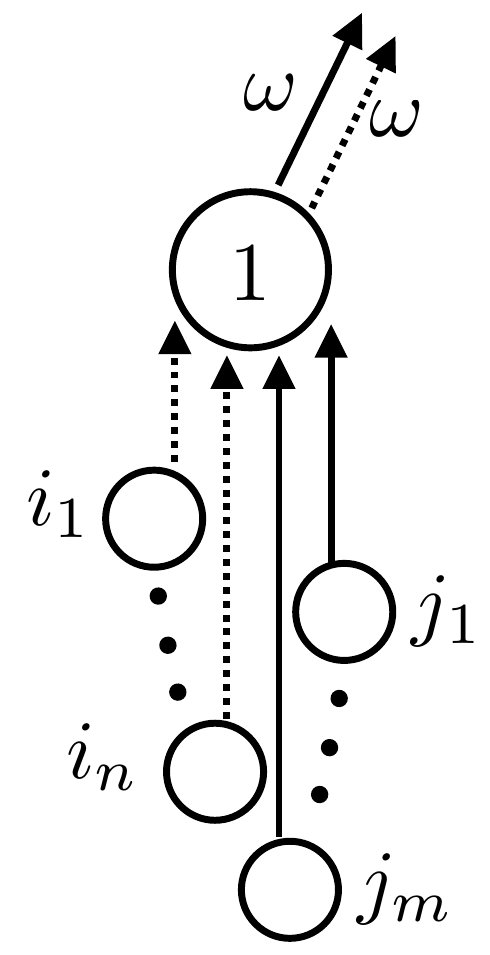}
\caption{The  fields radiated by atoms $i_1,\dots i_n$ and $j_1,\dots,j_m$ (carrying negative- and positive-frequency photons, respectively) interfere destructively with the  laser, and thus lead to an attenuation of the laser beam incident on atom 1. \label{fig:laser}}
\end{figure}

The last ingredient needed for a complete summation of all ladder diagrams is the attenuation of the incident laser beam due to scattering in the atomic medium. For this purpose, let us examine the diagram shown in Fig.~\ref{fig:laser}. Since we seek a result which is non-perturbative in the laser amplitude, we consider an arbitrary number $n+m$ of different atoms emitting negative- and positive-frequency photons, respectively, which, as shown in the following, interfere with the incident laser. Remember that each of the atoms $i_1,\dots,j_m$ is itself irradiated by the intensities emitted from other atoms, which, for simplicity, are not explicitly indicated in Fig.~\ref{fig:laser}. 
Again in a similar way as in Sec.~\ref{sec:classical}, it can be shown that 
this amounts to replacing the corresponding single-atom building blocks $s_{{\bf r}_i}^+$  and $s_{{\bf r}_j}^-$  by 
the stochastic field averages
$\overline{s_{{\bf r}_i}^+}^{\,({\rm cl})}$ and $\overline{s_{{\bf r}_j}^-}^{\,({\rm cl})}$. 
Thus, the contribution of the diagram shown in Fig.~\ref{fig:laser} reads:
\begin{eqnarray}
P_{11}^{({\rm F}\ref{fig:laser})}(\omega) & = & \sum_{n,m=0}^\infty \left(\prod_{k=1}^nT_{i_k1}
\overline{s_{{\bf r}_{i_k}}^+}^{\,({\rm cl})}  \right)\nonumber\\
& & \!\!\!\!\!\!\!\!\!\!\!\!\!\!\!\!\!\!\!\!\!\!\!\!\!\!\!\!\!\!\!\!\! \times \left(\prod_{l=1}^m T^*_{j_l1}  \overline{s_{{\bf r}_{j_l}}^-}^{\,({\rm cl})}\right) P_{{\bf r}_1}(0,\dots,0;\omega)^{(+\dots+-\dots-)}\label{eq:fig7}
\end{eqnarray}
We recognize, again, the occurrence of the single-atom building block $P_{{\bf r}_1}^{(\alpha_1,\dots,\alpha_{n+m})}$ carrying labels $\alpha_1,\dots,\alpha_{n+m}$ with 
$\alpha_1,\dots,\alpha_n=+1$ and $\alpha_{n+1},\dots,\alpha_{n+m}=-1$. The frequencies of all incident photons are equal to the laser frequency (i.e., frequency zero in the rotating frame). 
This can be traced back to the fact that, in the quasi-stationary regime,  the  atomic Bloch vector oscillates with the same frequency as the laser (i.e., it is time-independent in the rotating frame).  

Using Eqs.~(\ref{eq:Pclassp},\ref{eq:Pclassm}), we see that Eq.~(\ref{eq:fig7}) represents the complete Taylor series (ignoring multiple derivatives with respect to the same field component) of the spectrum of a single atom driven by the following time-independent (in the rotating frame) classical fields:
\begin{equation}
E^{+}=\frac{\hbar}{2d}\sum_{j=2}^N T_{j1}^*  \overline{s_{{\bf r}_j}^-}^{\,({\rm cl})},\ \ E^{-}=\frac{\hbar}{2d}\sum_{i=2}^N T_{i1}  \overline{s_{{\bf r}_i}^+}^{\,({\rm cl})}\label{eq:attlaser}
\end{equation}
which must be added to the positive- and negative-frequency amplitudes $E_L\exp(\pm i{\bf k}_L\cdot{\bf r}_1)/2$
 of the laser field at atom 1. After averaging over the positions ${\bf r}_2,\dots,{\bf r}_N$ (and assuming $N\gg 1$),
 the sum over all atoms except atom 1 in Eq.~(\ref{eq:attlaser}) can be represented as an integral over the atomic cloud: 
\begin{equation}
E^{+}=\frac{\hbar}{2d}\int_V{\rm d}{\bf r}~{\mathcal N}({\bf r})T^*(|{\bf r}-{\bf r}_1|)\overline{s_{\bf r}^-}^{\,({\rm cl})}\label{eq:attlaser2}
\end{equation}
Since the position-dependent phase of  $\overline{s_{\bf r}^-}^{\,({\rm cl})}$ is given by the phase of the incident laser $E_L\exp(i{\bf k}_L\cdot{\bf r})/2$, 
the above integral  yields in 
stationary phase approximation:
\begin{equation}
E^{+}  =  ik_L e^{i{\bf k}_L\cdot{\bf r}_1}
\int_0^\infty {\rm d}z'~ \left.\frac{d{\mathcal N}({\bf r})}{2\epsilon_0}~
\frac{\overline{s_{\bf r}^-}^{\,({\rm cl})}}{e^{i {\bf k}_L\cdot {\bf r}}}\right|_{{\bf r}={\bf r}_1-z' {\bf e}_L}\label{eq:attlaser3}
\end{equation}
where ${\bf e}_L={\bf k}_L/k_L$ is the unit vector pointing in the direction of the incident laser. Since ${\mathcal N}({\bf r})=0$ if ${\bf r}\notin V$ (i.e. if the point ${\bf r}$ lies outside the atomic medium $V$), the integral over $z'$ can be restricted to a finite interval corresponding to the distance the laser beam covers in the medium before reaching ${\bf r}_1$. 

The attenuated laser field at position ${\bf r}_1$ then results as:
\begin{equation}
E_L^+({\bf r}_1)  =  \frac{E_L e^{i{\bf k}_L\cdot{\bf r}_1}}{2}+E^+\label{eq:attlaser4}
\end{equation}
When evaluating the stochastic field average $\overline{s_{\bf r}^-}^{\,({\rm cl})}$ in Eq.~(\ref{eq:attlaser3}), we 
have to use the Rabi frequency $\Omega({\bf r})=2 d E_L^+({\bf r})/\hbar$  associated with the attenuated laser field (since, as mentioned above, the atoms $i_1,\dots,j_m$ are irradiated by the fields emitted from other atoms).
Eqs.~(\ref{eq:attlaser3},\ref{eq:attlaser4}) can be rewritten in the following, more intuitive form:
\begin{equation}
E^+_L({\bf r}) = \frac{E_L}{2} e^{i{\bf k}_L\cdot {\bf r}}\,\exp\left[i k_L \int_0^\infty {\rm d}z'~\Bigl(n_L({\bf r}-z' {\bf e}_L)-1\Bigr)\right] \ \label{eq:laser}
\end{equation}
with refractive index
\begin{equation}
n_L({\bf r})=1+\frac{d{\mathcal N}({\bf r})}{2\epsilon_0} \frac{\overline{s_{\bf r}^-}^{\,({\rm cl})}}{E_L^+({\bf r})}\label{eq:nL}
\end{equation}
The corresponding mean free path $\ell_L({\bf r})$ is obtained from the imaginary part of $n_L({\bf r})$ in the same way as in Eq.~(\ref{eq:ellom}).
The refractive index $n_L({\bf r})$ for the laser is different from the refractive index $n_\omega({\bf r})$ for the fields scattered between atoms, see Eq.~(\ref{eq:index}). This is not surprising, since the laser field is strong, whereas Eq.~(\ref{eq:index}) has been derived under the assumption of a  weak field (described by a single photon) propagating from one atom to another one. Only for a weak laser beam, where the quotient in Eq.~(\ref{eq:nL}) can be interpreted as a derivative, the two refractive indices $n_L({\bf r})$ and $n_\omega({\bf r})$ coincide  (for $\omega=0$).

Finally, let us remark that the above treatment of average propagation must be modified in the case 
of more than one incident laser beam. Then, additional coherent components are produced by four-wave mixing processes \cite{boyd}. These are described by diagrams which are neither ladder nor crossed diagrams, but nevertheless fulfill a phase-matching condition, such as, for example, 
$\exp[i({\bf k}_1+{\bf k}_2)\cdot {\bf r}]=1$ in the case ${\bf k}_1=-{\bf k}_2$ of two opposite incident laser beams.

\subsection{Ladder transport equations}
\label{sec:ladder_transport}
  
 We now have all ingredients at 
 hand to formulate our final result - which amounts to the complete summation of
 all ladder diagrams. First, we have shown in Sec.~\ref{sec:classical}, see Eq.~(\ref{eq:Pomfinal}), that the average spectral density  ${\mathcal P}(\omega,{\bf r})$ of the dipole correlation function of an atom placed at position ${\bf r}$ inside the atomic cloud is obtained as ${\mathcal P}(\omega,{\bf r})={\mathcal N}({\bf r})\overline{P_{\bf r}(\omega)}^{({\rm cl})}$. By virtue of Eq.~(\ref{eq:Pcl}),
 $\overline{P_{\bf r}(\omega)}^{({\rm cl})}$ represents the spectrum of an atom placed at ${\bf r}$ driven, both, by a classical stochastic field representing the radiation emitted from all other atoms, and by the laser with associated Rabi frequency $\Omega({\bf r})=2 d E_L^+({\bf r})/\hbar$. The latter,  
taking into account the attenuation due to scattering in the atomic medium, is given by Eqs.~(\ref{eq:laser},\ref{eq:nL}).
The stochastic properties of the classical field, see Eqs.~(\ref{eq:stochastic},\ref{eq:rayleigh}), are completely characterized by the average spectrum $\overline{I(\omega_n,{\bf r})}$ of the scattered field on a sufficiently fine, discrete grid of frequencies $\omega_n=n\Delta\omega$. This spectrum exhibits an inelastic and an elastic component,
\begin{equation}
\overline{I(\omega_n,{\bf r})}=\overline{I^{({\rm in})}(\omega_n,{\bf r})}+\frac{\delta_{n,0}}{\Delta\omega}\overline{I^{({\rm el})}({\bf r})}
\end{equation}
 which, in turn, are determined by the corresponding atomic dipole spectra
as follows:
\begin{eqnarray}
\overline{I^{({\rm in})}(\omega_n,{\bf r})} & = &   \frac{\hbar^2 c\epsilon_0}{2d^2}\int_V{\rm d}{\bf r}'| T_{\omega_n}({\bf r},{\bf r}')|^2 {\mathcal P}^{({\rm in})}(\omega_n,{\bf r}')\ \ \ \ \ \label{eq:Ibarin}\\
\overline{I^{({\rm el})}({\bf r})} & = &   \frac{\hbar^2 c\epsilon_0}{2d^2}\int_V{\rm d}{\bf r}'| T_{\omega=0}({\bf r},{\bf r}')|^2
{\mathcal P}^{({\rm el})}({\bf r}')\label{eq:Ibarel}
\end{eqnarray}
where, as compared to Eq.~(\ref{eq:Ibar}), the effect of the atomic medium has been taken into account through 
$T_\omega({\bf r},{\bf r}')$, see Eqs.~(\ref{eq:Tom},\ref{eq:index}). Furthermore, we have split ${\mathcal P}(\omega,{\bf r})$ into its inelastic and elastic components: 
\begin{equation}
{\mathcal P}(\omega,{\bf r})={\mathcal P}^{({\rm in})}(\omega,{\bf r})+\delta(\omega){\mathcal P}^{({\rm el})}({\bf r})
 \label{eq:Pinel}
\end{equation}
where, according to Eqs.~(\ref{eq:Pcl},\ref{eq:Pomfinal}), the latter is obtained as
\begin{equation}
{\mathcal P}^{({\rm el})}({\bf r})={\mathcal N}({\bf r}) \lim_{\tau\to+\infty} {\rm Re}\left\{ \overline{\langle\sigma^+(\tau)\sigma^-(0)\rangle}_{\bf r}^{({\rm cl})}\right\}
\end{equation}
The limit $\tau\to\infty$ exists only {\em after} taking the classical field average, since no truly stationary state is reached for a single realization of the polychromatic classical field, see the discussion after Eq.~(\ref{eq:blochclassical}).
The inelastic component ${\mathcal P}^{({\rm in})}(\omega,{\bf r})$ is obtained from Eq.~(\ref{eq:Pcl}), after subtracting from the dipole correlation functions
their asymptotic values reached at $\tau\to\infty$.

The above coupled system of equations can be solved numerically by an iterative procedure. Initially, there are no scattered fields, i.e., $\overline{I(\omega_n,{\bf r})}=0$ and the laser $E_L^+({\bf r})=E_L e^{i{\bf k}_L\cdot {\bf r}}/2$ is given by a plane wave. We then calculate, in a first iteration step, the spectra of the atomic dipoles ${\mathcal P}(\omega_n,{\bf r})$ at each position inside the atomic cloud. According to Eq.~(\ref{eq:Pomfinal}), this involves the solution of single-atom Bloch equations for a large number of realizations of the stochastic field, with subsequent averaging. In a similar way, the refractive indices 
$n_L({\bf r})$ and $n_{\omega_n}({\bf r})$ are obtained by solving Eqs.~(\ref{eq:index},\ref{eq:nL}). The laser amplitude follows through Eq.~(\ref{eq:laser}), and finally, the spectrum of the scattered light via Eqs.~(\ref{eq:Ibarin},\ref{eq:Ibarel}). This scheme is repeated until convergence is achieved.

Finally, the average normalized spectrum, see Eqs.~(\ref{eq:ID},\ref{eq:bistatic}), measured by a detector placed in the far field follows as:
\begin{eqnarray}
\gamma_L(\omega,{\bf e}_D) & = & \frac{\hbar\omega_0\Gamma}{I_L} \int_V\frac{{\rm d}{\bf r}}{{\mathcal A}}~{\mathcal P}(\omega,{\bf r}) \nonumber\\
&  & \!\!\!\!\!\!\!\!\!\!\!\!\!\!\!\!\!\!\!\!\!\!\!\!\!\!\!\!\!\!\!\!\!\!\!\! \times\exp\left[-\int_0^{\infty}{\rm d}s~\frac{1}{\ell_{\omega}({\bf r}+s {\bf e}_D)}\right]\label{eq:IDfinal}
\end{eqnarray}
Again, the integral over $s$ can be restricted to a finite interval
since the scattering mean free path $\ell_{\omega}({\bf r}+s{\bf e}_D)$ tends to infinity if the point ${\bf r}+s{\bf e}_D$ lies outside the scattering volume $V$. Without the exponential factor, Eq.~(\ref{eq:IDfinal}) reproduces the ensemble average of the diagonal terms ($i=l$) in Eq.~(\ref{eq:ID}), as can be seen from Eq.~(\ref{eq:Pav}). The exponential factor then takes into account additional (non-diagonal) terms arising in the ladder approximation due to  the 
final step of propagation from the last scattering event through the atomic medium towards the detector.

Taking into account Eq.~(\ref{eq:Pinel}), it is possible to extract  the elastic and inelastic components of the detected light from Eq.~(\ref{eq:IDfinal}):
\begin{equation}
\gamma_L(\omega,{\bf e}_D)=\gamma^{({\rm in})}_L(\omega,{\bf e}_D)+\delta(\omega)\gamma^{({\rm el})}_L({\bf e}_D)\label{eq:gammaLinel}
\end{equation}
As shown in Appendix~\ref{sec:flux}, $\gamma_L(\omega,{\bf e}_D)$ fulfills the property of flux conservation after integration over the frequency $\omega$ and the angles ${\bf e}_D$ and adding the flux of the coherently transmitted light. 

The coupled set of ladder transport equations (\ref{eq:Pcl}), (\ref{eq:Pomfinal}-\ref{eq:rayleigh}), (\ref{eq:Tom}-\ref{eq:ellom}) and (\ref{eq:laser}-\ref{eq:gammaLinel}) possesses a physically transparent structure, which can also be explained without using diagrams. The most important assumption is the one that the intensities emitted from different atoms are uncorrelated with each other. A light field of this form can be modelled as a classical field, since quantum properties of light become apparent only in the form of intensity-intensity correlations \cite{Kimble:1977aa}. Unlike the incoming laser field, this classical field is not purely coherent, but exhibits stochastic properties. The stochasticity can be traced back to two different physical reasons: First, the quantum-mechanical fluctuations of the atomic dipoles  induce a certain probabilistic frequency distribution of the scattered fields. Second, the classical average over the atomic positions  leads to a Rayleigh distribution
of the intensities at each single frequency component, see Eq.~(\ref{eq:rayleigh}). 
Finally, also the expressions for the refractive indices $n_\omega({\bf r})$ and $n_L({\bf r})$, see Eqs.~(\ref{eq:index},\ref{eq:nL}), can be understood in terms of the susceptibility of the atomic dipoles with respect to the  small scattered fields, and by the fact that the elastically forward-scattered light is phase-coherent with the incident laser, and thus attenuates the latter by destructive interference.

Nevertheless, the diagrammatic approach is useful for giving a more rigorous justification of the above heuristic arguments. Furthermore, it allows us to include, in a systematic way, the influence of nonlinear and inelastic scattering on interference effects leading  to weak localization and coherent backscattering (see the following Sec.~\ref{sec:crossed}), which, up to now, can only be explained within the diagrammatic approach.

\section{Summation of crossed diagrams}
\label{sec:crossed}

As discussed in Sec.~\ref{sec:laddercrossed} above, crossed diagrams describe the interference between fields emitted from different atoms, which gives rise to a coherent backscattering peak around the direction ${\bf k}_D\simeq -{\bf k}_L$ opposite to the incident laser beam. As evident from the example shown in Fig.~\ref{fig:ladder_meq}, crossed diagrams are constructed from the ladder diagrams discussed in Sec.~\ref{sec:ladder} by reversing a single photon line. In Fig.~\ref{fig:ladder_meq}(b), for example, the path of the positive-frequency photon (solid lines) propagating from atom 3 to 6 via the intermediate atoms 4 and 5 in Fig.~\ref{fig:ladder_meq}(a) is reversed. This leaves us with a pair of counterpropagating paths (solid and dotted lines pointing in opposite directions) between atoms 3 and 6, respectively. Due to the condition of energy conservation, see Eq.~(\ref{eq:omegap}), the frequencies $\omega_i'$ and $\omega_i$ of counterpropagating photons are related by $\omega_i'=\omega-\omega_i$. Here, and in the remainder of this section, $\omega$ always denotes the frequency of the detected photon.
As shown in the following, the sum of all crossed diagrams can be expressed as the solution of an integral  equation describing transport of a counterpropagating pair of amplitudes through the atomic medium. Diagrams with more than a single pair of counterpropagating amplitudes cannot occur due to our restriction to building blocks with at most one outgoing dashed and/or solid arrow, see Fig.~\ref{fig:bblock}.

\subsection{Crossed building blocks}

\begin{figure}
\includegraphics[width=0.45\textwidth]{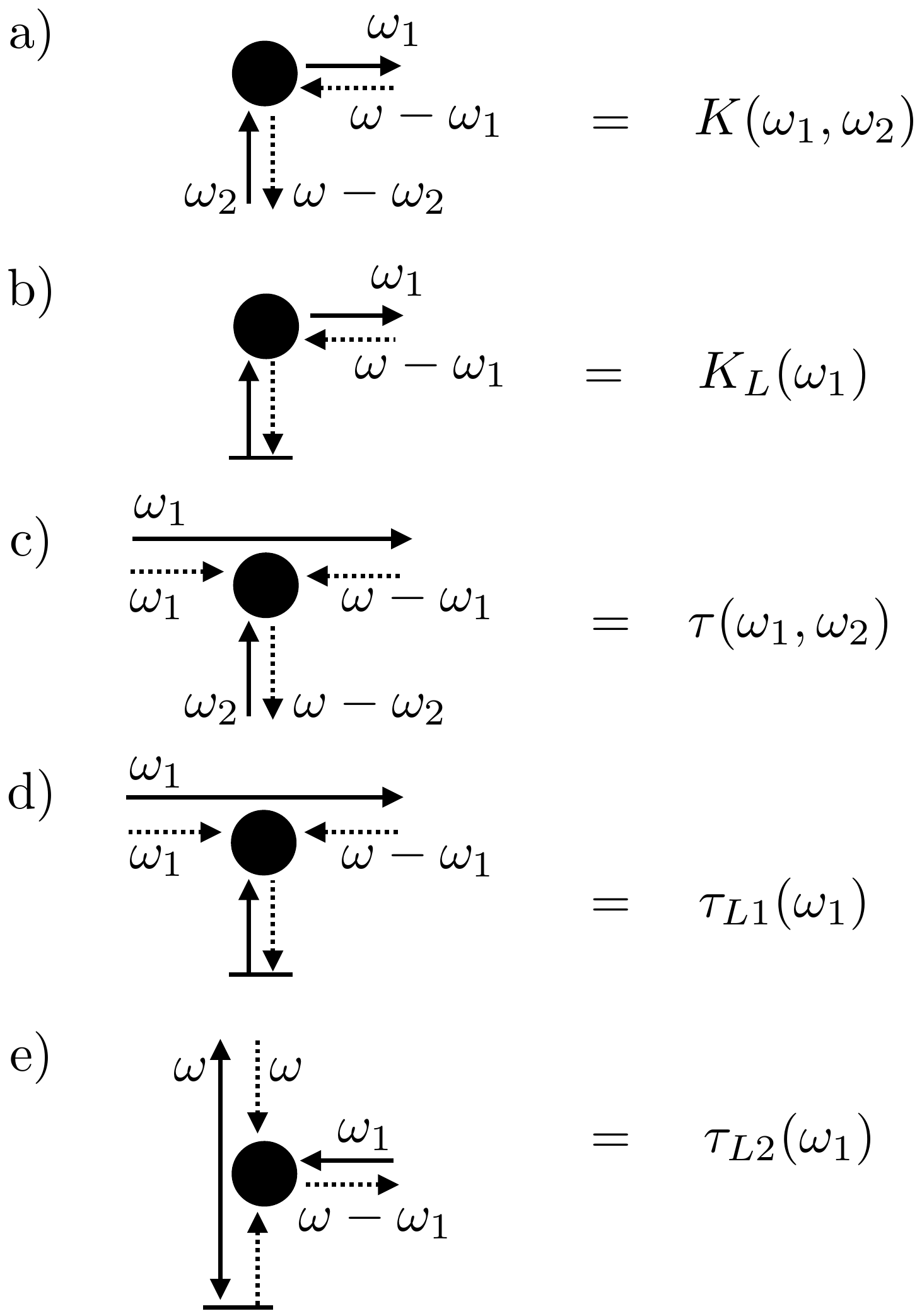}
\caption{Building blocks for crossed diagrams, describing scattering events for counterpropagating pairs of amplitudes. Arrows attached to the horizontal bar at the bottom of b), d), and e) describe photons originating from the laser mode or propagating towards the detector, respectively. In e), the solid line may propagate in either one of the two indicated directions. The black circles indicate atoms that are driven by the laser field and the stochastic classical field representing radiation emitted from other atoms. The corresponding complex conjugate building blocks (not shown) are obtained by exchanging solid with dotted lines (while keeping the arrows' directions).
\label{fig:crossedbb}}
\end{figure}

Let us first identify the building blocks which any crossed diagram is composed of. Since we are interested in the counterpropagating pair of amplitudes, the diagrams presented below only indicate the photon exchanges associated with these amplitudes. In addition, each atom may be \lq dressed\rq\ by an arbitrary number of incoming ladder intensities. In a similar way as above, it can again be proven that these incoming ladder intensities 
may be represented by the classical stochastic field introduced in Sec.~\ref{sec:classical}. For example, in Fig.~\ref{fig:ladder_meq}(b), we see that atom 6 is subject to a ladder intensity emitted from atom 2. 
Atoms driven by this stochastic field (in addition to the laser) are represented by a filled circle in the following. Then, any crossed diagram describing counterpropagating amplitudes can be constructed from the building blocks depicted in Fig.~\ref{fig:crossedbb}.

The corresponding equations for these building blocks are the following:
\begin{eqnarray}
K(\omega_1,\omega_2,{\bf r}) & = & \frac{\hbar^2{\mathcal N}({\bf r})}{4d^2}\int_0^\infty\frac{{\rm d}\tau}{2\pi} \Biggl(\Biggr.\nonumber\\
& & 
e^{-i\omega_1\tau} \overline{\frac{\partial^2 \langle \sigma^+(\tau)\sigma^-(0)\rangle_{\bf r}}{\partial E^+_{\omega_2}(\tau)\partial E^-_{\omega-\omega_1}(\tau)}}^{\,({\rm cl})} \nonumber\\
&  & \!\!\!\!\!\!\!\!\!\!\!\!\!\!\! + \Biggl. e^{i(\omega-\omega_2)\tau} \overline{\frac{\partial^2 \langle \sigma^+(0)\sigma^-(\tau)\rangle_{\bf r}}{\partial E^+_{\omega_2}(\tau)\partial E^-_{\omega-\omega_1}(\tau)}}^{\,({\rm cl})}\Biggr)\label{eq:K}\\
 K_L(\omega_1,{\bf r}) & = & \frac{\hbar{\mathcal N}({\bf r})}{2d E_L^+({\bf r})}\int_0^\infty\frac{{\rm d}\tau}{2\pi} \Biggl(\Biggr.\nonumber\\
& &  e^{-i\omega_1\tau} \overline{\frac{\partial \langle \sigma^+(\tau)\sigma^-(0)\rangle_{\bf r}}{\partial E^-_{\omega-\omega_1}(\tau)}}^{\,({\rm cl})}\nonumber\\
 & & \Biggl.+ e^{i\omega\tau} \overline{\frac{\partial \langle \sigma^+(0)\sigma^-(\tau)\rangle_{\bf r}}{\partial E^-_{\omega-\omega_1}(\tau)}}^{\,({\rm cl})}\Biggr)\label{eq:KL}
 \end{eqnarray}
\begin{eqnarray}
 \tau(\omega_1,\omega_2,{\bf r}) & = & -\frac{i\hbar^2 k_L{\mathcal N}({\bf r})}{8\epsilon_0 d}\nonumber\\
& & \times\overline{\frac{\partial^3 s_{\bf r}^+(t)}{~\partial E_{\omega_2}^+(t)\partial E_{\omega_1}^-(t)\partial E_{\omega-\omega_1}^-(t)}}^{\,({\rm cl})}\ \ \ \ \label{eq:tau}\\
 \tau_{L1}(\omega_1,{\bf r}) & = & -\frac{i\hbar k_L{\mathcal N}({\bf r})}{4\epsilon_0  E_L^+({\bf r})}\overline{\frac{\partial^2 s_{\bf r}^+(t)}{\partial E_{\omega_1}^-(t)\partial E_{\omega-\omega_1}^-(t)}}^{\,({\rm cl})}\label{eq:tauL1}\\
 \tau_{L2}(\omega_1,{\bf r}) & = & -\frac{i\hbar k_L{\mathcal N}({\bf r})}{4\epsilon_0  E_L^-({\bf r})}
 \overline{\frac{\partial^2 s_{\bf r}^+(t)}{\partial E_{\omega_1}^+(t)\partial E_{\omega}^-(t)}}^{\,({\rm cl})}\label{eq:tauL2}
\end{eqnarray}
As discussed in Sec.~\ref{sec:medium}, each incoming arrow leads to a partial derivative with respect to the corresponding field, together with a prefactor $\hbar/(2d)$. For example, in Fig.~\ref{fig:crossedbb}(a), there is an incoming positive frequency photon $\omega_2$ and a negative-frequency photon $\omega-\omega_1$, which turn into partial derivatives $\partial/\partial E^+_{\omega_2}$ and $\partial/\partial E^-_{\omega-\omega_1}$ in Eq.~(\ref{eq:K}). In addition, there is a prefactor ${\mathcal N}({\bf r})$ taking into account the probability to find an atom at ${\bf r}$. The building blocks denoted by $\tau$, see Fig.~\ref{fig:crossedbb}(c-e) and Eqs.~(\ref{eq:tau}-\ref{eq:tauL2}), obtain an additional factor $-2 \pi i\Gamma/k_L^2=-ik_Ld^2/(\hbar\epsilon_0)$, see Eq.~(\ref{eq:gamma}), originating from the integral over ${\bf r}$ evaluated in stationary phase approximation, see Eq.~(\ref{eq:TT2}). In the following transport equations, the integral over ${\bf r}$ is then restricted to a straight line defined by the positions of other building blocks to which the building blocks $\tau$ are attached. 

Photons originating from the laser mode are explicitly indicated in diagrams Fig.~\ref{fig:crossedbb}(b), (d) and (e) by an arrow attached to a horizontal bar. These photons do not represent partial derivatives, but, instead, lead to a denominator $1/E_L^+({\bf r})$ or $1/E_L^-({\bf r})$ (for solid and dotted arrows, respectively) in Eqs.~(\ref{eq:KL},\ref{eq:tauL1}) and (\ref{eq:tauL2}), where $E_L^-({\bf r})=[E_L^+({\bf r})]^*$.
This turns the quantities $K_L$, $\tau_{L1}$ and $\tau_{L2}$ into smoothly varying  functions of ${\bf r}$.
In the following transport equations, these denominators must be cancelled by corresponding multiplications with $E_L^+({\bf r})$ or $E_L^-({\bf r})$, which, in turn, compensate the phases of the corresponding counterpropagating photons propagating towards the detector (if the latter is placed in exact backscattering direction).

Furthermore, each of the diagrams shown in Fig.~\ref{fig:crossedbb} exhibits a complex conjugate counterpart obtained by exchanging solid with dotted lines. In case of Fig.~\ref{fig:crossedbb}(a), the complex conjugate diagram is identical to the original one with relabeled frequencies, i.e., $K^*(\omega_1,\omega_2)=K(\omega-\omega_2,\omega-\omega_1)$. The complex conjugate counterparts of Fig.~\ref{fig:crossedbb}(b-e), however, give rise to new building blocks $K_L^*(\omega_1),\dots,\tau_{L2}^*(\omega_1)$ which must be taken into account separately in
the transport equations derived hereafter.

Finally, we note that the building block $\tau$ depicted in Fig.~\ref{fig:crossedbb}(c) (and, similarly, $\tau_{L1}$ and $\tau_{L2}$) can be interpreted as an optical phase conjugation \cite{Kravtsov:1990aa},
where the two dotted arrows $\omega_1$ and $\omega-\omega_1$ play the role of two counterpropagating pump beams, which reverse the
phase and direction of the incident photon $\omega_2$.

\begin{figure}
\includegraphics[width=0.45\textwidth]{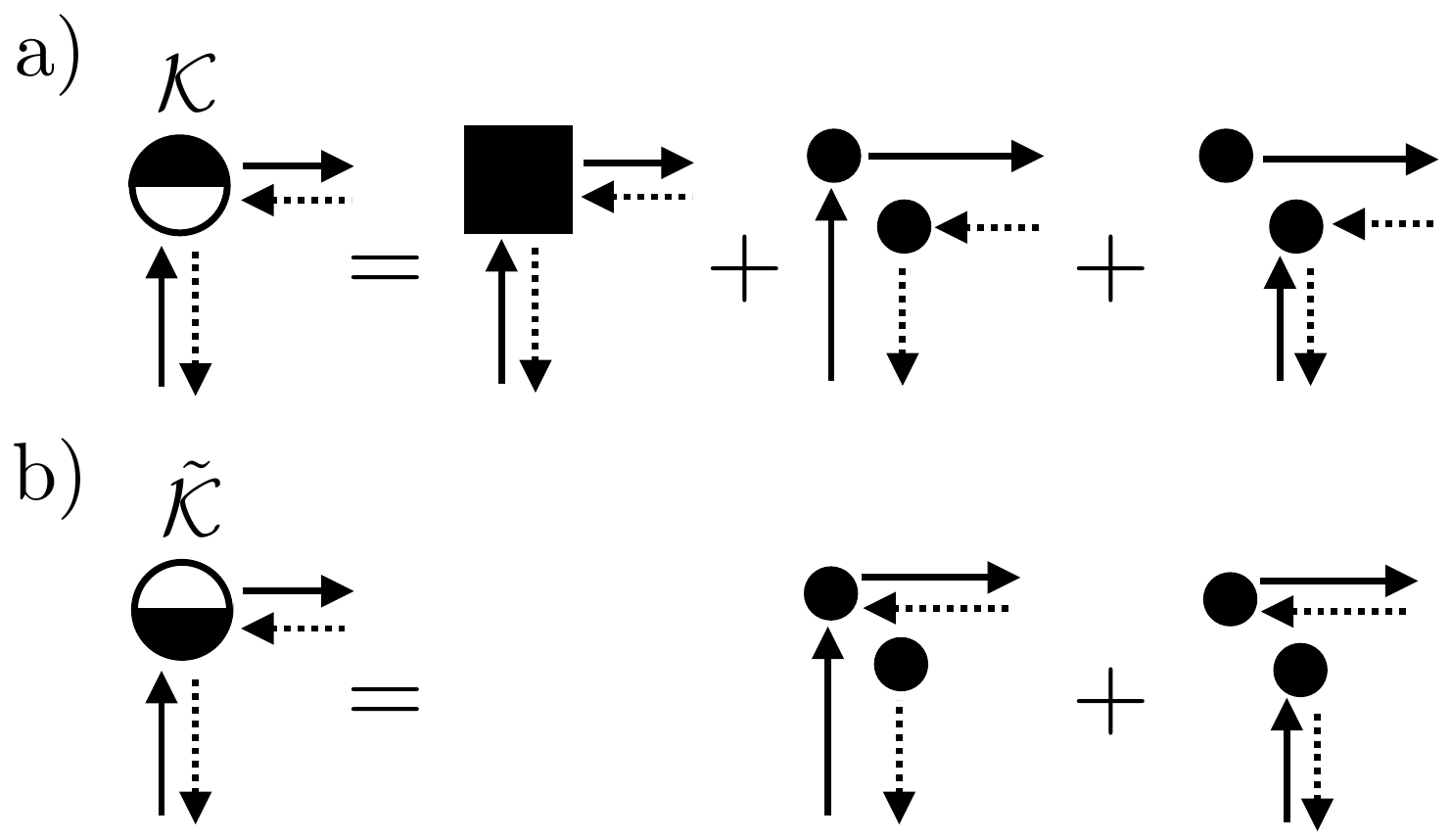}
\caption{Diagrammatic representation of the crossed building blocks  a) ${\mathcal K}$ and b) $\tilde{\mathcal K}$, see Eqs.~(\ref{eq:calK}) and (\ref{eq:calKtilde}), obtained after splitting the correlation function appearing in $K$, see Fig.~\ref{fig:crossedbb}(a) and Eq.~(\ref{eq:K}), into a term arising from quantum mechanical fluctuations (square) plus the product of averages
(two dots).
The diagrams for ${\mathcal K}_L$ and $\tilde{\mathcal K}_L$ (not shown) are obtained by adding a horizontal bar at the bottom, 
 in strict analogy to the changes from 
Fig.~\ref{fig:crossedbb}(a) to \ref{fig:crossedbb}(b).
\label{fig:crossedbb2}}
\end{figure}

In the following, it will be useful to split the quantum mechanical expectation value $\langle\sigma^+(\tau)\sigma^-(0)\rangle$ appearing in the expressions of $K$ and $K_L$, see
Eqs.~(\ref{eq:K},\ref{eq:KL}), into the product $\langle\sigma^+(\tau)\rangle\langle\sigma^-(0)\rangle=s^+(\tau)s^-(0)$ of expectation values plus a remaining term $\langle\sigma^+(\tau)\sigma^-(0)\rangle-s^+(\tau)s^-(0)$ describing the effect of quantum mechanical fluctuations. 
Evaluating the second derivative of the product $s^+(\tau)s^-(0)$ with respect to the two probe fields, the product rule 
yields in total four different terms. Thereby, $K$ contains in total five different terms, which we group as follows:
\begin{equation}
K(\omega_1,\omega_2,{\bf r})  =  {\mathcal K}(\omega_1,\omega_2,{\bf r})+\tilde{\mathcal K}(\omega_1,\omega_2,{\bf r})\label{eq:KKtilde}
\end{equation}
where ${\mathcal K}$ (the sum of the first three terms) and $\tilde{\mathcal K}$ (the remaining two terms) are defined as indicated in Fig.~\ref{fig:crossedbb2}, see also  Eqs.~(\ref{eq:calK},\ref{eq:calKtilde}) in Appendix~\ref{sec:crossedbb}.
Similarly,
\begin{equation}
K_L(\omega_1,{\bf r})  =  {\mathcal K}_L(\omega_1,{\bf r})+\tilde{\mathcal K}_L(\omega_1,{\bf r})\label{eq:KKtildec}
\end{equation}
 see Eqs.~(\ref{eq:calKL},\ref{eq:calKLtilde}). In Fig.~\ref{fig:crossedbb2}, the square represents the term
originating from quantum fluctuations (as explained above), whereas the small circles with outgoing dotted (or solid) arrow stand for
$s^+$ (or $s^-$), with incoming arrows indicating probe field derivatives  acting either on $s^+$ or on $s^-$.
The building blocks $\tilde{\mathcal K}$ and $\tilde{\mathcal K}_L$, see Fig.~\ref{fig:crossedbb2}(b), have the property that the incoming dashed arrow is associated with $s^-$ (outgoing solid arrow) and not with $s^+$ (outgoing dotted arrow). This will be relevant in the context of forbidden diagrams to be discussed in the following subsection. 

\subsection{Forbidden diagrams}
\label{sec:forbidden}

\begin{figure}
\includegraphics[width=0.45\textwidth]{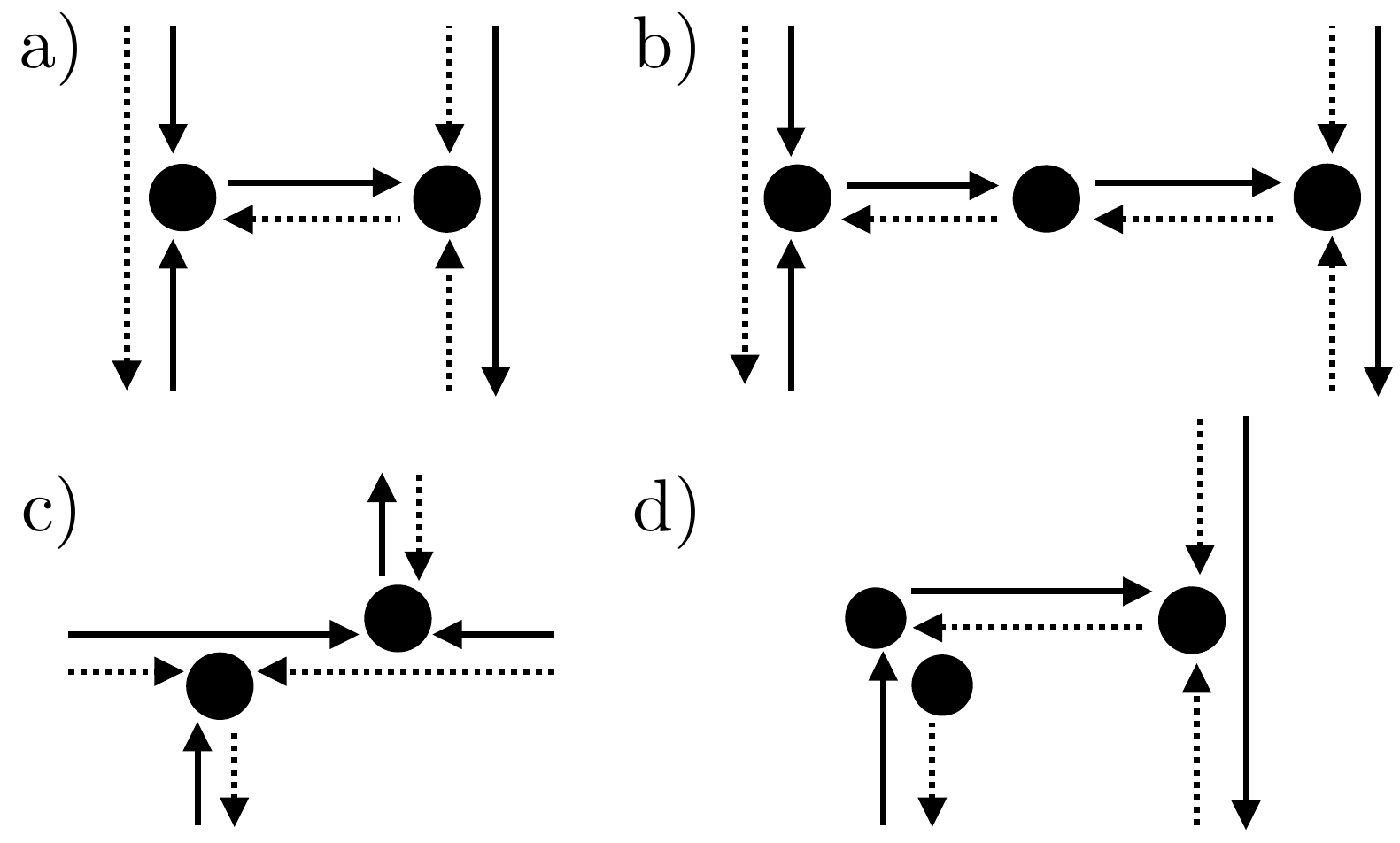}
\caption{a) Forbidden combination of the building block $\tau$ shown in Fig.~\ref{fig:crossedbb}(c) with its complex conjugate $\tau^*$. b) Combinations of the type shown in a) remain forbidden, if other building blocks -- here, e.g. Fig.~\ref{fig:crossedbb}(a) --  are inserted in between. c) Allowed combination of the same two building blocks $\tau$ and $\tau^*$ as in a), but connected in a different order. d) Forbidden combination of building block $\tilde{\mathcal K}$ -- here represented by one of the two terms appearing in Fig.~\ref{fig:crossedbb2}(b) -- with $\tau$.
\label{fig:forbidden}}
\end{figure}

Transport of a counterpropagating pair of amplitudes through the atomic medium is now described by connecting the crossed building blocks displayed in Fig.~\ref{fig:crossedbb} in all possible ways. In general, the connection between two building blocks is achieved by identifying an \lq outgoing\rq\ pair of counterpropagating arrows of one building block with the \lq incoming\rq\ pair of the other building block (where we define \lq outgoing\rq\ or \lq incoming\rq\ by the direction of the solid arrow). In addition, the building blocks shown in Fig.~\ref{fig:crossedbb}(c-e) exhibit an incoming ladder pair, which, as described in Sec.~\ref{sec:ladder}, is described by the spectrum ${\mathcal P}(\omega_1,{\bf r})$ of the dipole correlation function, see Eq.~(\ref{eq:Pav}).

Some combinations of crossed building blocks give rise to \lq forbidden diagrams\rq\ which yield a vanishing contribution. These forbidden diagrams are those where the outgoing pair of counterpropagating arrows of the building blocks $\tau^*$, $\tau_{L1}^*$, $\tau_{L2}^*$, $\tilde{\mathcal K}$ or $\tilde{\mathcal K}_L$ is identified with an incoming pair of $\tau$, $\tau_{L1}$, $\tau_{L2}$, 
$\tilde{\mathcal K}^*$ or $\tilde{\mathcal K}_L^*$.
Two examples are shown in Fig.~\ref{fig:forbidden}(a) and (d). These combinations are forbidden for the following reason: first, we note that all of these terms ($\tau$, $\tau_{L1}$, $\tau_{L2}$, $\tilde{\mathcal K}$, $\tilde{\mathcal K}_L$ and their complex conjugates) contain single-atom building blocks with only one outgoing arrow.
As explained in Sec.~\ref{sec:decomposition}, 
the photon exchange associated with the outgoing arrow must then occur {\em after} the exchanges associated with the incoming arrows. In case of the diagram shown in Fig.~\ref{fig:forbidden}(a), this condition cannot be fulfilled for both atoms at the same time, since the outgoing arrow of one atom serves as incoming arrow for the other atom, and vice versa. A similar argument holds if arbitrary additional building blocks are inserted in between, see Fig.~\ref{fig:crossedbb}(b). 
In contrast, the Fig.~\ref{fig:crossedbb}(c) shows an example of an \lq allowed\rq\ combination, where no conflict of orderings appears, and which therefore must be taken into account in  the transport equations which we will formulate further down.

All three forbidden diagrams, Fig.~\ref{fig:forbidden}(a), (b) and (d), correspond to diagrams with conflicting local orderings, the contribution of which vanishes as discussed in Sec.~\ref{sec:bblocks}. In principle, they would not change the final result even if they were included in the transport equations.
From a numerical perspective, however, this is true only if the integration over the frequencies of the exchanged photon is performed with perfect accuracy, yielding exactly zero for a forbidden combination. 
To minimize sources for numerical errors, we therefore explicitly exclude these combinations from our subsequent calculations.

\subsection{Crossed transport equations}

In principle, two different strategies can be pursued to
describe transport of counterpropagating amplitudes through the atomic medium: the first one consists in following the propagation from one end of the crossed scattering sequence to the other one. This strategy was employed in 
\cite{Wellens:2008aa,Wellens:2009aa}. The second considers two crossed propagators $C$ and $C^*$ starting from both ends in opposite directions, and joins them at a particular point within the scattering medium \cite{Hartmann:2012aa}. In this paper, we will adopt the second approach, since it leads to a more compact and physically transparent form of the transport equations. 

\begin{figure}
\includegraphics[width=0.45\textwidth]{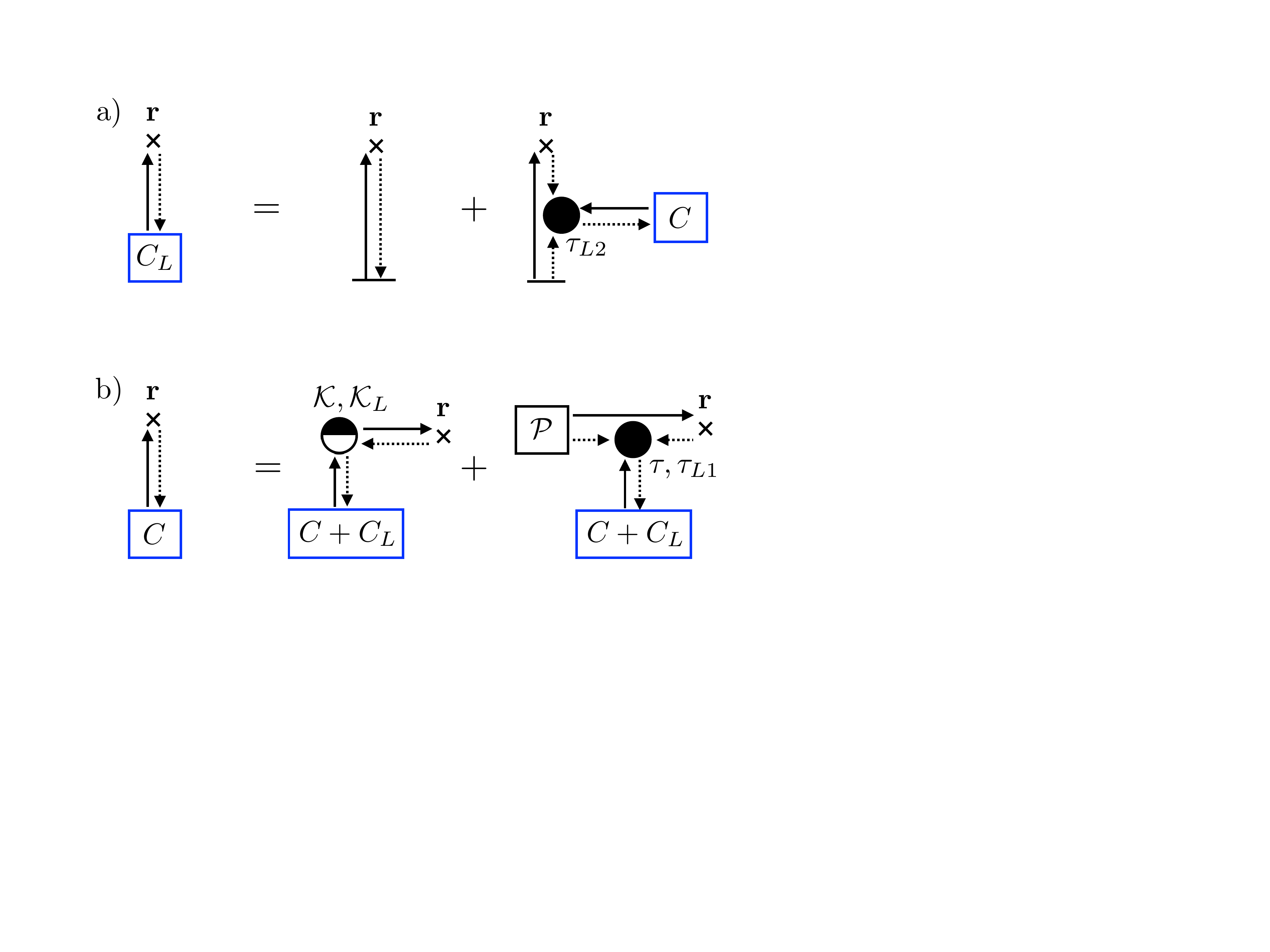}
\caption{Graphical representation of the transport equations for the crossed propagators a) $C_L$ and b) $C$, see Eqs.~(\ref{eq:CL},\ref{eq:Ccrossed}).
The coupled set of equations a) and b) describes all possible combinations of the crossed building blocks 
$\mathcal K$, ${\mathcal K}_L$, $\tau$, $\tau_{L1}$ and $\tau_{L2}$.
\label{fig:crossed1}}
\end{figure}

To define $C$, we choose the propagation direction defined by the solid arrow. The first step consists of a positive-frequency (solid line) laser photon propagating towards a point ${\bf r}$ within the atomic medium, and the corresponding negative-frequency photon (dotted line) propagating  from ${\bf r}$ towards the detector. Graphically, this term is represented by the first diagram on the right-hand side of Fig.~\ref{fig:crossed1}(a), and contributes to the quantity $C_L({\bf r})$ describing a counterpropagating pair of amplitudes originating from the laser mode. After that, the counterpropagating amplitudes may be scattered through an arbitrary sequence composed of the building blocks ${\mathcal K}_L$, $\mathcal K$, $\tau$ and $\tau_{L1}$. This gives rise to the quantity $C(\omega_1,{\bf r})$, describing a scattered pair of counterpropagating amplitudes (with frequency $\omega_1$ for the solid line, and $\omega-\omega_1$ for the dotted line), see Fig.~\ref{fig:crossed1}(b). Concerning $C_L({\bf r})$, we also have to take into account the possible occurrence of building block $\tau_{L2}$, see the second term on the right-hand side of Fig.~\ref{fig:crossed1}(a).

The corresponding equations read as follows:
\begin{eqnarray}
C_L({\bf r}) & = & E_L^+({\bf r})
\Bigl(
{\mathcal E}_D^-({\bf r},{\bf e}_D)+H_{\tau_2}({\bf r})\Bigr)\label{eq:CL}\\
C(\omega_1,{\bf r}) & = & \int_V {\rm d}{\bf r'}~T_{\omega_1}({\bf r},{\bf r}') T^*_{\omega-\omega_1}({\bf r}',{\bf r})\nonumber\\
& & \!\!\!\!\!\!\!\!\!\!\!\!\!\! \times 
\Bigl[
 H_K(\omega_1,{\bf r}')
 +
 H_{\tau 1}(\omega_1,{\bf r}')
Q\left(\omega_1,{\bf r}',{\bf r}\right)
\Bigr]\label{eq:Ccrossed}
\end{eqnarray}
where
\begin{eqnarray}
{\mathcal E}_D^-({\bf r},{\bf e}) &  = & e^{i k_L {\bf e}\cdot{\bf r}}\nonumber\\
& & \!\!\!\!\!\!\!\!\!\!\!\!\!\!\!\!\!\!\!\!\!\! \times 
\exp\left[-ik_L\int_0^\infty {\rm d}z'~\Bigl(n^*_\omega({\bf r}+z'{\bf e})-1\Bigr)\right]\label{eq:Omd}
\end{eqnarray}
with ${\bf e}={\bf e}_D$ or ${\bf e}=-{\bf e}_L$ represents an \lq outgoing\rq\ (with respect to the solid arrow) pair of counterpropagating amplitudes, and
\begin{eqnarray}
H_{\tau 1}(\omega_1,{\bf r}') & = & \tau_{L1}(\omega_1,{\bf r}')C_L({\bf r}') \nonumber\\
& & + \int_{-\infty}^\infty{\rm d}\omega_2~\tau(\omega_1,\omega_2,{\bf r}')
C(\omega_2,{\bf r}')\label{eq:Htau1}\\
H_{\tau 2}({\bf r}) & = &  \int_0^\infty{\rm d}z'~ \frac{E_L^-({\bf r}-z'{\bf e}_L){\mathcal E}_D^-({\bf r},-{\bf e}_L)}{{\mathcal E}_D^-({\bf r}-z'{\bf e}_L,-{\bf e}_L)}\nonumber\\
& &  \!\!\!\!\!\!\!\!\!\!\!\!\!\!\!\!\!\!\!\!\!\!\!\!\!\!\!\!\!\!\!\! \times \int_{-\infty}^\infty {\rm d}\omega_1~
      \tau_{L2}(\omega_1,{\bf
    r}-z'{\bf e}_L) C(\omega_1,{\bf r}-z'{\bf e}_L) \label{eq:Htau2}
\end{eqnarray}
\begin{eqnarray}
H_K(\omega_1,{\bf r}') & = & {\mathcal K}_{L}(\omega_1,{\bf r}')C_L({\bf r}') \nonumber\\
& & \!\!\!\!\! + \int_{-\infty}^\infty{\rm d}\omega_2~{\mathcal K}(\omega_1,\omega_2,{\bf r}')
C(\omega_2,{\bf r}')\label{eq:HK}\\
Q(\omega_1,{\bf r}',{\bf r}) & = & \int_0^{\infty} {\rm d}\rho
~{\mathcal P}\left(\omega_1,{\bf r}'+\rho\frac{{\bf r}'-{\bf r}}{|{\bf r}'-{\bf r}|}\right)\nonumber\\
& & \!\!\!\!\!\!\!\!\!\!\!\!\!\!\!\!\!\! \times 
\exp\left[-\int_0^\rho{\rm d}\rho'/\ell_{\omega_1}\left({\bf r}'+\rho'\frac{{\bf r}'-{\bf r}}{|{\bf r}'-{\bf r}|}\right)
\right]
\label{eq:Q}
\end{eqnarray}
In Eqs.~(\ref{eq:Omd},\ref{eq:Htau2}) and (\ref{eq:Q}), the integrations over $z'$ and $\rho$ are again restricted to 
finite intervals where the corresponding points ${\bf r}+z'{\bf e}$, ${\bf r}-z'{\bf e}_L$, and ${\bf r}'+\rho ({\bf r}'-{\bf r})/|{\bf r}'-{\bf r}|$  
lie inside the atomic medium. Moreover, Eqs.~(\ref{eq:Htau2}) and (\ref{eq:Q}) involve a stationary phase approximation (i.e., the building block $\tau_{L2}$ is placed on the line pointing from ${\bf r}$ in direction $-{\bf e}_L$, and the building blocks
$\tau$ and $\tau_{L1}$ on the line connecting ${\bf r}$ with ${\mathcal P}$.) 

The coupled set of equations (\ref{eq:CL},\ref{eq:Ccrossed}) can now be solved numerically by an iterative procedure. The quantities $C_L$ and $C$  contain all combinations of the building blocks 
${\mathcal K}$, ${\mathcal K}_L$, $\tau$, $\tau_{L1}$ and $\tau_{L2}$. The remaining building blocks (i.e., 
$\tau^*$, $\tau_{L1}^*$, $\tau_{L2}^*$,
$\tilde{\mathcal K}$, $\tilde{\mathcal K}_L$ and $K^*_L={\mathcal K}_L^*+\tilde{\mathcal K}_L^*$) are then used to connect $C_L$ and $C$ to the complex conjugate quantities $C_L^*$ and $C^*$. Thereby, all the forbidden combinations discussed in Sec.~\ref{sec:forbidden} (and only those!) are excluded.

\begin{figure}
\includegraphics[width=0.48\textwidth]{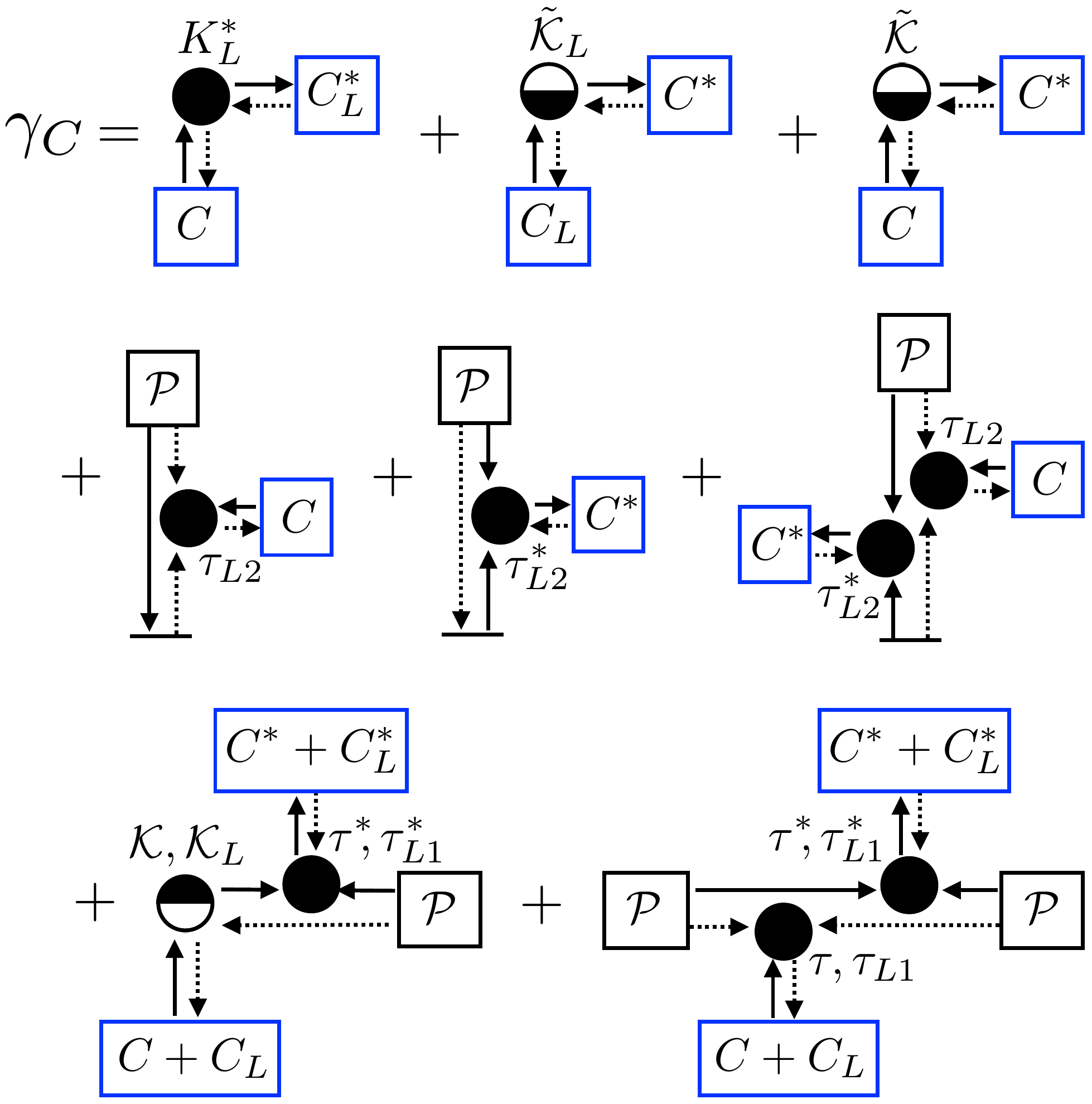}
\caption{The crossed contribution $\gamma_C(\omega,{\bf e}_D)$ to the photodetection signal 
results from connecting scattering sequences described by $C_L$ and $C$, see Fig.~\ref{fig:crossed1}, with 
the corresponding complex conjugate sequences  $C_L^*$ and $C^*$. \label{fig:crossed2}}
\end{figure}

The resulting crossed contribution $\gamma_C$ to the photodetection signal, graphically depicted in Fig.~\ref{fig:crossed2}, reads as follows:
\begin{widetext}
\begin{eqnarray}
\gamma_C(\omega,{\bf e}_D) & = & \frac{\hbar\omega_0\Gamma}{I_L} \int_V\frac{{\rm d}{\bf r}}{\mathcal A}~C_L^*({\bf r})\int_{-\infty}^\infty{\rm d}\omega_1~K_L^*(\omega-\omega_1,{\bf r})C(\omega_1,{\bf r})\nonumber\\
& + & \frac{\hbar\omega_0\Gamma}{I_L} \int_V\frac{{\rm d}{\bf r}}{\mathcal A}\int_{-\infty}^\infty{\rm d}\omega_1~C^*(\omega-\omega_1,{\bf r})\left(\tilde{\mathcal K}_L(\omega_1,{\bf r})C_L({\bf r})
+\int_{-\infty}^\infty{\rm d}\omega_2~\tilde{\mathcal K}(\omega_1,\omega_2,{\bf r})C(\omega_2,{\bf r})\right)
\nonumber\\
& + &  \frac{\hbar\omega_0\Gamma}{I_L}\int_V\frac{{\rm d}{\bf r}}{\mathcal A}~ {\mathcal P}\left(\omega,{\bf r}\right) 
\Bigl({\mathcal E}_D^+({\bf r},{\bf e}_D) H_{\tau 2}({\bf r})+{\mathcal E}_D^-({\bf r},{\bf e}_D) H_{\tau 2}^*({\bf r})+|H_{\tau 2}({\bf r})|^2\Bigr)
\nonumber\\
& + & \frac{\hbar\omega_0\Gamma}{I_L}\int_V\frac{{\rm d}{\bf r}}{\mathcal A}~ \int_V{\rm d}{\bf r}' \int_{-\infty}^\infty{\rm d}\omega_1~
H^*_{\tau 1}(\omega-\omega_1,{\bf r}) Q(\omega-\omega_1,{\bf r},{\bf r'}) T_{\omega_1}({\bf r},{\bf r}') T^*_{\omega-\omega_1}({\bf r}',{\bf r})\nonumber\\
& & \times \Bigl(H_K(\omega_1,{\bf r}')+H_{\tau 1}(\omega_1,{\bf r}') Q(\omega_1,{\bf r}',{\bf r})\Bigr)\label{eq:IDcrossed}
\end{eqnarray}
\end{widetext}
where ${\mathcal E}_D^+({\bf r},{\bf e}_D)=[{\mathcal E}_D^-({\bf r},{\bf e}_D)]^*$.
Again, the crossed contribution can be divided into an elastic and an inelastic component:    
\begin{equation}
\gamma_C(\omega,{\bf e}_D)=\gamma^{({\rm in})}_C(\omega,{\bf e}_D)+\delta(\omega)\gamma^{({\rm el})}_C({\bf e}_D)\label{eq:gammaCinel}
\end{equation}    
Numerically, we first calculate the total contribution $\gamma_C(\omega,{\bf e}_D)$ given by Eq.~(\ref{eq:IDcrossed}) on a discrete grid of frequencies $\omega_i=i\Delta\omega$. Then, we determine the inelastic component at $i=0$ by an interpolation between neighbouring points $i=\pm 1$ 
 where the elastic component vanishes, i.e. $\gamma_C^{({\rm in})}(0,{\bf e}_D)=[\gamma_C(-\Delta\omega,{\bf e}_D)+\gamma_C(\Delta\omega,{\bf e}_D)]/2$.
    
\section{Results}
\label{sec:results}

After having exposed our general theory for multiple scattering of laser light by a disordered cloud of two-level atoms, valid under the assumptions of large distances between the atoms, we now present numerical solutions of the ladder and crossed transport equations derived in the previous chapters \ref{sec:ladder} and \ref{sec:crossed}.
 
 \begin{figure}
\includegraphics[width=0.45\textwidth]{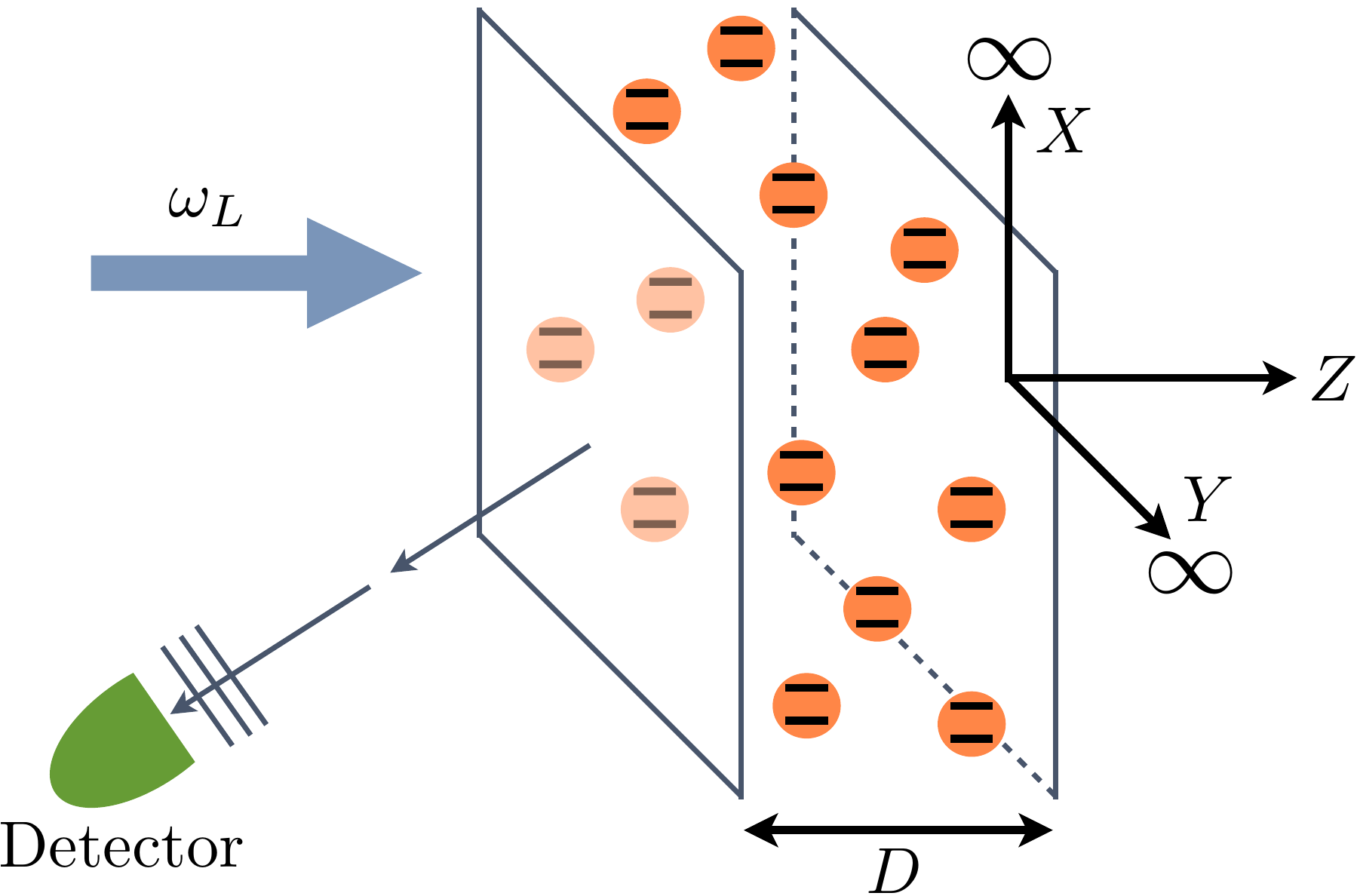}
\caption{Scattering geometry used for the numerical solution of the ladder and crossed transport equations. The atomic scattering medium is confined to a slab with thickness $D$ in $z$-direction, and infinite extension in $x$- and $y$-direction. The incoming laser beam is perpendicular to the surface of the slab.\label{fig:geometry}}
\end{figure}

\subsection{Scattering geometry}

We consider a one-dimensional slab as atomic scattering medium, with thickness $D$ in $z$-direction and infinite extension in $x$- and $y$-direction, see Fig.~\ref{fig:geometry}. We quantify the thickness $D$ in terms of the optical thickness $b=D/\ell_{\rm lin}$, where $\ell_{\rm lin}$ denotes the linear scattering mean free path, i.e., the scattering mean free path in the limit of low laser intensity:
\begin{equation}
\ell_{\rm lin}=\frac{k_L^2(1+4\delta^2/\Gamma^2)}{4\pi{\mathcal N}}
\end{equation}
It depends on the atom-laser detuning $\delta$, the radiative decay rate $\Gamma$, the wavenumber $k_L$ of the incoming laser and the density of atoms $\mathcal N$, which we assume to be constant within the slab. The intensity of the incoming laser (with Rabi-frequency $\Omega$) is measured in terms of the saturation parameter
 \begin{equation}
 s=\frac{2|\Omega|^2}{4\delta^2+\Gamma^2}
 \end{equation}
of a single atom driven by the laser field.

The advantage of the slab geometry is that all averaged quantities appearing in the transport equations, which depend on a single position variable ${\bf r}$ (e.g.  the average spectrum $\overline{I(\omega,{\bf r})}$, the mean free path $\ell_\omega({\bf r})$, the crossed building blocks $K(\omega_1,\omega_2,{\bf r})$ and $\tau(\omega_1,\omega_2,{\bf r})$, etc.) are independent of $x$ and $y$, due to translational invariance after disorder averaging. Only the terms $T_\omega({\bf r},{\bf r}')$, see Eq.~(\ref{eq:Tom}) and $Q(\omega,{\bf r},{\bf r}')$, see Eq.~(\ref{eq:Q}), depend on the transverse distance between ${\bf r}$ and ${\bf r}'$, over which the integral can be performed analytically, e.g., in Eqs.~(\ref{eq:Ibarin},\ref{eq:Ibarel}):
\begin{eqnarray}
\int_{-\infty}^\infty{\rm d}x\int_{-\infty}^\infty {\rm d}y \left|T_\omega({\bf r},{\bf r}')\right|^2 & = & \nonumber\\
& & \!\!\!\!\!\!\!\!\!\!\!\!\!\!\!\!\!\!\!\!\!\!\!\!\!\!\!\!\!\!\!\!\!\!\!\!\!\!\!\!\!\!\!\!\!\!\!\!\!\!\!\!\!\!\!\!\!\!\!\!\!\!\!\!\!\!\!\!\!\!\!\!\!\!\!\!\!\!\!\!
-\frac{\pi\Gamma^2}{2k_L^2}{\rm Ei}\left[-|z-z'|\int_0^1{\rm d}s~\frac{1}{\ell_\omega(sz+(1-s)z')}\right]\label{eq:1d}
\end{eqnarray}
where ${\rm Ei}(-t)=-\int_t^\infty{\rm d}t' \exp(-t')/t'$ (with $t>0$) denotes the exponential integral function. 
Eq.~(\ref{eq:1d}) is valid if the average value of the inverse mean free path $1/\ell_\omega$ between $z$ and $z'$ is positive -- a point which will be further discussed below.
Thereby, all position integrals  can be transformed into one-dimensional integrations along $z$, which enormously reduces the numerical effort.

 Nevertheless, the latter is still considerable, especially for the calculation of the crossed component. The most time-consuming part  is the calculation of the crossed building blocks $K(\omega_1,\omega_2,z)$ and $\tau(\omega_1,\omega_2,z)$, see Eqs.~(\ref{eq:K},\ref{eq:tau}). With a frequency grid of size $128$, and a position grid of size $100$ (which, as we have checked, is adequate for achieving well-converged results for a slab with optical thickness $b=5$), we have to calculate, for each desired value of the detected frequency $\omega$,  about $3\times 10^6$ different complex numbers ($K$ and $\tau$ for all values of $\omega_1$, $\omega_2$ and $z$). Each of these requires the solution of the three-dimensional single-atom optical Bloch equations from time $t_i=-40/\Gamma$ to $t_f=50/\Gamma$  for approximately 2000 realizations of the stochastic driving fields. 
 Using parallel computation on a large computer grid (bwGRiD and bwUniCluster), the solution of the ladder and crossed transport equations
 for each set of parameters $(b,s,\delta)$ takes approximately one week.
 
 \subsection{Diffusive transport}
 \label{sec:results_diffusive}
 
\begin{figure}
\includegraphics[width=0.5\textwidth]{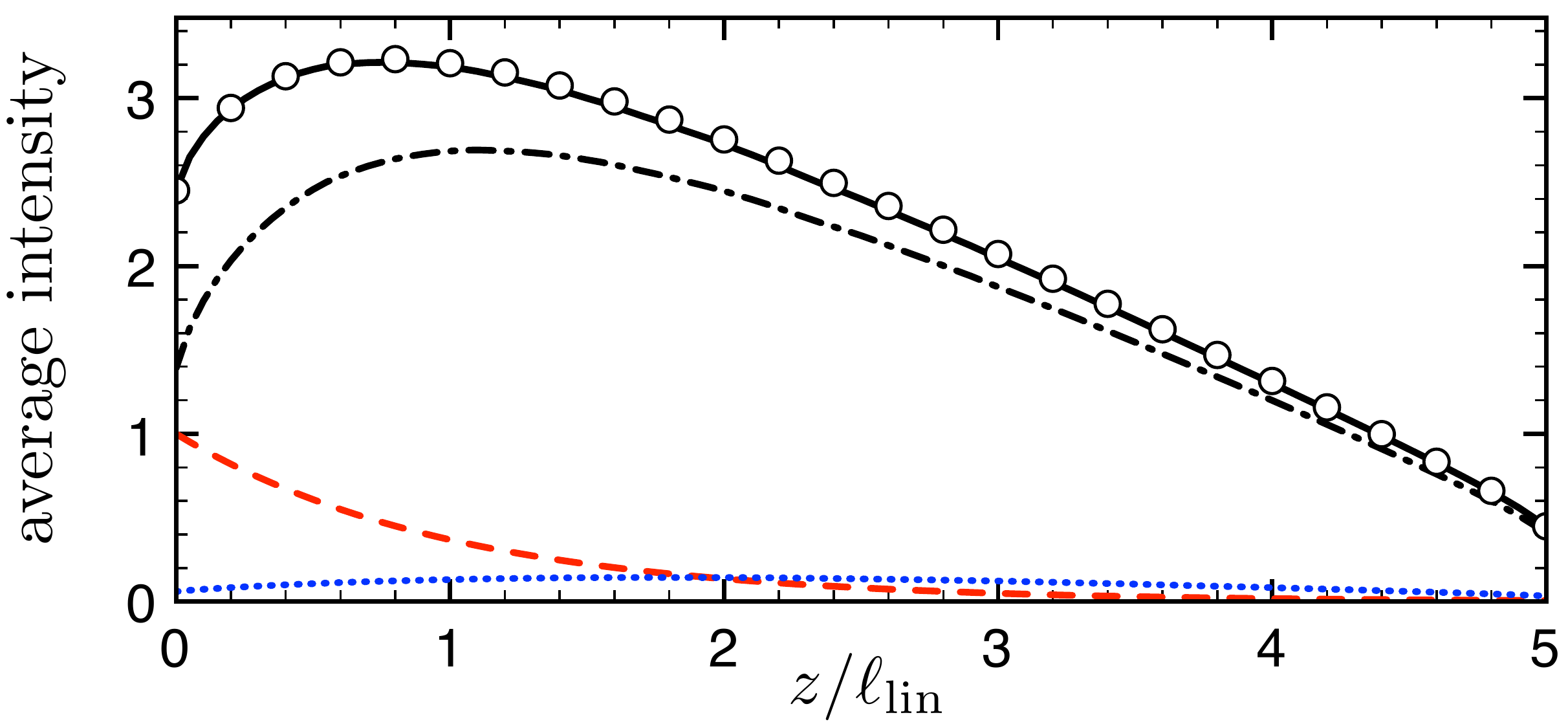}
\caption{Average light intensity $\overline{I^{({\rm tot})}(z)}$ (black solid line) in units of the incident laser intensity $I_L$  as a function of the position $z$ inside a 
slab with length $D=5\ell_{\rm lin}$, for detuning $\delta=0$ and saturation $s=1/1000$. As expected for small saturation, the curve coincides with the solution $\overline{I^{(0)}(z)}$ of the linear transport equation (\ref{eq:milne}) with $\ell^{(0)}=\ell_{\rm lin}$ (circles). 
For large $z$ (but not too close to the boundary at $z=5\ell_{\rm lin}$), it exhibits an approximately linear decay characteristic for diffusive transport. For small $z$, the intensity is up to three times larger than the incident laser intensity, as a consequence of multiple scattering.
The remaining curves show the three different components of $\overline{I^{({\rm tot})}(z)}$, see Eq.~(\ref{eq:Itot}):
 incident laser light $I_L(z)$ (red dashed line), 
elastically scattered light $\overline{I^{({\rm el})}(z)}$ (black dash-dotted line) and inelastically scattered light $\overline{I^{({\rm in})}(z)}$ (blue dotted line), all in units of $I_L$.
\label{fig:linprofile}}
\end{figure}

Let us start by verifying that our approach reproduces the well-known results of linear radiation transport in the limit of small saturation $s\ll 1$. For isotropic scattering with mean free path $\ell^{(0)}$, the average intensity inside a slab of thickness $D$ is obtained as the solution of the following linear transport equation \cite{Rossum:1999aa}:
\begin{equation}
\overline{I^{(0)}(z)}=I_L e^{-z/\ell^{(0)}}-\int_0^D{\rm d}z'~\frac{{\rm Ei}\left(-|z-z'|/\ell^{(0)}\right)}{2\ell^{(0)}}\overline{I^{(0)}(z')}\label{eq:milne}
\end{equation}
In the following, we compare this  solution with the solution of our nonlinear ladder transport equations describing radiation transport in a dilute and cold cloud of two-level atoms.
Here, the total average intensity consists of the incident laser intensity
$I_L(z)=2\epsilon_0 c |E_L^+(z)|^2$, with $E_L^+(z)$ given by Eq.~(\ref{eq:laser}), the intensity of elastically scattered light, see Eq.~(\ref{eq:Ibarel}), and the intensity of inelastically scattered light:
\begin{equation}
\overline{I^{({\rm tot})}(z)}=I_L(z) + \overline{I^{({\rm el})}(z)}+\overline{I^{({\rm in})}(z)}\label{eq:Itot}
\end{equation}
where the latter results from integrating the inelastic spectrum $\overline{I^{({\rm in})}(\omega,z)}$, see Eq.~(\ref{eq:Ibarin}), over the frequency $\omega$:
\begin{equation}
\overline{I^{({\rm in})}(z)}=\int_{-\infty}^\infty {\rm d}\omega~\overline{I^{({\rm in})}(\omega,z)}\label{eq:Iin}
\end{equation}
For very small saturation, e.g. $s=1/1000$,  the solution of our ladder transport equations, see Sec.~\ref{sec:ladder_transport} indeed reproduces the linear solution defined by Eq.~(\ref{eq:milne}) with $\ell^{(0)}=\ell_{\rm lin}$, as expected, see Fig.~\ref{fig:linprofile}.
 
\begin{figure}
\includegraphics[width=0.5\textwidth]{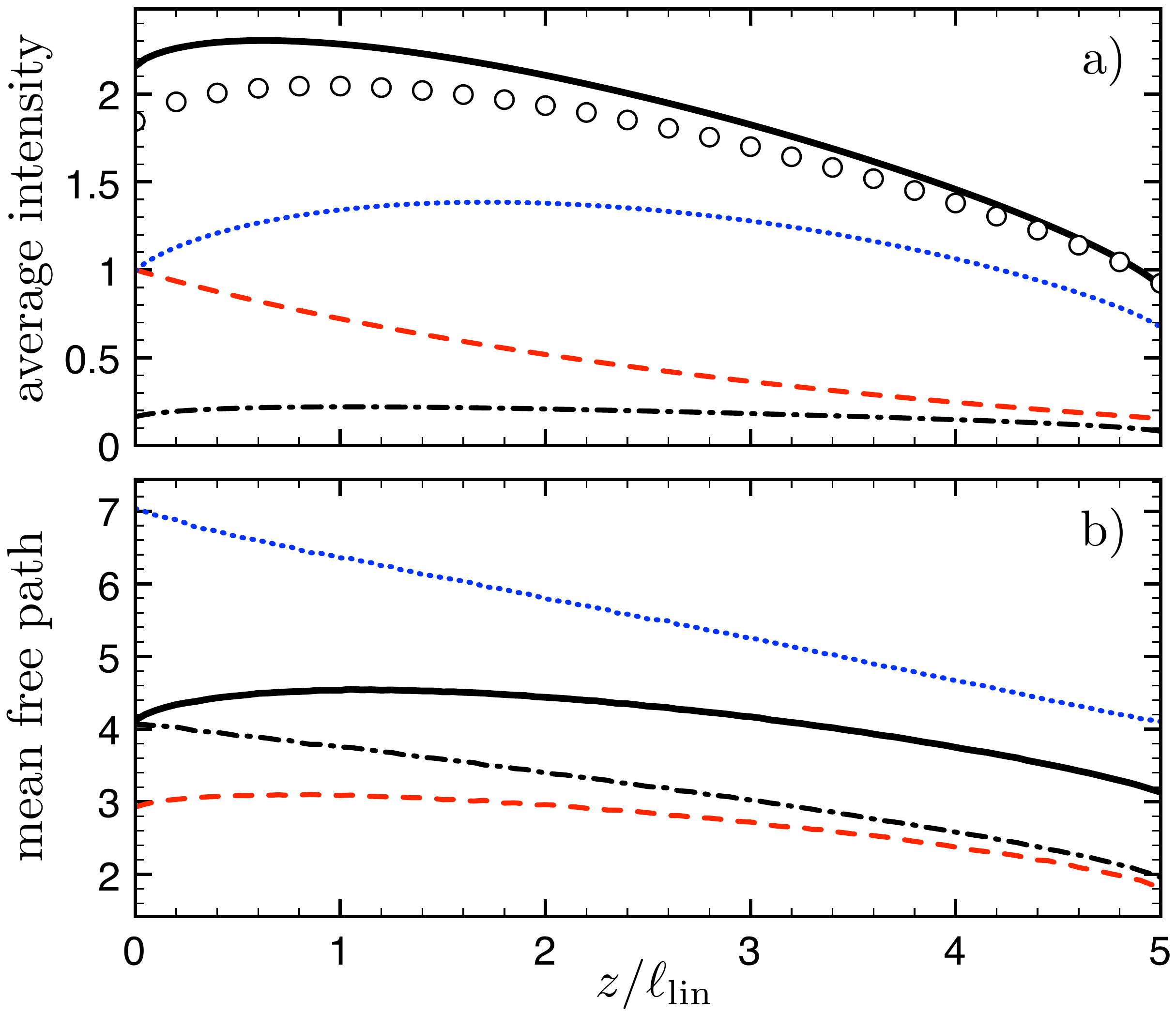}
\caption{a) Average total intensity $\overline{I^{({\rm tot})}(z)}$ (black solid line) and its three components $I_L(z)$ (red dashed line), $\overline{I^{({\rm el})}(z)}$ (black dash-dotted line) and $\overline{I^{({\rm in})}(z)}$ (blue dotted line) representing incident laser light, elastically and inelastically scattered light, respectively,
as a function of the position $z$ inside a 
slab with length $D=5\ell_{\rm lin}$, all in units of the incident laser intensity $I_L$, for detuning $\delta=0$ and saturation $s=1$. As compared to the case $s=1/1000$ of weak saturation (Fig.~\ref{fig:linprofile}), the overall intensity (black solid line) is reduced. The intensity profile is similar to the solution $\overline{I^{(0)}(z)}$ of the linear transport equation (\ref{eq:milne})  with increased mean free path $\ell^{(0)}=4.10 \ell_{\rm lin}$, corresponding to a reduced optical thickness $b=D/\ell^{(0)}=1.22$ (circles). b) Scattering mean free paths $\ell^{({\rm tot})}(z)$ for the total light (black solid line), see Eq.~(\ref{eq:elltot}), 
$\ell_L(z)$ for coherent light (red dashed line), $\ell_{\omega=0}(z)$ for elastically scattered light (black dash-dotted line) and $\ell^{({\rm in})}(z)$ for inelastically scattered light (blue dotted line), see Eq.~(\ref{eq:ellin}), all in units of $\ell_{\rm lin}$. Due to saturation, all mean free paths are increased as compared to $\ell_{\rm lin}$. The mean free path $\ell^{({\rm in})}(z)$ for inelastically scattered light (blue dotted line) is largest, since off-resonant light is scattered less efficiently than resonant light.
 \label{fig:profile}}
\end{figure}

The situation changes for larger saturation.
For a single atom driven by a laser with saturation parameter $s$, the excited state is populated with probability $s/(2+2s)$, and the intensity of scattered light divided by the incident light intensity is proportional to $1/(1+s)$. In other words, for $s=1$, the scattering cross section of a single atom  is only half as large as for $s\to 0$. In addition, each atom is not only driven by the incident laser, but also by the light emitted from all other atoms. 
Fig.~\ref{fig:profile}(a) shows the intensity profile for $s=1$ resulting from our ladder transport equations, where all these effects are taken into account. As most prominent difference with respect to the case of small saturation, we note that the intensity is now dominated by inelastically scattered light (blue dotted line). Moreover, we see that, in accordance with the above expectation, the amount of multiple scattering is reduced as compared to the case of small saturation, since the total intensity (black solid line) is smaller, whereas the decay of the incident laser light (red dashed line) with increasing $z$ is slower than in Fig.~\ref{fig:linprofile}.
  
\begin{figure}
\includegraphics[width=0.5\textwidth]{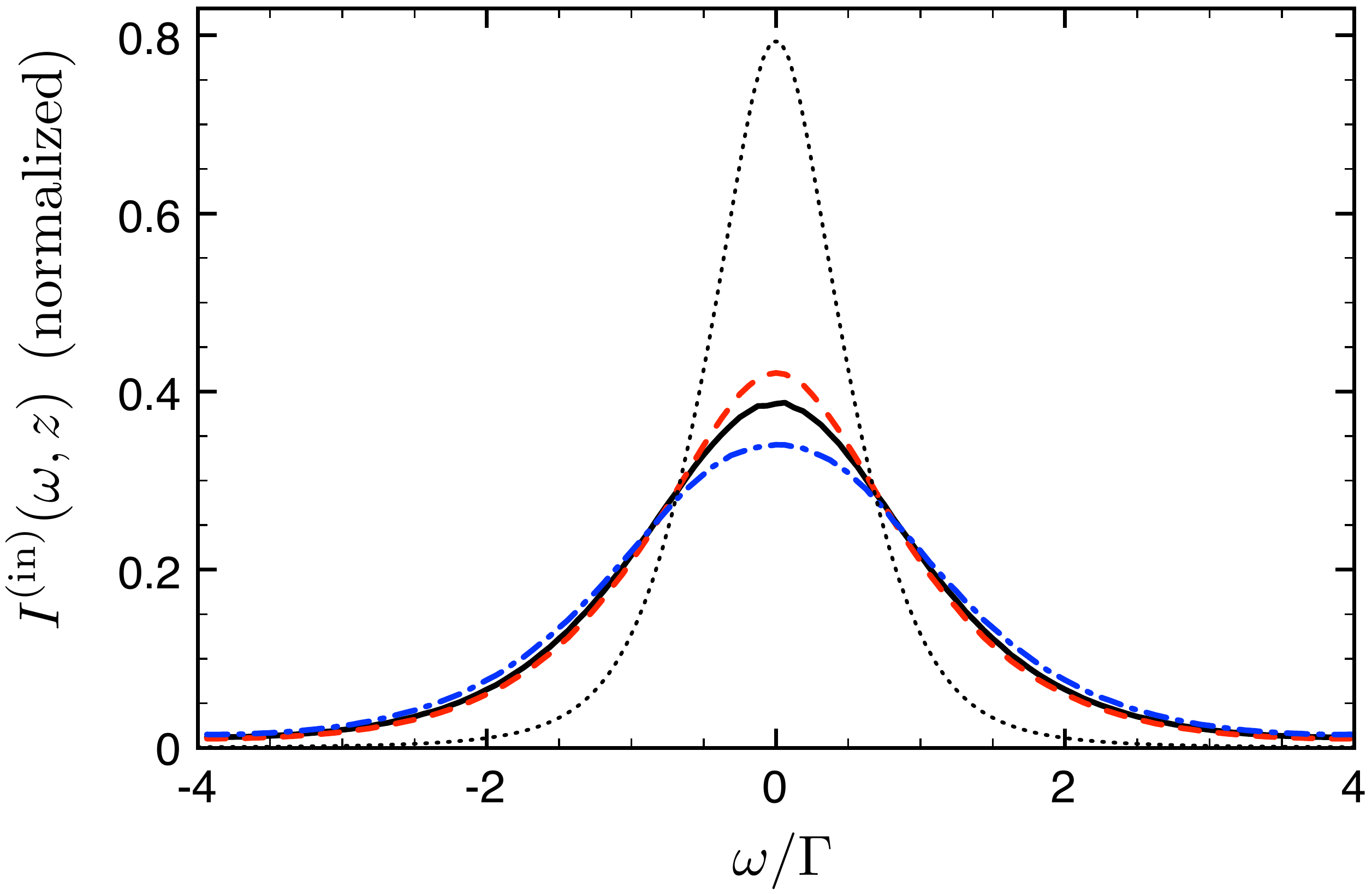}
\caption{Spectra $\overline{I^{({\rm in})}(\omega,z)}$ of inelastically scattered light 
for $z=0$ (black solid line), $z=D/2$ (red dashed line) and $z=D$ (blue dash-dotted line)
and otherwise the same parameters as in Fig.~\ref{fig:profile} ($D=5\ell_{\rm lin}$, $s=1$, $\delta=0$).
 All spectra are normalized such that $\int{\rm d}\omega~\overline{I^{({\rm in})}(\omega,z)}=\Gamma$. For comparison, the thin black dotted line shows the spectrum emitted by a single laser-driven atom ($s=1$ and $\delta=0$). 
\label{fig:spectrum}}
\end{figure}

These findings indicate that the  mean free paths characterizing scattering of light in a saturated atomic medium are larger than in the linear case (i.e. for small saturation). This is demonstrated in Fig.~\ref{fig:profile}(b), which shows the mean free path $\ell_{\omega=0}$ for elastically scattered light (black dash-dotted line) and $\ell_L$ for the incident laser light (red dashed line) as a function of the position $z$ inside the atomic slab. 
Concerning inelastically scattered light, we  
define an effective mean free path $\ell^{({\rm in})}$ (blue dotted line) 
by averaging its inverse over the spectrum (see also the corresponding discussion in Appendix~\ref{sec:flux}): 
\begin{equation}
\frac{\overline{I^{({\rm in})}(z)}}{\ell^{({\rm in})}(z)}= \int_{-\infty}^\infty{\rm d}\omega~\frac{\overline{I^{({\rm in})}(\omega,z)}}{\ell_{\omega}(z)}\label{eq:ellin}
\end{equation}
 Similarly, we define an effective mean free path $\ell^{({\rm tot})}$ for the total light (black solid line) as follows:
 \begin{equation}
 \frac{\overline{I^{({\rm tot})}(z)}}{\ell^{({\rm tot})}(z)}=\frac{I_L(z)}{\ell_L(z)}+\frac{\overline{I^{({\rm el})}(z)}}{\ell_{\omega=0}(z)}+\frac{\overline{I^{({\rm in})}(z)}}{\ell^{({\rm in})}(z)}\label{eq:elltot}
 \end{equation}
From  Fig.~\ref{fig:profile}(b), we see that $\ell^{({\rm tot})}$ is between three and five times larger than the mean free path $\ell_{\rm lin}$ in the  limit of very small saturation. In Fig.~\ref{fig:profile}(a), we therefore compare the
total light intensity (black solid line) with the solution of the linear transport equation (\ref{eq:milne}) with inverse mean free path $1/\ell^{(0)}=\int_0^D {\rm d}z/[D\ell^{({\rm tot})}(z)]\simeq 1/(4.10~\ell_{\rm lin})$ and find rough agreement.
 
The spectrum of inelastically scattered light is plotted in Fig.~\ref{fig:spectrum}, for three different positions $z=0$, $z=D/2$ and $z=D$ inside the atomic slab. For comparison, the thin dotted line shows the spectrum emitted by a single laser-driven atom, which is approximately half as broad as the other three spectra. The broadening of these spectra has the following two reasons: first, each atom not only sees the incident laser light, but also the light emitted from all other atoms. This increases the saturation of each single atom, and thus leads to a broader spectrum. Second, the spectrum is broadened by multiple scattering, since the frequencies of photons emitted by one atom may again be shifted due to subsequent scattering by other atoms.

As already mentioned above, the description of propagation in an infinitely extended medium (in the $x$- and $y$-direction) requires a positive mean free path (since, otherwise, the intensity diverges). For the results presented in this article, $\ell_\omega$ indeed remains positive for all frequencies and all positions inside the slab. For a strong laser, however, $\ell_\omega$ may also assume negative values for certain frequencies, an effect known as \lq Mollow gain\rq\ \cite{Mollow:1972aa}. In this case, the incident strong laser light is used to amplify a weak probe beam. For a single atom (with detuning $\delta=0$), this effect occurs if $s\geq 3$. In the case of an atomic medium with thickness $D=5\ell_{\rm lin}$, we have verified that, due to  spectral  broadening discussed above, frequency windows with $\ell_\omega<0$ are smeared out, such that
$\ell_\omega$ remains positive (for all frequencies and everywhere inside the slab) up to $s\leq 14.3$ \cite{tobithesis}. For even larger $s$, Mollow gain occurs in the atomic medium, and our assumption that all scattered fields are weak (and therefore only single photons are exchanged between each pair of atoms) breaks down. An extension of our theory taking into account Mollow gain will therefore be an interesting task for  future work.

 \subsection{Coherent backscattering}
 
 As discussed in Sec.~\ref{sec:laddercrossed}, the effect of coherent backscattering becomes apparent when measuring the average  backscattered intensity $\gamma({\bf e}_D)$, see Eq.~(\ref{eq:gammaDtot}).
 This intensity consists of a (weakly angle-dependent) diffusive background
 $\gamma_L$, see Eq.~(\ref{eq:IDfinal}), and an interference contribution $\gamma_C$, see Eq.~(\ref{eq:IDcrossed}), which is strongly peaked around  ${\bf e}_D=-{\bf e}_L$. In the following, we restrict ourselves to the case of exact backscattering direction (${\bf e}_D=-{\bf e}_L$), i.e., we investigate the height of the coherent backscattering cone and the corresponding coherent backscattering enhancement factor
 \begin{equation}
 \eta=\frac{\gamma_L+\gamma_C}{\gamma_L}\label{eq:eta}
 \end{equation}
In the limit $s\to 0$, the backscattered intensities  $\gamma_L$ and $\gamma_C$ converge to the values  \cite{Rossum:1999aa}
\begin{eqnarray}
\gamma_L^{(0)} & = & \int_0^D\frac{{\rm d}z}{\ell^{(0)}}~e^{-z/\ell^{(0)}}~\frac{\overline{I^{(0)}(z)}}{I_L}\label{eq:gammaL0}\\
\gamma_C^{(0)} & = & \int_0^D\frac{{\rm d}z}{\ell^{(0)}}~e^{-z/\ell^{(0)}}~\left(\frac{\overline{I^{(0)}(z)}}{I_L}-e^{-z/\ell^{(0)}}\right)\label{eq:gammaC0}
\end{eqnarray}
obtained from the solution of the linear transport equation (\ref{eq:milne}) with $\ell^{(0)}=\ell_{\rm lin}$. Note that 
$\gamma_C^{(0)}<\gamma_L^{(0)}$ (and hence $\eta^{(0)}<2$) since single scattering contributes only to $\gamma_L$, but not to $\gamma_C$. 

 \begin{figure}
\includegraphics[width=0.5\textwidth]{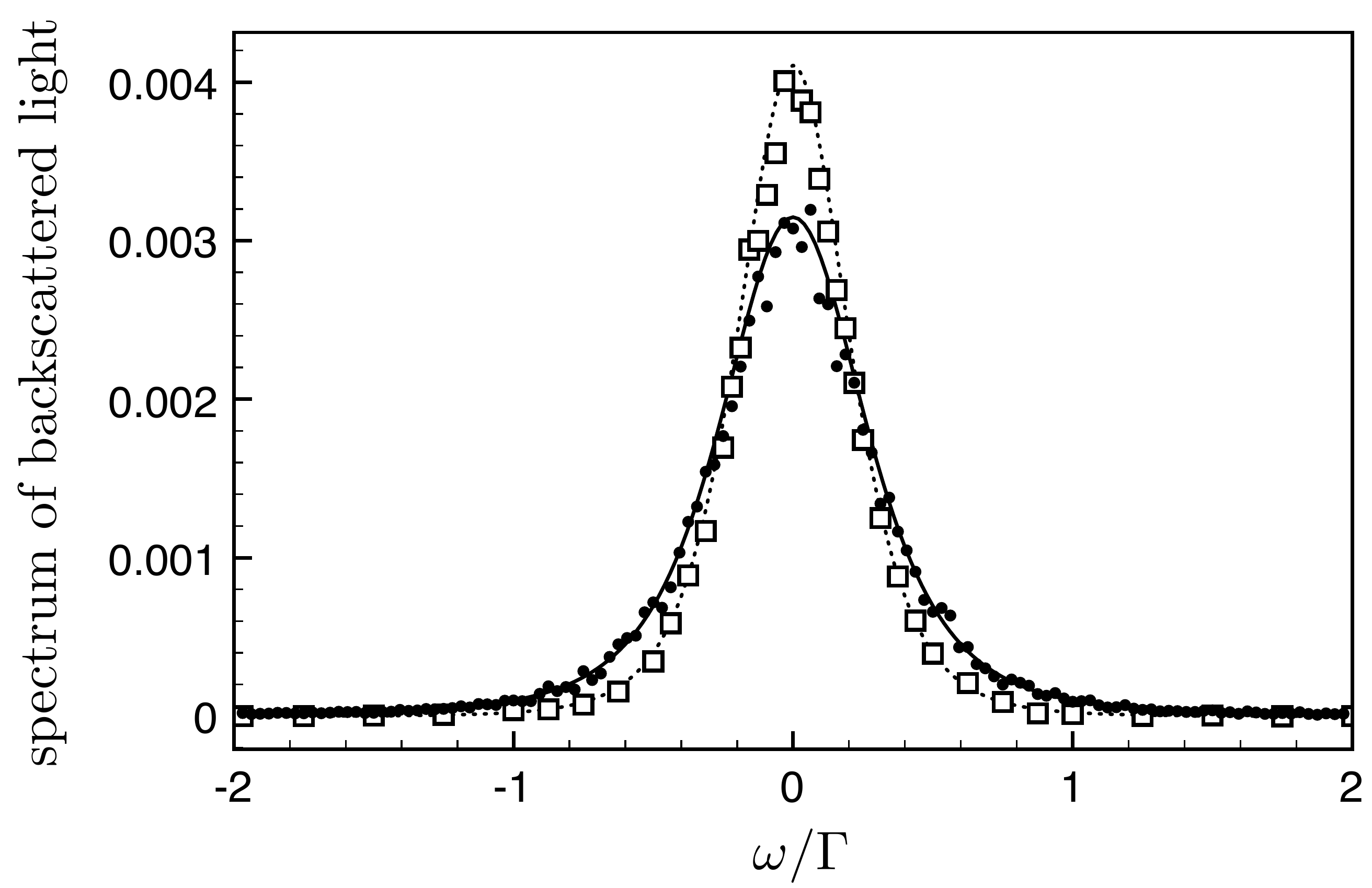}
\caption{Inelastic ladder and crossed spectra $\gamma_L^{({\rm in})}(\omega)$ (filled circles) and 
$\gamma_C^{({\rm in})}(\omega)$ (open squares)
of backscattered light for $s=0.001$, $\delta=0$ and $D=0.5\ell_{\rm lin}$. The results agree with the perturbative predictions 
of two-photon scattering theory (solid and dotted line, respectively). For $\omega\simeq 0$, the height of the coherent backscattering cone $\gamma_C^{({\rm in})}(\omega)$  is larger than the diffusive background $\gamma_L^{({\rm in})}(\omega)$, corresponding to a coherent backscattering enhancement factor $\eta(\omega)>2$. 
\label{fig:spec_pert}}
\end{figure}

Corrections to $\gamma_L^{(0)}$ and $\gamma_C^{(0)}$ in first order of $s$ can be calculated in terms of the
two-photon scattering matrix of the dilute atomic sample \cite{Wellens:2006aa}. Let us first verify that our transport equations reproduce the results of this two-photon scattering approach for small saturation, see Fig.~\ref{fig:spec_pert}. An intriguing prediction of this approach is the fact that the coherent backscattering enhancement factor originating from inelastically scattered photons with frequencies 
$\omega\simeq 0$ close to the frequency of the incident light exceeds the value of two, such that $\gamma_C^{({\rm in})}(0)>\gamma_L^{({\rm in})}(0)$.
In the linear case, this is not possible since any scattering path of a single photon exhibits only one reversed counterpart with which it interferes in the backscattering direction. For  two photons, however, coherent backscattering originates from interference between three reversed amplitudes \cite{Wellens:2006aa}. In other words, in the presence of nonlinear scattering, a ladder diagram may give rise to more than one crossed diagram. (For example, from the ladder diagram shown in Fig.~\ref{fig:ladder_meq}(a), we may construct an additional crossed diagram apart from the one shown in Fig.~\ref{fig:ladder_meq}(b), where the scattering sequence $2\to 6$ is reversed.)

The reversed amplitudes, however, are able to interfere fully constructively only if the detected frequency $\omega$ is close to the laser frequency $\omega_L=0$ (in the rotating frame). If this is not the case, the crossed component is suppressed due to dephasing  induced by the different frequencies of the counterpropagating photons. Consequently, we can see in Fig.~\ref{fig:spec_pert} that $\gamma_C^{({\rm in})}(\omega)<\gamma_L^{({\rm in})}(\omega)$ for larger frequencies. 

 \begin{figure}
\includegraphics[width=0.5\textwidth]{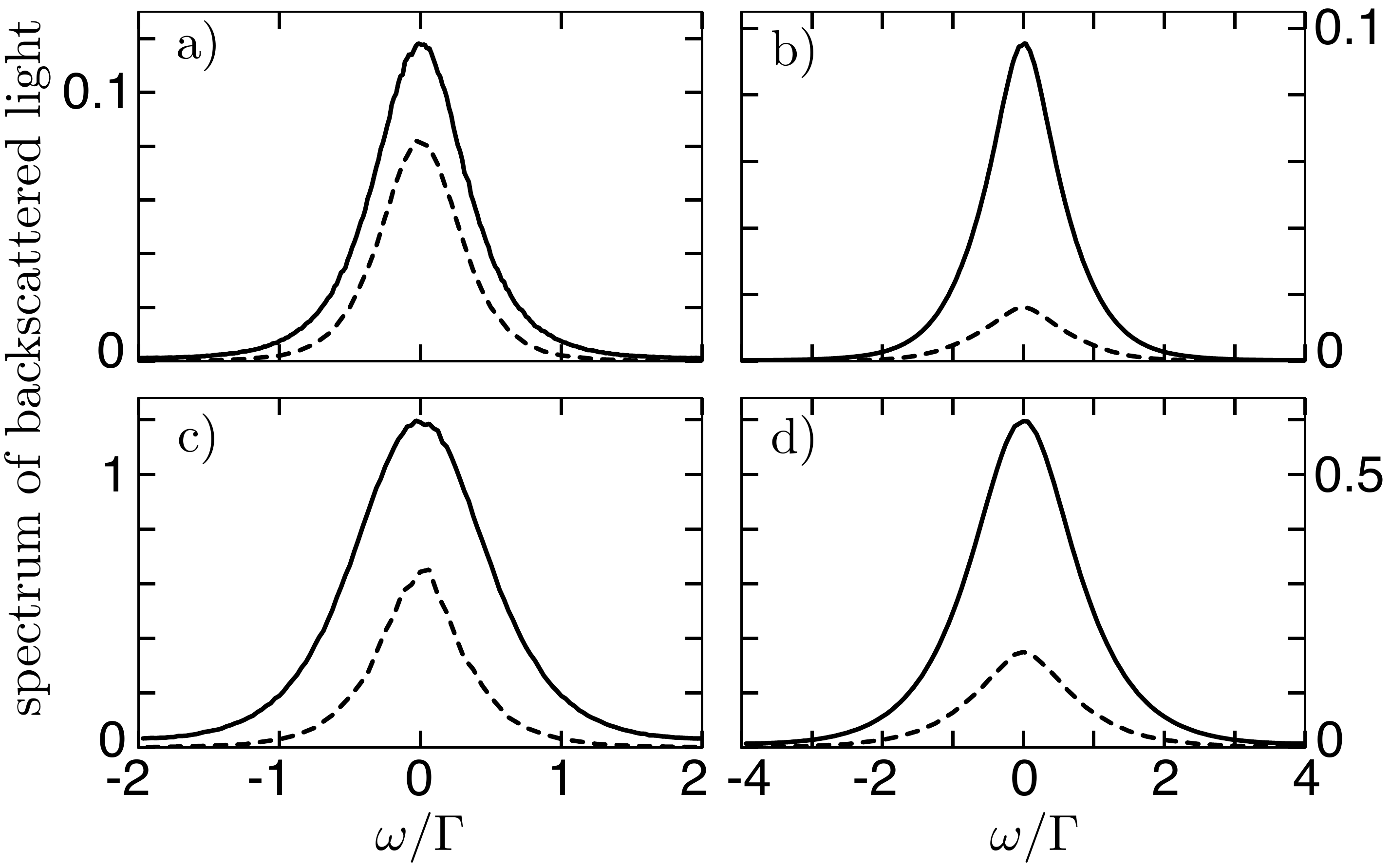}
\caption{Spectra $\gamma_L^{({\rm in})}(\omega)$ (solid lines) and $\gamma_C^{({\rm in})}(\omega)$ (dashed lines) of the ladder and crossed component of inelastically backscattered light for detuning $\delta=0$ and a) $(s,D)=(0.1,0.5\ell_{\rm lin})$, b) $(s,D)=(1,0.5\ell_{\rm lin})$, c) $(s,D)=(0.1,5\ell_{\rm lin})$ and d) $(s,D)=(1,5\ell_{\rm lin})$. In contrast to the limit of very small saturation, see Fig.~\ref{fig:spec_pert}, larger values of $s$ lead to an increasing suppression of the crossed spectrum, especially for the slab with smaller thickness (b).
\label{fig:spec_LC}}
\end{figure}

An interesting question, which we can now answer using the ladder and crossed transport equations derived in this article, concerns the behaviour for larger values of the saturation parameter beyond the validity of two-photon scattering theory: is it possible to achieve an even stronger amplification of the 
inelastic coherent backscattering enhancement factor? In Fig.~\ref{fig:spec_LC}, we see that this is not the case. Here, we show the ladder and crossed spectra (solid and dotted lines) for $s=0.1$ (a,c) and $s=1$ (b,d) for two different thicknesses $D=0.5\ell_{\rm lin}$ (a,b) and   $D=5\ell_{\rm lin}$ (c,d). In all cases, the crossed component is smaller than the ladder component for all frequencies. 
The crossed component is increasingly suppressed with stronger saturation, especially for the slab with smaller thickness, see Fig.~\ref{fig:spec_LC}(b). This has two different reasons: first, the dephasing due to the change of frequencies induced by inelastic scattering becomes stronger for larger $s$. For example, in the case of two inelastic scattering events, the frequency of the intermediate photon (between the first and the second inelastic event) may differ from $\omega_L$ even if the frequency $\omega$ of the detected photon is close to $\omega_L$.  Second, contributions to the backscattered spectrum of higher order in $s$ may also carry a negative weight. 
For example, for the well-known case of a single atom driven by a monochromatic laser, the ratio of inelastically scattered intensity divided by the incident intensity is proportional to $s/(1+s)^2$.
Expanding this result in powers of $s$, we get
\begin{equation}
\frac{s}{(1+s)^2}=s-2s^2+\dots
\end{equation}
i.e. the second order in $s$ (resulting from scattering of three photons) carries a negative sign. In this case, the above-mentioned effect of interference between more than two reversed-path amplitudes supresses the corresponding crossed component more strongly than the ladder component.

\begin{figure}
\includegraphics[width=0.5\textwidth]{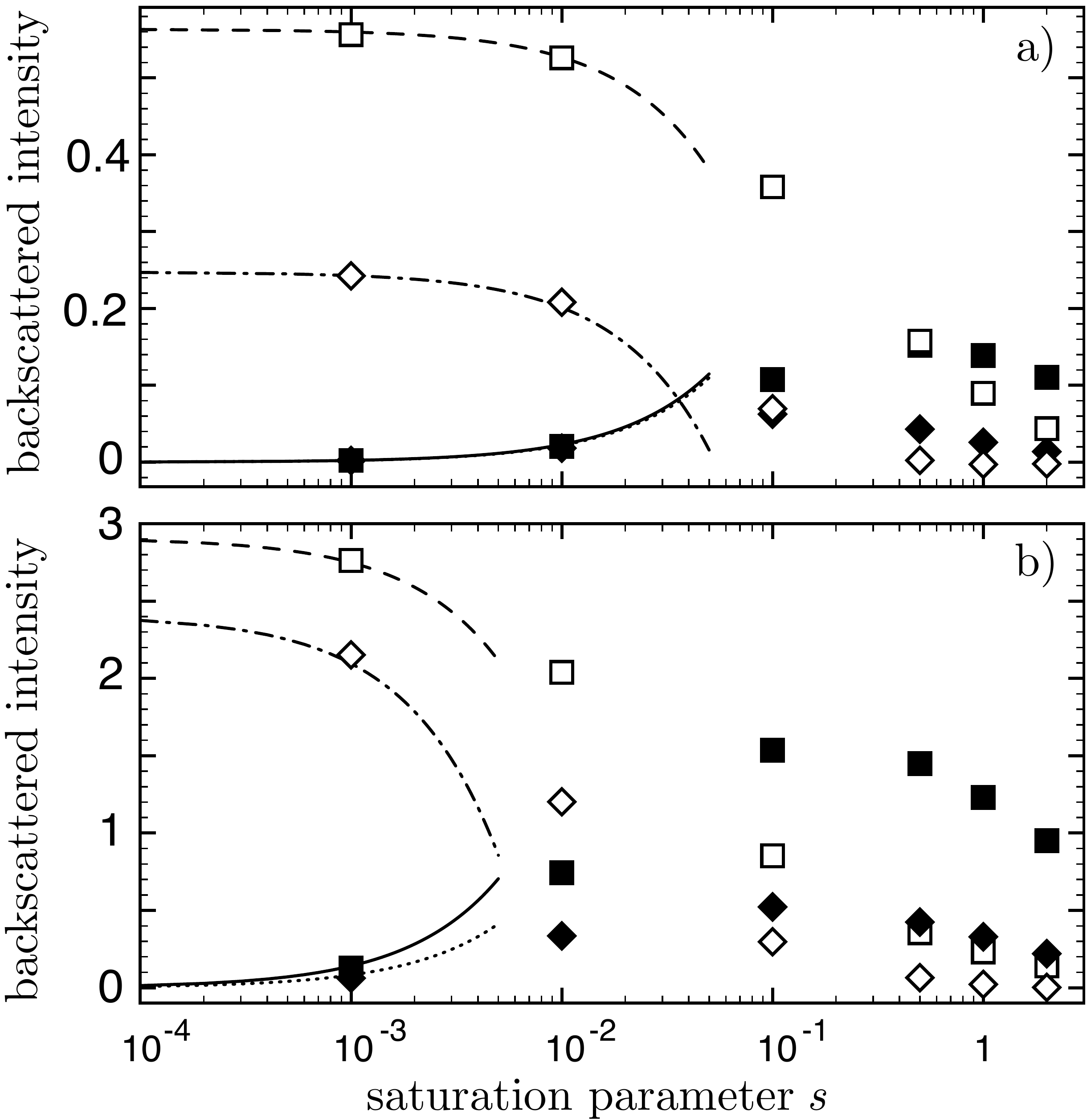}
\caption{Elastic and inelastic ladder and crossed components $\gamma_L^{({\rm el})}$ (open squares and dashed line),
$\gamma_C^{({\rm el})}$ (open diamonds and dash-dotted line),
$\gamma_L^{({\rm in})}$ (filled squares and solid line) and
$\gamma_C^{({\rm in})}$ (filled diamonds and dotted line) of the backscattered intensity, as a function of the saturation parameter $s$, for zero detuning ($\delta=0$) and thickness a) $D=0.5\ell_{\rm lin}$ and b) $D=5\ell_{\rm lin}$. The symbols result from the numerical solution of the ladder and crossed transport equations derived in this article, whereas the lines denote the corresponding perturbative predictions of two-photon scattering theory. In the atomic cloud with smaller (larger) thickness $D=0.5\ell_{\rm lin}$ ($D=5\ell_{\rm lin}$),
the perturbative theory is valid for $s<10^{-2}$ ($s<10^{-3}$). 
\label{fig:LC}}
\end{figure}

In summary, the height $\gamma_C$ of the coherent backscattering cone results from an interplay between two effects: (i) dephasing due to random changes of the frequency induced by inelastic scattering events (which always reduce $\gamma_C$ as compared to $\gamma_L$) and (ii) interference between many reversed-path amplitudes (which may increase or decrease $\gamma_C$, depending on the sign of the respective contributions). 

In Fig.~\ref{fig:LC}, we show the total inelastic ladder and crossed contributions $\gamma_L^{({\rm in})}$ and $\gamma_C^{({\rm in})}$ (obtained from integrating the inelastic spectra, see Fig.~\ref{fig:spec_LC}, over the frequency $\omega$ of the detected photon) together with the corresponding elastic contributions $\gamma_L^{({\rm el})}$ and $\gamma_C^{({\rm el})}$. The upper plot, Fig.~\ref{fig:LC}(a) refers to a slab with thickness $D=0.5\ell_{\rm lin}$, 
whereas $D=5\ell_{\rm lin}$ in the lower one, Fig.~\ref{fig:LC}(b). The saturation parameter is varied on a logarithmic scale from $s=10^{-4}$ to $s=2$. The lines correspond to the prediction of  two-photon scattering theory. We see that the regime of validity of the latter depends on the thickness of the medium. This is not surprising since the number of scattering events is larger in a thicker medium.

In both cases, Fig.~\ref{fig:LC}(a) and (b), we first observe that the elastic ladder contribution (open squares) decreases as a function of $s$. This has two reasons: first, with increasing saturation, the ratio of elastic vs. inelastic scattering decreases and, second, also the total amount of scattering (elastic plus inelastic) decreases, as already discussed in Sec.~\ref{sec:results_diffusive}. 
For the same reasons, the inelastic ladder component (filled squares), which starts at zero for $s=0$, first increases, then assumes a maximum and decreases again for large $s$. The elastic crossed component (open diamonds) also decreases as a function of $s$, and it does so faster than the ladder component. This is a consequence of interference between 
many reversed-path amplitudes, which, in this case, give predominantly negative contributions (due to the fact that
$\gamma^{({\rm el})}_L(s)<\gamma_L^{({\rm el})}(0)$ for $s>0$). For $b=0.5\ell_{\rm lin}$ and $s\geq 1$, the elastic crossed contribution even assumes values below zero ($\gamma_C^{({\rm el})}\simeq-0.003$ for $s=1$ and $-0.002$ for $s=2$), 
corresponding to destructive instead of constructive coherent backscattering interference. This effect has already been observed for
classical nonlinear coherent backscattering \cite{Hartung:2008aa,Wellens:2009ab,Hartmann:2012aa}. In the present case, it is less pronounced, due to the strong suppression of elastically backscattered photons at large $s$. The inelastic crossed component (filled diamonds) 
is smaller than the ladder component (filled squares)  for all values of the saturation parameter, and also exhibits a maximum as a function of $s$.

\begin{figure}
\includegraphics[width=0.5\textwidth]{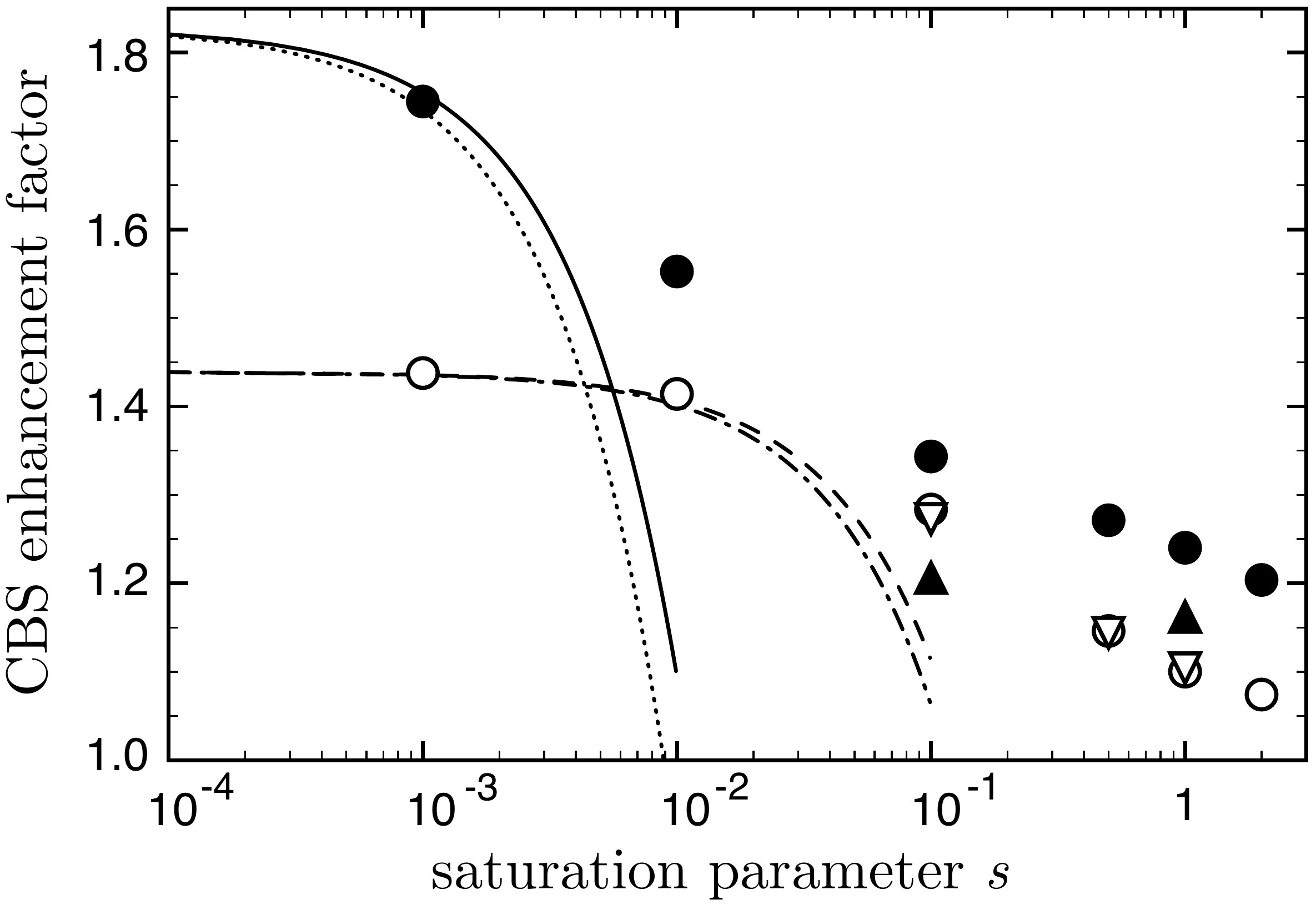}
\caption{Coherent backscattering enhancement factor $\eta=1+[\gamma_C^{({\rm el})}+\gamma_C^{({\rm in})}]/[\gamma_L^{({\rm el})}+\gamma_L^{({\rm in})}]$ as a function of saturation $s$ for the same parameters as in
 Fig.~\ref{fig:LC},
i.e., for vanishing detuning ($\delta=0$) and two different values of the thickness
$D=0.5\ell_{\rm lin}$ (open circles: numerical data, dashed line: perturbative theory) and $D=5\ell_{\rm lin}$ (filled circles and solid line).
For comparison, we also show a few data points corresponding to non-vanishing detuning:
$(D,\delta)=(0.5\ell_{\rm lin},0.5 \Gamma)$ (open triangles: numerical data for $s=0.1,0.5$ and $1$, dash-dotted line: perturbative prediction for small $s$) and
$(D,\delta)=(5\ell_{\rm lin},\Gamma)$ (filled triangles: numerical data for $s=0.1$ and $1$, dotted line: perturbative prediction).
In all cases, the coherent backscattering enhancement factor $\eta$ decreases with increasing saturation.
For $D=0.5\ell_{\rm lin}$, a detuning of $\delta=0.5\Gamma$ has almost no effect on $\eta$ (open triangles for $\delta=0.5\Gamma$ vs. open circles for $\delta=0$), whereas, for $D=5\ell_{\rm lin}$, the coherent backscattering enhancement factor for $\delta=\Gamma$ (filled triangles) is considerably smaller than for $\delta=0$ (filled circles), if the saturation is not too small ($s=0.1$ and $s=1$).
\label{fig:enh}}
\end{figure}

From the elastic and inelastic ladder and crossed components shown in Fig.~\ref{fig:LC},
we finally calculate the 
coherent backscattering enhancement factor  $\eta=1+[\gamma_C^{({\rm el})}+\gamma_C^{({\rm in})}]/[\gamma_L^{({\rm el})}+\gamma_L^{({\rm in})}]$, see Fig.~\ref{fig:enh}. In the limit $s\to 0$, the coherent backscattering enhancement factor converges to the value $\eta^{(0)}=1+\gamma_C^{(0)}/\gamma_L^{(0)}$ predicted by Eqs.~(\ref{eq:gammaL0},\ref{eq:gammaC0}). As already mentioned above, $\eta^{(0)}<2$ due to single scattering. With increasing saturation, the coherent backscattering enhancement factor $\eta$ decreases. In case of the slab with larger thickness ($D=5\ell_{\rm lin}$, filled symbols), a significant decrease of $\eta$ (from $1.83$ to $1.74$) can be observed already for $s=10^{-3}$, in agreement with the perturbative prediction (solid line). This can be traced back mainly to the decrease of the elastic crossed component, as discussed above. For $s=2$, the coherent backscattering enhancement factor has dropped to $\eta=1.20$ for the medium with larger thickness ($D=5\ell_{\rm lin}$), and to
$\eta=1.07$ for the one with smaller thickness ($D=0.5\ell_{\rm lin}$). In this case, the remaining coherent backscattering enhancement originates mainly from inelastically backscattered photons. For comparison, we also included in Fig.~\ref{fig:LC} a few data points for the case of nonvanishing detuning. For $D=0.5\ell_{\rm lin}$, a detuning of $\delta=0.5\Gamma$ has almost no effect on $\eta$ (open triangles for $\delta=0.5\Gamma$ vs. open circles for $\delta=0$), whereas, for $D=5\ell_{\rm lin}$, the coherent backscattering enhancement factor for $\delta=\Gamma$ (filled triangles) is considerably smaller than for $\delta=0$ (filled circles), if the saturation is not too small ($s=0.1$ and $s=1$).

Remember that, in the present article, we concentrate on the scalar field model. For a comparison with experimental data 
 \cite{Chaneliere:2004aa,Balik:2005aa}, it will be necessary to take into account the vectorial character of the light field: The latter makes it possible to filter out single scattering such that $\eta\to 2$ can be observed for $s\to 0$ in the case of atoms with non-degenerate ground state \cite{Bidel:2002aa}. Also nonlinear crossed scattering processes are influenced by the vectorial character. For example, the initial decrease of $\eta$ for small $s$ predicted by two-photon scattering theory is less steep than in the scalar case \cite{Wellens:2006aa}.
 
\section{Conclusion}
\label{sec:conclusions}
 
 This work was dedicated to the solution of a problem that, due to its exponential complexity, was deemed intractable: the propagation of intense laser light across a dilute, disordered ensemble of cold atoms. 
 
Our solution is based, on the one hand, on quantum optical methods which provide an accurate account of the individual atomic responses to a saturating laser field. On the other hand, it relies on diagrammatic methods whereby multiple scattering signals can be expressed in terms of single-atom responses. We developed a combination of these methods under the approximation  
that the intensities emitted from different atoms are uncorrelated with each other.
This approximation is valid for a dilute atomic medium (where the distances between atoms are larger than the wave length of the scattered light) and makes it possible to
represent the photons exchanged between the atoms by a classical field.
For a small number of atoms  ($N=2$ and $N=3$), the latter property has already been proven and applied in previous work \cite{Wellens_Phys_Rev_A_2010,Shatokhin_Phys_Rev_A_2012}, whereas the present article provides the generalization to an arbitrary number of atoms.

To achieve 
this goal,
we started
from a microscopic quantum optical master equation for $N$ laser-driven atoms exchanging photons via the far-field dipole-dipole interaction. Thereafter, we obtained a formal 
solution of the master equation in the form of a diagrammatic series, and performed the complete summation of diagrams surviving the disorder average, the so-called ladder and crossed diagrams. We thereby derived transport equations for the diffusive and coherent backscattering intensities which we solved numerically for the case of an atomic medium confined to a slab. In this way, we were able to determine the local spectral irradiance of light propagating inside the slab and to demonstrate how  
increasing the incident laser intensity leads to a broadening of the spectrum of backscattered light and to a reduction of the height of the coherent backscattering interference peak.

To reduce the technical overload, the present theory was developed for two-level atoms and scalar electromagnetic fields. However, the ideas lying at the basis of our method are equally valid for atoms with degenerate dipole transitions and for vector fields \cite{shatokhin14,Ketterer:2014aa}. The generalisation to this more realistic scenario is possible and necessary for achieving a satisfactory agreement with the experimental results 
on coherent backscattering of strong laser light by cold atoms \cite{Chaneliere:2004aa,Balik:2005aa}. In principle, we expect that our theory can be generalized to dilute media composed out of quantum mechanical scatterers with an arbitrary level structure for which the interaction of the incident field with a single scatterer can be treated by Bloch equations, e.g., atoms with three or four levels as a microscopic model for random lasing \cite{Cao:2003aa} (see also the discussion at the end of Sec.~\ref{sec:results_diffusive}), or with a $\Lambda$-type level structure suitable for electromagnetically induced transparency \cite{Fleischhauer:2000aa}. 
In these and similar cases, our quantum-optical multiple scattering approach provides the possibility to access new regimes which cannot be treated by  presently availabe theories, in particular to account for
nonlinear effects occurring at high field strengths (such as saturation or inelastic scattering induced by quantum fluctuations) in combination with multiple scattering.

\acknowledgements

We would like to acknowledge the use of the computing resources provided by bwGRiD (http://www.bw-grid.de), member of the German D-Grid initiative, funded by the Ministry for Education and Research (Bundesministerium f\"ur Bildung und Forschung) and the Ministry for Science, Research and Arts Baden-Wuerttemberg (Ministerium f\"ur Wissenschaft, Forschung und Kunst Baden-W\"urttemberg). Furthermore, 
the authors acknowledge support by the state of Baden-W\"urttemberg through bwHPC. V.N.S. and A.B. acknowledge support by the Deutsche Forschungsgemeinschaft under Grant No. DFG BU 1337/17-1. Finally, we thank Heinz-Peter Breuer for illumating discussions on master equations and the uniqueness of stationary states.

\appendix

\section{Uniqueness of the stationary state}
\label{sec:appendixstat}
To prove that the $N$-atom master equation has a unique stationary state, we first 
include the reversible part of the atom-atom interaction into an effective atomic Hamiltonian $\tilde{H}_A$
\begin{eqnarray}
\tilde{H}_A & = & -\hbar \sum_{j=1}^N\left[\delta \sigma_j^+\sigma_j^-+\frac{1}{2}(\Omega_j\sigma_j^++\Omega_j^*\sigma_j^-)\right]\nonumber\\
& & -\frac{\hbar\Gamma}{2} \sum_{j\neq k=1}^N \frac{\cos(k_L r_{jk})}{k_L r_{jk}} \sigma_j^+\sigma_k^-\label{eq:H}
\end{eqnarray}
such that Eq.~(\ref{eq:master}) turns into:
\begin{eqnarray}
\langle\dot{Q}\rangle  &  = &    \frac{i}{\hbar} \left<[\tilde{H}_A,Q]\right>+ \frac{\Gamma}{2} \sum_{j,k=1}^N W_{jk}\nonumber\\
& & \times \left<2\sigma_j^+Q\sigma_k^--\sigma_j^+\sigma_k^-Q-Q \sigma_j^+\sigma_k^-\right>\label{eq:master2}
\end{eqnarray}
with the irreversible part given by the coupling matrix:
\begin{equation}
W_{jk}=\left\{\begin{array}{cl} 1 & \text{if }j=k\\ \frac{\sin(k_Lr_{jk})}{k_Lr_{jk}} & \text{if }j\neq k\end{array}\right.\label{eq:W}
\end{equation}
Switching from the Heisenberg to the Schr\"odinger picture, we rewrite 
Eq.~(\ref{eq:master2}) as an equation for the atomic density matrix $\rho$:
\begin{equation}
\dot{\rho}   =   -\frac{i}{\hbar} [\tilde{H}_A,\rho] +
  \frac{\Gamma}{2} \sum_{j,k=1}^N W_{jk} \left(2\sigma_k^-\rho\sigma_j^+-\rho\sigma_j^+\sigma_k^--\sigma_j^+\sigma_k^-\rho\right)\label{eq:masterrho}
\end{equation}

We first show that all eigenvalues of $W$ are positive, provided that the distances between all pairs of atoms are larger than zero. For this purpose, we note that:
\begin{equation}
W_{jk}=\int\frac{{\rm d}\Omega}{4\pi}~e^{i{\bf k}_\Omega\cdot {\bf r}_j} e^{-i{\bf k}_\Omega\cdot {\bf r}_k}\label{eq:Wij}
\end{equation} 
where $\Omega$ denotes the angular variables of ${\bf k}_\Omega$, and $|{\bf k}_\Omega|=k_L$.
All eigenvalues of $W$ are positive if and only if $\sum_{jk} c_j c_k^* W_{jk}>0$ for all coefficients $(c_1,\dots,c_N)\neq (0,\dots,0)$.
Due to Eq.~(\ref{eq:Wij}), however, we have. 
\begin{equation}
\sum_{jk} c_j W_{jk} c_k^*=\int\frac{{\rm d}\Omega}{4\pi} \left|\sum_j c_j e^{i{\bf k}_\Omega\cdot{\bf r}_j}\right|^2\geq 0\label{eq:Wgeq0}
\end{equation}
Eq.~(\ref{eq:Wgeq0}) is equal to zero  if and only if
\begin{equation}
\sum_j c_j e^{i{\bf k}_\Omega\cdot{\bf r}_j}=0\label{eq:ci}
\end{equation}
for all $\Omega$. This, in turn, is possible only if there are at least two atoms $j\neq k$ with identical positions ${\bf r}_j={\bf r}_k$.
A solution of Eq.~(\ref{eq:ci}) with non-zero coefficients would then be given by $c_j=-c_k$ (and $c_i=0$ for $i\neq j,k$).
The corresponding eigenvalue $0$ of the coupling matrix $W$ then gives rise to a \lq dark state\rq, i.e., a subradiant state with infinite lifetime.
If all positions ${\bf r}_i$ differ from each other, however, the only solution of Eq.~(\ref{eq:ci}) with the same set of coefficients $c_j$ for all $\Omega$ is given by $c_j=0$ for all $j$. This implies that, in the absence of the driving laser, all atoms finally decay to the ground state with a finite (i.e. non-zero) rate.  

We can now bring the master equation (\ref{eq:masterrho}) into the diagonal form:
\begin{equation}
\dot{\rho}   =   -\frac{i}{\hbar} [\tilde{H}_A,\rho] +\sum_{j=1}^N \gamma_j \left(2a_j\rho a_j^\dagger-\rho a_j^\dagger a-a_j^\dagger a_j\rho\right)
\end{equation}
where the rates $\gamma_j>0$ are the eigenvalues of $W$ times $\Gamma/2$, and the operators $a_j$ are linear combinations of $\sigma_1^-,\dots,\sigma_N^-$ (determined by the corresponding eigenstates of $W$). As shown in \cite{Frigerio:1977aa} and \cite{Evans:1977aa},
the solution of this equation relaxes towards a unique stationary state if there exists no operator different from a multiple of the identity operator that commutes with all operators $a_j$ and $a_j^\dagger$. Since each $\sigma_j^-$ can be expressed as a linear combination of $a_1,\dots,a_N$ (due to the orthogonality and completeness of eigenstates of $W$), this, in turn, is the case if only multiples of the identity operator commute with all operators $\sigma_j^-$ and $\sigma_j^+$. Finally, the latter condition is fulfilled since, for each single atom $j$, multiples of the identity operator (in the subspace of atom $j$) are the only operators that commute with, both,   $\sigma_j^-$ and $\sigma_j^+$.

To show that the stationary state is given by Eq.~(\ref{eq:Sstat}), we first note that the condition $L\vec{S}=0$ is equivalent to:
\begin{equation}
\vec{S}=\left(\frac{1}{\epsilon-L}V+{\mathbbm 1}\right)\vec{S}_0+\frac{\epsilon}{\epsilon-L}\left(\vec{S}-\vec{S}_0\right)\label{eq:Sstatproof}
\end{equation}
as can be seen by applying $(\epsilon-L)$ to both sides of the equation, and using $L\vec{S}_0=(A+V)\vec{S}_0=V\vec{S}_0$, see Eq.~(\ref{eq:AS0}). The first term on the right-hand side of Eq.~(\ref{eq:Sstatproof}) is the same one which occurs in Eq.~(\ref{eq:Sstat}). It remains to be shown that the second term vanishes in the limit $\epsilon\to 0$. 
For this purpose, we first observe that, due to the normalization conditions for $\vec{S}$ and $\vec{S}_0$, the vector $\vec{S}-\vec{S}_0$ is orthogonal to the left-eigenvector of $L$ associated with the eigenvalue $0$, i.e., $(1,0,\dots,0)(\vec{S}-\vec{S}_0)=1-1=0$. Therefore, the eigenvalue $0$ of $L$ does not contribute in the second term on the right-hand side of Eq.~(\ref{eq:Sstatproof}). In other words, in the limit $\epsilon\to 0$, the norm of the vector $(\epsilon-L)^{-1} (\vec{S}-\vec{S}_0)$  is bounded from above, i.e. $|(\epsilon-L)^{-1} (\vec{S}-\vec{S}_0)|\leq |\epsilon-\lambda_2|^{-1} |\vec{S}-\vec{S}_0|$, where $\lambda_2$ is the second smallest singular value of $L$.  The latter is strictly larger than zero, since  the eigenvalue $0$ of  $L$  is non-degenerate due to the uniqueness of the stationary state shown above. Therefore,
\begin{equation}
\lim_{\epsilon\to 0}\left|\frac{\epsilon}{\epsilon-L}\left(\vec{S}-\vec{S}_0\right)\right|=0
\end{equation}
and Eq.~(\ref{eq:Sstat}) follows from Eq.~(\ref{eq:Sstatproof}).

\section{Derivatives with respect to probe fields}
\label{sec:deriv}

To determine the refractive index $n_\omega({\bf r})$, see Eq.~(\ref{eq:index}), and the crossed building blocks
$\tau(\omega_1,\omega_2,{\bf r})$, $\tau_{L1}(\omega_1,{\bf r})$ and $\tau_{L2}(\omega_1,{\bf r})$,
see Eqs.~(\ref{eq:tau}-\ref{eq:tauL2}), we need to calculate partial derivatives 
of $s^\pm_{\bf r}(t)$ 
with respect to small probe fields. For this purpose, let us consider a single realization of the classical stochastic field
$E^\pm(t)$ 
(representing the radiation emitted from other atoms), see Eqs.~(\ref{eq:stochastic},\ref{eq:rayleigh}), and define:
\begin{equation}
A({\bf r},t)=A({\bf r})+C^+\frac{2d}{\hbar}E^-(t)+C^-\frac{2d}{\hbar}E^+(t)
\end{equation}
Then, the single-atom Bloch vector in the presence of this stochastic field fulfills the optical Bloch equation:
\begin{equation}
\frac{{\rm d}}{{\rm d}t}\vec{s}_{\bf r}(t)=A({\bf r},t)\vec{s}_{\bf r}(t)
\end{equation}
see also Eq.~(\ref{eq:blochclassical}). Using a numerical integration routine, we first solve this equation with an arbitrary initial condition at time $t_0\ll -1/\Gamma$, such that a quasi-stationary state is reached at time $t=0$.
In a second step, we then determine  the derivative with respect to an additional probe field (with frequency $\omega$) by solving the following equation:
\begin{equation}
\frac{{\rm d}}{{\rm d}t}\frac{\partial \vec{s}_{\bf r}(t)}{\partial E_\omega^\pm(t)}=\left[A({\bf r},t)\pm i  \omega\right]\frac{\partial \vec{s}_{\bf r}(t)}{\partial E_\omega^\pm(t)}+C^{\mp} \frac{2d}{\hbar}\vec{s}_{\bf r}(t)\label{eq:deriv}
\end{equation}
with initial condition $\partial \vec{s}_{\bf r}(t_0)/\partial E_\omega^\pm(t_0)=0$.
Note that, in the absence of the stochastic field, i.e., if $E^\pm(t)=0$, the solution of Eq.~(\ref{eq:deriv}) reproduces the single-atom building blocks $s^+_{\bf r}(\omega)^{(\mp)}$ and $s^-_{\bf r}(\omega)^{(\mp)}$, see Eqs.~(\ref{eq:bbsigma},\ref{eq:sigmaclass}). For $E^\pm(t)\neq 0$, however, it is necessary to solve Eq.~(\ref{eq:deriv}) again by numerical integration.
Finally, in order to perform the ensemble average (denoted by the overbar $\overline{\phantom{xx}}^{({\rm cl})}$), these steps must be repeated for many different realizations of the
stochastic field.

Higher derivatives are obtained recursively as follows:
\begin{eqnarray}
\frac{{\rm d}}{{\rm d}t}\frac{\partial^n \vec{s}_{\bf r}(t)}{\partial E_{\omega_1}^{\alpha_1}(t)\dots\partial E_{\omega_n}^{\alpha_n}(t)} & = & \nonumber\\
& & \!\!\!\!\!\!\!\!\!\!\!\!\!\!\!\!\!\!\!\!\!\!\!\!\!\!\!\!\!\!\!\!\!\!\!\!\!\!\!\!\!\!\!\!\!\!\!\!\!\!\!\!\!\!\!\!\!\!\!\!\!\!\! \left[A({\bf r},t)+i \sum_{m=1}^n \alpha_m \omega_m\right] \frac{\partial^n \vec{s}_{\bf r}(t)}{\partial E_{\omega_1}^{\alpha_1}(t)\dots\partial E_{\omega_n}^{\alpha_n}(t)}\label{eq:deriv2}\\
& & \!\!\!\!\!\!\!\!\!\!\!\!\!\!\!\!\!\!\!\!\!\!\!\!\!\!\!\!\!\!\!\!\!\!\!\!\!\!\!\!\!\!\!\!\!\!\!\!\!\!\!\!\!\!\!\!\!\!\!\!\!\!\!\!\!\!\!\!\!\!\!\!\!\!\!\!\! +\sum_{m=1}^n
C^{-\alpha_m}\frac{2d}{\hbar} \frac{\partial^{n-1} \vec{s}_{\bf r}(t)}{\partial E_{\omega_1}^{\alpha_1}(t)\dots\partial E_{\omega_{m-1}}^{\alpha_{m-1}}(t)\partial E_{\omega_{m+1}}^{\alpha_{m+1}}(t)\dots\partial E_{\omega_{n}}^{\alpha_{n}}(t)}\nonumber
\end{eqnarray}
with initial condition $\partial^n \vec{s}_{\bf r}(t_0)/\partial E_{\omega_1}^{\alpha_1}(t_0)\dots\partial E_{\omega_n}^{\alpha_n}(t_0)=0$.
For the crossed building blocks $K(\omega_1,\omega_2,{\bf r})$ and $K_L(\omega_1,{\bf r})$, see Eqs.~(\ref{eq:K},\ref{eq:KL}), we also need the derivatives of the correlation functions $\langle \sigma^+(\tau)\sigma^-(0)\rangle$ and 
$\langle \sigma^+(0)\sigma^-(\tau)\rangle$. According to the quantum regression theorem \cite{Lax:1963aa}, these fulfill the same optical Bloch equation, i.e.,
\begin{eqnarray}
\frac{{\rm d}}{{\rm d}\tau} \langle \vec{\sigma}(\tau)\sigma^-(0)\rangle_{\bf r} & = & A({\bf r},\tau) \langle \vec{\sigma}(\tau)\sigma^-(0)\rangle_{\bf r}\label{eq:regression1}\\
\frac{{\rm d}}{{\rm d}\tau} \langle \sigma^+(0)\vec{\sigma}(\tau)\rangle_{\bf r} & = & A({\bf r},\tau) 
 \langle \sigma^+(0)\vec{\sigma}(\tau)\rangle_{\bf r}\label{eq:regression2}
\end{eqnarray}
but with a different initial condition at time $\tau=0$:
\begin{eqnarray}
\langle \vec{\sigma}(0)\sigma^-(0)\rangle_{\bf r}& = & B^- \vec{s}_{\bf r}(0)\label{eq:initial1}\\
\langle \sigma^+(0)\vec{\sigma}(0)\rangle_{\bf r} & = & B^+ \vec{s}_{\bf r}(0)\label{eq:initial2}
\end{eqnarray} 
Derivatives with respect to probe fields are calculated in the same way as above, see Eqs.~(\ref{eq:deriv},\ref{eq:deriv2}), where $\vec{s}_{\bf r}(t)$ is replaced by $\langle\vec{\sigma}(t)\sigma^-(0)\rangle_{\bf r}$ or $\langle \sigma^+(0)\vec{\sigma}(\tau)\rangle_{\bf r}$, respectively. According to Eqs.~(\ref{eq:initial1},\ref{eq:initial2}), the initial condition at time $\tau=0$ is then given by the corresponding derivative of $\vec{s}_{\bf r}(0)$.

\section{Flux conservation}
\label{sec:flux}

Within the ladder approximation valid for the case of large distances between the atoms, recurrent scattering is neglected (see Sec.~\ref{sec:laddercrossed}), such that
different scattering events can be regarded independently of each other. Flux conservation in multiple scattering then follows from the conservation of flux at each single scattering event. For linear scatterers, the latter condition is guaranteed by the optical theorem, which, in turn, leads to the relation ${\mathcal N}\sigma_{\rm tot}=1/\ell$ between the total scattering cross section $\sigma_{\rm tot}$ of a single scatterer, the density ${\mathcal N}$ of scatterers  and the mean free path $\ell$
\cite{Tiggelen:1990aa}.
Let us now generalize this expression to our case of nonlinear quantum scatterers: 

(i) The cross section $\sigma_{\rm tot}$ is defined by the total light intensity radiated by a single atom (integrated over all angles) divided by the intensity of the incident light. The intensity radiated by a single atom, in turn,  is proportional to the dipole spectrum $\overline{P_{\bf r}(\omega)}^{({\rm cl})}$, see Eq.~(\ref{eq:Pomfinal}), integrated over all frequencies $\omega$. 

(ii) The mean free path $\ell$ is identified with the effective mean free path $\ell^{({\rm tot})}$ defined by Eqs.~(\ref{eq:ellin},\ref{eq:elltot}). This is justified since the incident laser light and the different frequency components of the scattered  light do not interfere with each other. Therefore, the total incident light intensity is obtained by summing the intensities of each component, see Eqs.~(\ref{eq:Itot},\ref{eq:Iin}), and, due to scattering, each component is attenuated by a factor proportional to $1/\ell_L$ (for the laser light) or $1/\ell_\omega$ (for scattered light with frequency $\omega$),
see Eqs.~(\ref{eq:ellin},\ref{eq:elltot}).

In total, the condition expressing the conservation of flux at each single scattering event reads:
\begin{equation}
\hbar\omega_0\Gamma \int_{-\infty}^\infty{\rm d}\omega~{\mathcal P}(\omega,{\bf r})=
\frac{\overline{I^{({\rm tot})}({\bf r})}}{\ell^{({\rm tot})}({\bf r})}\label{eq:b1}
\end{equation}
We first show that Eq.~(\ref{eq:b1}) is indeed fulfilled by our ladder transport equations.
Due to $\sigma^+\sigma^-=(1+\sigma^z)/2$, we get
\begin{equation}
\int_{-\infty}^\infty{\rm d}\omega~{\mathcal P}(\omega,{\bf r}) = {\mathcal N}({\bf r})\frac{1+\overline{s_{\bf r}^z(t)}^{({\rm cl})}}{2}\label{eq:b2}
\end{equation}
On the other hand, the expressions (\ref{eq:index},\ref{eq:ellom},\ref{eq:nL}) for the mean free paths $\ell_\omega$ and $\ell_L$ yield:
\begin{eqnarray}
\frac{\overline{I^{({\rm tot})}({\bf r})}}{\ell^{({\rm tot})}({\bf r})}& =& 2d \omega_L{\mathcal N}({\bf r})
{\rm Im}\Biggl\{\overline{s_{\bf r}^-(t)}^{({\rm cl})}E_L^-({\bf r})+\Biggr.\nonumber\\
& & + \frac{1}{2\epsilon_0c}\Biggl.\int_{-\infty}^\infty{\rm d}\omega \frac{\overline{\partial s_{\bf r}^-(t)}^{({\rm cl})}}{\partial E^+_\omega(t)} \overline{I(\omega,{\bf r})}\Biggr\}
\end{eqnarray}
In the quasi-stationary regime (see Sec.~\ref{sec:bloch}), averages over the stochastic classical field (see Sec.~\ref{sec:classical}) are time-independent, in particular $\overline{\dot{s}_{\bf r}^z(t)}^{({\rm cl})}=0$. From the single-atom Bloch equation, see Eq.~(\ref{eq:blochclassical}), we therefore deduce:
\begin{equation}
\Gamma \frac{\overline{s_{\bf r}^z(t)}^{({\rm cl})}+1}{2} = \frac{2d}{\hbar}{\rm Im} \left\{\overline{s_{\bf r}^-(t) \left(E_L^-({\bf r})+E^-(t)\right)}^{({\rm cl})}\right\}
\end{equation}
The laser amplitude $E_L^-({\bf r})$ is a non-fluctuating quantity and therefore can be taken out of the classical field average. To average the product $s_{\bf r}^-(t) E^-(t)$, we represent the field $E^-(t)$ as a sum over many components $E_{jk}^-(t)$ with random phases, see Eqs.~(\ref{eq:stochastic1},\ref{eq:components}).
Since the phase of $E_{jk}^-(t)$ can be compensated only by $E_{jk}^+(t)$, we obtain 
for a small field  $E_{jk}^+(t)$:
\begin{equation}
\overline{s_{\bf r}^-(t) E_{jk}^-(t)}^{({\rm cl})}=\overline{\frac{\partial s_{\bf r}^-(t)}{\partial E^+_{jk}(t)}}^{({\rm cl})} |E_{jk}|^2\label{eq:b5}
\end{equation}
Taking into account Eqs.~(\ref{eq:stochastic1},\ref{eq:components}), Eq.~(\ref{eq:b1}) follows from Eqs.~(\ref{eq:b2}-\ref{eq:b5}). Finally, from the definition of $\gamma_L(\omega,{\bf e}_D)$, see Eq.~(\ref{eq:IDfinal}), and the bistatic coefficient $\gamma_L({\bf e}_D)=\int{\rm d}\omega~\gamma_L(\omega,{\bf e}_D)$, together with the ladder transport equations (\ref{eq:Ibarin},\ref{eq:Ibarel}), it is possible to derive the following equation:
\begin{equation}
\int\frac{{\rm d}{\bf e}_D}{4\pi}\gamma_L({\bf e}_D)+\int_{\mathcal A}\frac{{\rm d}x{\rm d}y}{\mathcal A}
\frac{I_L(x,y,L)}{I_L}=1
\end{equation}
expressing flux conservation of multiply scattered light. The two terms on the left-hand side 
represent the flux of scattered light (with integral over the two angles characterizing the outgoing direction ${\bf e}_D$) 
and the flux of coherently transmitted light (where ${\bf e}_L$ is parallel to the $z$-axis, and $L$ is chosen such that $z\leq L$ for all points $(x,y,z)\in V$ inside the scattering volume $V$), respectively. Their sum equals the normalized flux of the incident laser through the transverse area $\mathcal A$ (defined by the projection of $V$ onto the $xy$-plane).

\section{Expressions for the crossed building blocks ${\mathcal K}$ and $\tilde{\mathcal K}$}
\label{sec:crossedbb}

The diagrammatic representation shown in Fig.~\ref{fig:crossedbb2} leads to the following expressions for the crossed building blocks ${\mathcal K}$ and $\tilde{\mathcal K}$:
\begin{widetext}
\begin{eqnarray}
{\mathcal K}(\omega_1,\omega_2,{\bf r}) & = & 
\frac{\hbar^2{\mathcal N}({\bf r})}{4d^2}\int_0^\infty\frac{{\rm d}\tau}{2\pi}\left[
e^{-i\omega_1\tau}\overline{\frac{\partial^2 \Bigl(\langle\sigma^+(\tau)\sigma^-(0)\rangle_{\bf r}-s_{\bf r}^+(\tau)s_{\bf r}^-(0)\Bigr)}{\partial E_{\omega_2}^+(\tau)\partial E_{\omega-\omega_1}^-(\tau)}}^{({\rm cl})}
\right.\nonumber\\ 
& &  
\ \ \ \ \ \ +e^{i(\omega-\omega_2)\tau}\overline{\frac{\partial^2 \Bigl(\langle\sigma^+(0)\sigma^-(\tau)\rangle_{\bf r}-s_{\bf r}^+(0)s_{\bf r}^-(\tau)\Bigr)}{\partial E_{\omega_2}^+(\tau)\partial E_{\omega-\omega_1}^-(\tau)}}^{({\rm cl})}
\nonumber\\
& & \ \ \ \ \  + \left.e^{-i\omega_1\tau}
\overline{\left(
 \frac{\partial^2 s_{\bf r}^+(\tau)}{\partial E^+_{\omega_2}(\tau)\partial E^-_{\omega-\omega_1}(\tau)}\right)
 s^-_{\bf r}(0)}^{({\rm cl})}+
e^{i\omega_1\tau} 
 \overline{
 \left(\frac{\partial^2 s_{\bf r}^+(0)}{\partial E^+_{\omega_2}(0)\partial E^-_{\omega-\omega_1}(0)}\right)s_{\bf r}^-(\tau)}^{\,({\rm cl})}
 \right.\nonumber\\
& & \ \ \ \ \ \left.+
e^{i(\omega_2-\omega_1)\tau}
\overline{\left(\frac{\partial s_{\bf r}^+(\tau)}{\partial E^-_{\omega-\omega_1}(\tau)}\right)\left(\frac{\partial s^-_{\bf r}(0)}{\partial E^+_{\omega_2}(0)}\right)}^{({\rm cl})}
+e^{i(\omega_1-\omega_2)\tau} 
 \overline{\left(\frac{\partial s_{\bf r}^+(0)}{\partial E^-_{\omega-\omega_1}(0)}\right)\left(\frac{\partial s_{\bf r}^-(\tau)}{\partial E^+_{\omega_2}(\tau)}\right)}^{\,({\rm cl})} \right]
\label{eq:calK}
\end{eqnarray}
\begin{eqnarray}
\tilde{\mathcal K}(\omega_1,\omega_2,{\bf r}) & = & 
\frac{\hbar^2{\mathcal N}({\bf r})}{4d^2}\int_0^\infty\frac{{\rm d}\tau}{2\pi}\left[
e^{-i\omega\tau}
\overline{\left(
 \frac{\partial s_{\bf r}^+(\tau)}{\partial E^+_{\omega_2}(\tau)}\right)
 \left(\frac{\partial
 s^-_{\bf r}(0)}{\partial E^-_{\omega-\omega_1}(0)}\right)}^{({\rm cl})}+
e^{i\omega\tau} 
 \overline{
 \left(\frac{\partial s_{\bf r}^+(0)}{\partial E^+_{\omega_2}(0)}\right)\left(\frac{\partial s_{\bf r}^-(\tau)}{\partial E^-_{\omega-\omega_1}(\tau)}\right)}^{\,({\rm cl})}
 \right.\nonumber\\
& & \ \ \ \ \ \left.+
e^{i(\omega_2-\omega)\tau}
\overline{s_{\bf r}^+(\tau)\left(\frac{\partial^2 s^-_{\bf r}(0)}{\partial E^+_{\omega_2}(0)\partial E^-_{\omega-\omega_1}(0)}\right)}^{({\rm cl})}
+e^{i(\omega-\omega_2)\tau} 
 \overline{s_{\bf r}^+(0)\left(\frac{\partial^2 s_{\bf r}^-(\tau)}{\partial E^+_{\omega_2}(\tau)\partial E^-_{\omega-\omega_1}(\tau)}\right)}^{\,({\rm cl})} \right]
\label{eq:calKtilde}
\end{eqnarray}
\begin{eqnarray}
{\mathcal K}_L(\omega_1,{\bf r}) & = & 
\frac{\hbar{\mathcal N}({\bf r})}{2dE_L^+({\bf r})}\int_0^\infty\frac{{\rm d}\tau}{2\pi}\left[
e^{-i\omega_1\tau}\frac{\overline{\partial\Bigl(\langle\sigma^+(\tau)\sigma^-(0)\rangle_{\bf r}-s_{\bf r}^+(\tau)s_{\bf r}^-(0)\Bigr)}^{({\rm cl})}}{\partial E_{\omega-\omega_1}^-(\tau)}+
e^{i\omega\tau}\frac{\overline{\partial\Bigl(\langle\sigma^+(0)\sigma^-(\tau)\rangle_{\bf r}-s_{\bf r}^+(0)s_{\bf r}^-(\tau)\Bigr)}^{({\rm cl})}}{\partial E_{\omega-\omega_1}^-(\tau)}
\right.\nonumber\\
& & \ \ \ \ \ \ \ \ \ \ \ \ \left.+e^{-i\omega_1\tau}
\overline{\left(\frac{\partial s^+_{\bf r}(\tau)}{\partial E^-_{\omega-\omega_1}(\tau)}\right)s_{\bf r}^-(0)}^{({\rm cl})}+e^{i\omega_1\tau} 
 \overline{ 
\left( \frac{\partial s_{\bf r}^+(0)}{\partial E^-_{\omega-\omega_1}(0)}\right)s_{\bf r}^-(\tau)}^{\,({\rm cl})} \right]
\label{eq:calKL}\\
\tilde{\mathcal K}_L(\omega_1,{\bf r}) & = & 
\frac{\hbar{\mathcal N}}{2dE_L^+({\bf r})}\int_0^\infty\frac{{\rm d}\tau}{2\pi}\left[
e^{-i\omega\tau}
\overline{s_{\bf r}^+(\tau)\left(\frac{\partial s^-_{\bf r}(0)}{\partial E^-_{\omega-\omega_1}(0)}\right)}^{({\rm cl})}+e^{i\omega\tau} 
 \overline{ s_{\bf r}^+(0)
\left( \frac{\partial s_{\bf r}^-(\tau)}{\partial E^-_{\omega-\omega_1}(\tau)}\right)}^{\,({\rm cl})} \right]
\label{eq:calKLtilde}
\end{eqnarray}
\end{widetext}
To obtain the correct frequencies in the exponential factors, we must take into account that
$E^+_{\omega_2}(\tau)=e^{-i\omega_2\tau}E^+_{\omega_2}(0)$ and $E^-_{\omega-\omega_1}(\tau)=e^{i(\omega-\omega_1)\tau}E^-_{\omega-\omega_1}(0)$,
which implies a shift of frequency if a derivative with respect to a probe field is evaluated at time $0$ instead of time $\tau$.

\bibliography{nonlinear_QT}

\end{document}